\documentclass[a4page]{article}
\pdfoutput=1
\usepackage[singlespacing]{setspace}
\usepackage{float}
\usepackage{fullpage}
\usepackage{titlesec}
\usepackage{wrapfig}
\usepackage{rotating}
\usepackage{amsfonts}
\usepackage{amsmath}
\usepackage{url}
\usepackage[affil-it]{authblk}
\usepackage{lscape}
\usepackage{graphicx}
\usepackage{cite}
\usepackage{chngcntr}
\usepackage[all]{nowidow}
\usepackage{notoccite}
\usepackage[binary-units]{siunitx}
\usepackage{upgreek}
\usepackage{tikz}
\usetikzlibrary{positioning}
\usepackage{feynman}
\usepackage{epstopdf}
\usepackage{mathtools}
\usepackage[utf8]{inputenc}
\usepackage{textcomp}
\usepackage{hyperref}
\usepackage{bm}
\usepackage{subcaption}
\usepackage[nottoc,numbib]{tocbibind}
\usepackage{shorttoc}
\usepackage[titletoc]{appendix}

\numberwithin{figure}{section}
\numberwithin{equation}{section}
\numberwithin{table}{section}

\graphicspath{ {images/} }
\titleformat*{\subparagraph}{\bfseries\itshape}
\tikzset{basic/.style={draw,fill=blue!20,text width=1em,text badly centered}}
\tikzset{input/.style={basic,circle}}
\tikzset{weights/.style={basic,rectangle}}
\tikzset{functions/.style={basic,circle,fill=blue!10}}
\newcommand{\LiZnS}{$\prescript{6}{}{\mathrm{LiF:ZnS(Ag)}}$}
\newcommand{\nue}{\bar{\nu}_e}

\title{
	  \includegraphics[width=0.595\textwidth]{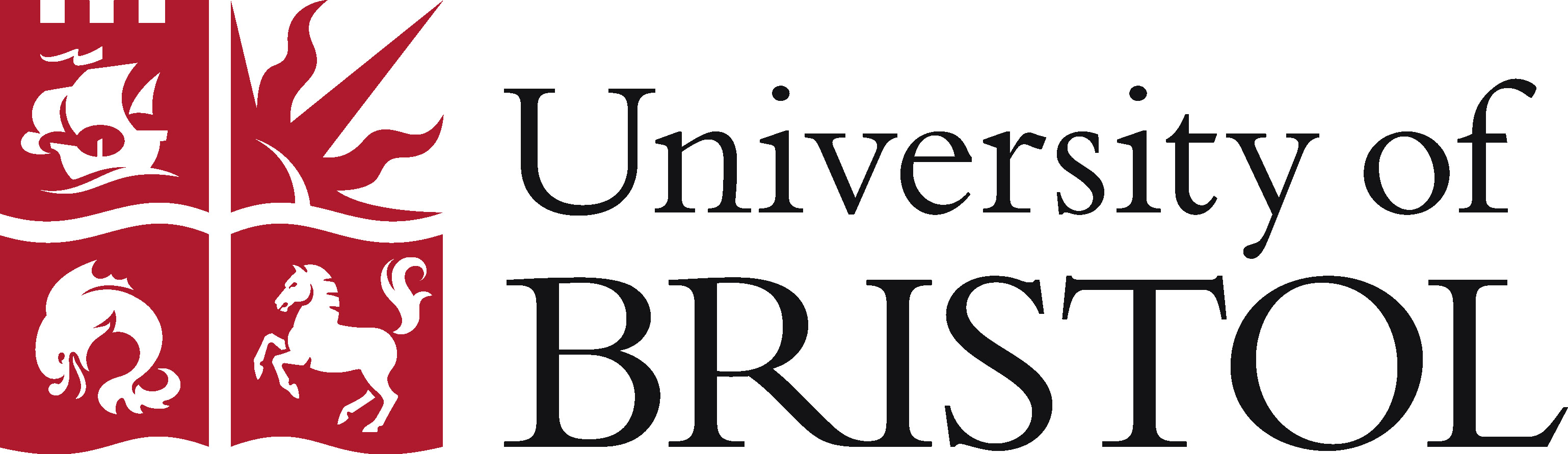}\\\vspace{1cm}
\Huge{Development \& Implementation of the Trigger for a Short-baseline Reactor Antineutrino Experiment (SoLid)}
\vspace{3cm}}
\author{\LARGE{Lukas On Arnold}
\vspace{4cm}

\Large{A dissertation submitted to the University of Bristol in accordance with the requirements for award of the degree of Master of Science by Research in Physics in the Faculty of Science in March 2017}\vspace{2cm}}
\date{\hspace{12cm}\raggedleft{$\sim20000$  words}
}

\begin{document}
\maketitle
\thispagestyle{empty}

\clearpage
\onehalfspacing
\begin{abstract}
SoLid, located at SCK-CEN in Mol, Belgium, is a reactor antineutrino experiment at a very short baseline of 5.5 -- 10m aiming at the search for sterile neutrinos and for high precision measurement of the neutrino energy spectrum of Uranium-235. It uses a novel approach using Lithium-6 sheets and PVT cubes as scintillators for tagging the Inverse Beta-Decay products (neutron and positron).

Being located overground and close to the BR2 research reactor, the experiment faces a large amount of backgrounds.  Efficient real-time background and noise rejection is essential in order to increase the signal-background ratio for precise oscillation measurement and decrease data production to a rate which can be handled by the online software. Therefore, a reliable distinction between the neutrons and background signals is crucial. This can be performed online with a dedicated firmware trigger. A peak counting algorithm and an algorithm measuring time over threshold have been identified as performing well both in terms of efficiency and fake rate, and have been implemented onto an FPGA.

After having introduced the experimental and theoretical background of neutrino oscillation physics, as well as SoLid's detector technology, read-out system and trigger scheme, the thesis presents the design of the firmware neutron trigger implemented by applying machine learning methods.
\end{abstract}
\clearpage

\renewcommand{\abstractname}{Acknowledgements}
\begin{abstract}
I would like to thank my supervisor, Dave Newbold, whose expertise and support in the field of trigger and firmware development has been of essential value in conducting the present research.\\
I also would like to express my thanks to David Cussans whose advice and competence in electronics and detector technology has been of greatest benefit.\\
I want to thank my other collaborators on SoLid, namely Nick Ryder, Dan Saunders, Steve Manley, Kostas Petridis, Jonas Rademacker, Sakari Ihantola, Antonin Vacheret, Simon Vercaemer and Yamiel Abreu, as well as Jules Desjardin, who was of great help as a summer student.\\
I also would like to acknowledge the scholarship provider, \textsc{Z\"{u}hlke AG}.
 \end{abstract}
\clearpage

\renewcommand{\abstractname}{Author's declaration}
\begin{abstract}
I declare that the work in this dissertation was carried out in accordance with the requirements of the University's Regulations and Code of Practice for Research Degree Programmes and that it has not been submitted for any other academic award. Except where indicated by specific reference in the text, the work is the candidate's own work. Work done in collaboration with, or with the assistance of, others, is indicated as such. Any views expressed in the dissertation are those of the author.
 \newline
 \newline
 
Signed:\\
\rule[1em]{25em}{0.5pt}  
 
Date:\\
\rule[1em]{25em}{0.5pt}  
 \end{abstract}
\clearpage
\shorttoc{Contents}{1} 
\clearpage
\tableofcontents
\clearpage
\listoffigures
\clearpage
\listoftables
\doublespacing
\clearpage

\section{Introduction}
The research presented in this dissertation covers two fields, both of vital and substantial interest for particle physics and beyond: Neutrino Physics, and Triggering, or, as an engineer might call it, real-time classification.

Recent discoveries in neutrino physics have not only advanced the knowledge on neutrinos themselves, but have also raised questions on physics beyond the current state of the Standard Model. One of these questions is what SoLid aims to answer: Might there be a fourth type of neutrino, a type not yet observed, and not included within the three flavours of the Standard Model? A type which does not interact weakly with any of the leptons? A type, however, which could give explanation to recently discovered anomalies?  Might there be a \textit{sterile} neutrino?

SoLid -- short for \textbf{S}earch for \textbf{o}scillations with a \textbf{Li}thium-6 \textbf{d}etector -- aims to answer these questions by measuring reactor neutrinos at a very short distance from the nuclear core. It employs a novel technique by using solid Lithium-6 as the neutron detector material and  provides a high degree of granularity. 

However, being located overground in the very close proximity of a reactor core means that a large amount of background signals is faced. In fact, the data produced by background processes exceed the capability of storage and software by several orders of magnitude. It is therefore a crucial task to reduce the data to what are possibly meaningful data. This is achieved by the \textit{trigger}: Signals are classified as either possibly meaningful or not, and sent for further processing only if they are potentially useful.

The decision -- or classification -- whether or not a set of data can be considered as potentially useful has to be made as data arrive, i.e. in real time. Field-Programmable Gate Arrays (FPGAs), re-configurable digital microelectronics devices, are ideal for the trigger to be implemented onto as they are programmable, fast and cover a wide range of application.

Designing and implementing the trigger is the research topic of the present thesis. This includes the creation of feature extraction algorithms, their evaluation, as well as their implementation on an FPGA.

From the methodological point of view, the decision on the feature is based on the evaluation of the classification performance of each feature in terms of efficiency and fake rate. The implementability into an FPGA in terms of logic resources and power consumption has to be taken into consideration, as these form limiting factors on the algorithm's complexity.

The field of triggering can be considered as essential not only for this particular experiment, but for particle physics in general. As particle physics experiments grow bigger, reach higher resolution and better sensitivity, their limits are often set by their capability to handle very high data rates, rather than by their sensor technology.

Also looking on a wider level beyond particle physics, triggers -- or real-time classifiers -- will get more important as well, as data volume is massively increasing by the societal change to a data-driven mode of production.


The dissertation gives an introduction to theoretical and experimental considerations of neutrinos and neutrino oscillations in general, and reactor neutrinos and sterile neutrinos in particular in section~\ref{sec:sterileneutrinos}. Other applications than the search for sterile neutrinos, namely investigation of the $5\si{\mega\electronvolt}$ excess and non-proliferation applications, are discussed in subsection~\ref{sec:otherapplications}. Section~\ref{sec:detector} explains SoLid's detector technology and its deployment, and the read-out system including firmware is described in section~\ref{sec:readout}. Section~\ref{sec:trigger} -- the main part and representing the research carried out by me -- discusses methodology, concept, design and implementation of SoLid's trigger system extensively. A conclusion is made in section~\ref{sec:conclusion}.

\clearpage
\section{Searching for Sterile Neutrinos}
\label{sec:sterileneutrinos}
After a theoretical overview on neutrinos in and beyond the Standard Model, the methods and history of neutrino oscillation experiments are discussed, with an emphasis on reactor antineutrino experiments. Finally the question is answered: Which anomalies have been observed in these experiments, and for what reason do they suggest the existence of \textit{sterile neutrinos}?

\subsection{Neutrinos in the Standard Model}
\label{sec:sm}
The Standard Model (SM) gives a very accurate description of elementary particles -- point-like particles without substructure -- and their interactions. The SM divides particles into matter particles that are fermions with half-integer spin $\frac{1}{2}$, and gauge bosons with integer spin $1$. SM particles gain mass through the interaction with the Higgs field, whose excitation is the Higgs boson. Fermions are classified into quarks and leptons. Leptons ($\ell$) comprise electrons ($e^-$), muons ($\mu^-$), and taus ($\tau^-$) with charge $Q=-1$, and their corresponding neutrinos ($\nu_e, \nu_\mu$, $\nu_\tau$), with no charge. An overview of the leptons is given on table~\ref{tab:leptons}. Generally, antiparticles with the same mass, but opposite charges to the respective particles are assumed for all fermions. \cite{Agashe:2014kda} \cite{Langacker:2009my} \cite{Herrero:1998eq} 

\begin{table}[htbp]
\centering
\begin{tabular}{lllll}
\hline\hline
\textbf{Name} & \textbf{Symbol} & \textbf{Mass $[\si{\mega\electronvolt\per\square\clight}]$} & \textbf{Charge} $[\si{\elementarycharge}]$ & \textbf{Weak Isospin} \\ \hline\hline
Electron & $e^-$ & $0.510998928\pm(1.1\times 10^{-8})$&$-1$&$-1/{2}$ \\
Electron neutrino & $\nu_e$ & $\neq 0,<0.002$&$0$&$+1/{2}$ \\ \hline
Muon & $\mu^-$ & $105.6583715\pm(3.5\times 10^{-6})$&$-1$&$-1/{2}$ \\
Muon neutrino & $\nu_\mu$ & $\neq 0, <0.002$&$0$&$+{1}/{2}$ \\ \hline
Tau & $\tau^-$ & $1776.82 \pm(0.16)$&$-1$&$-{1}/{2}$ \\
Tau neutrino & $\nu_\tau$ & $\neq 0, <0.002$&$0$&$+{1}/{2}$ \\
\hline\hline
\end{tabular}
\caption{Overview of properties of leptons. \textit{Source:} \cite{Agashe:2014kda} \cite{Fukuda:1998fd}.}
\label{tab:leptons}
\end{table}
Neutrinos are outstanding particles in many ways: Not only do they have a mass that is a tiny fraction of that of other fermions, they also interact only weakly, coupling solely to the weak bosons ($W$ and $Z$ bosons). Despite the fact that they occur abundantly -- as an example, the cosmological neutrino density is $336 \nu\si{\per\cubic\centi\meter}$ \cite{Zhang:2015wua} -- they are elusive: The only fundamental forces neutrinos are prone to are weak interaction and gravitation, which makes them challenging to detect.

There are two types of neutrino interactions: Coupling to a $Z^0$ boson  which leaves their identity unchanged, but changes their 4-momentum, and coupling to a $W^\pm$ boson. The flavour of the neutrino will be the same type as the charged lepton that is connected to the reaction. Negatively charged leptons and neutrinos have lepton number $+1$, while positively charged leptons and antineutrinos have lepton number $-1$; all other particles have lepton number 0. At all interactions in the SM, lepton number is conserved.

This should be illustrated by the following example of $W$ decay reactions:
\begin{equation}
W^+\rightarrow \ell^+ + \nu_\ell
\label{eq:Wminus}
\end{equation}
and
\begin{equation}
W^-\rightarrow \ell^- + \bar{\nu}_\ell,
\label{eq:Wplus}
\end{equation}
with $\ell\in\{e,\mu,\tau\}$. \cite{Lipari:2001is} 

What is the difference between neutrinos and antineutrinos apart from their mode of production? According to the classical Standard Model, neutrinos have their spin anti-parallel to their momentum, i.e. have negative helicity and are \textit{left-handed}, while antineutrinos have a parallel spin, i.e. have positive helicity and are \textit{right-handed}. \cite{Dobrynina:2016rwy} \cite{Goldhaber:1958nb} This, however, is not entirely correct, as will be discussed in section~\ref{sec:sterileneutrino}.

The Standard Model until recently assumed zero mass for neutrinos, and assumed no neutrino flavour transitions \cite{King:2008vg}; this has been proven wrong in the not-too-distant past. 


\subsection{Neutrinos beyond the Standard Model}
\label{sec:osc}
A remarkable physical phenomenon, originally predicted by Bruno Pontecorvo in 1967 \cite{Pontecorvo:1967fh}, was first confirmed at the SNO \cite{Ahmad:2002jz} and Super-Kamiokande \cite{Fukuda:1998mi} experiments in the early 2000s: neutrino flavour oscillations. As these require non-zero neutrino mass, as opposed to assumptions of the Standard Model, they are considered as going beyond the Standard Model \cite{King:2008vg}.
\subsubsection{Mass Mixing}
Neutrinos produced in association with a charged lepton do not have a well-defined mass, but are a linear combination of mass eigenstates $\nu_i$:
\begin{equation}
\begin{pmatrix}\nu_e \\\nu_\mu \\ \nu_\tau \end{pmatrix}= U^* \begin{pmatrix}\nu_1 \\\nu_2 \\ \nu_3 \end{pmatrix},
\label{eq:massmixing}
\end{equation}
or
\begin{equation}
\left|\nu _{\ell}\right\rangle =\sum _{i}U_{\ell i}^{*}\left|\nu _{i}\right\rangle \,
\label{eq:massmixing2}
\end{equation}
for neutrinos and
\begin{equation}
\begin{pmatrix}\bar{\nu}_e \\ \bar{\nu}_\mu \\ \bar{\nu}_\tau \end{pmatrix}= U \begin{pmatrix}\bar{\nu}_1 \\\bar{\nu}_2 \\ \bar{\nu}_3 \end{pmatrix},
\end{equation}
or
\begin{equation}
\left|\bar{\nu} _{\ell}\right\rangle =\sum _{i}U_{\ell i}\left|\bar{\nu} _{i}\right\rangle \,
\end{equation}
for antineutrinos, where $U$ is the Pontecorvo–Maki–Nakagawa–Sakata (PMNS) matrix \cite{Maki:1962mu}. The PMNS matrix can be seen as analogous for neutrino mixing to the Cabibbo–Kobayashi–Maskawa (CKM) matrix \cite{Kobayashi:1973fv} for quark mixing. If neutrinos are their own antiparticles, i.e. Majorana particles, $U$ is commonly expressed in the three-flavour neutrino model as \cite{EIDELMAN20041}:
\begin{equation}
{\begin{aligned}U&={\begin{pmatrix}U_{e1}&U_{e2}&U_{e3}\\U_{\mu 1}&U_{\mu 2}&U_{\mu 3}\\U_{\tau 1}&U_{\tau 2}&U_{\tau 3}\end{pmatrix}}\\&={\begin{pmatrix}1&0&0\\0&c_{23}&s_{23}\\0&-s_{23}&c_{23}\end{pmatrix}}{\begin{pmatrix}c_{13}&0&s_{13}e^{-i\delta }\\0&1&0\\-s_{13}e^{i\delta }&0&c_{13}\end{pmatrix}}{\begin{pmatrix}c_{12}&s_{12}&0\\-s_{12}&c_{12}&0\\0&0&1\end{pmatrix}}{\begin{pmatrix}e^{i\alpha _{1}/2}&0&0\\0&e^{i\alpha _{2}/2}&0\\0&0&1\\\end{pmatrix}}\\&={\begin{pmatrix}c_{12}c_{13}&s_{12}c_{13}&s_{13}e^{-i\delta }\\-s_{12}c_{23}-c_{12}s_{23}s_{13}e^{i\delta }&c_{12}c_{23}-s_{12}s_{23}s_{13}e^{i\delta }&s_{23}c_{13}\\s_{12}s_{23}-c_{12}c_{23}s_{13}e^{i\delta }&-c_{12}s_{23}-s_{12}c_{23}s_{13}e^{i\delta }&c_{23}c_{13}\end{pmatrix}}{\begin{pmatrix}e^{i\alpha _{1}/2}&0&0\\0&e^{i\alpha _{2}/2}&0\\0&0&1\\\end{pmatrix}}\end{aligned}},
\label{eq:pmns}
\end{equation}
 where $c_{ij} = \cos(\theta_{ij})$ and $s_{ij} = \sin(\theta_{ij})$. If neutrinos are Dirac particles, i.e. neutrinos and antineutrinos are different \cite{Czakon:1999ed}, this would simplify to \cite{xing}
\begin{equation}
{\begin{aligned}U&={{\begin{pmatrix}c_{12}c_{13}&s_{12}c_{13}&s_{13}e^{-i\delta }\\-s_{12}c_{23}-c_{12}s_{23}s_{13}e^{i\delta }&c_{12}c_{23}-s_{12}s_{23}s_{13}e^{i\delta }&s_{23}c_{13}\\s_{12}s_{23}-c_{12}c_{23}s_{13}e^{i\delta }&-c_{12}s_{23}-s_{12}c_{23}s_{13}e^{i\delta }&c_{23}c_{13}\end{pmatrix}}}\end{aligned}},
\label{eq:dirac}
\end{equation}
as the matrix containing the phase factors $\alpha_1$ and $\alpha_2$ in equation~\ref{eq:pmns} can be left out, since the neutrino's left- and right-handed fields are not correlated in this case. The matrix in equation~\ref{eq:dirac} is unitary. Whether or not neutrinos are Majorana particles or Dirac fermions is impossible to determine based on past and current experiments \cite{Olive:2016xmw}.

The PMNS matrix can be parameterised using three mixing angles, $\theta_{12}$, $\theta_{13}$ and $\theta_{23}$, the CP-violating phase $\delta_{CP}$, and, if neutrinos are Majorana-neutrinos, the phase factors $\alpha_1$ and $\alpha_2$. In addition, parameters for the mass eigenstates are relevant, namely \cite{Capozzi2016218}
\begin{equation}
\Delta m^2=m^2_3-\frac{m^2_1+m^2_2}{2},
\end{equation}
which in case it is positive would indicate a mass hierarchy $m_1<m_2<m_3$ and in case it is negative indicates $m_3<m_1<m_2$, and the mass splitting factors
\begin{equation}
\Delta m^2_{ij} = (m_i-m_j)^2.
\end{equation}
$\Delta m_{21}$ is most often fit for in three-flavour mixing experiments. 
It is these parameters that neutrino experiments aim to measure. Current data provide evidence that the masses $m_1$, $m_2$ and $m_3$ are light and different from each other \cite{Olive:2016xmw}.

\subsubsection{Neutrino Oscillations}
\label{sec:nosc}
The propagation of neutrinos can be described by
\begin{equation}
|\nu _{i}(t)\rangle =U^*_{\ell i}e^{-i(E_{i}t-{\vec {p}}_{i}\cdot {\vec {x}})}\mid \nu _{i}(0)\rangle ,
\label{eq:prop}
\end{equation}
where $E_i$ and $\vec{p}_i$ are the energy and momentum associated with the mass eigenstate $i$, $t$ denotes time and $\vec{x}$ location. Natural units $\si{\clight}=\hbar=1$ are used. As neutrinos have very small masses, taking
\begin{equation}
E_{i}={\sqrt {p_{i}^{2}+m_{i}^{2}}}\sim p_{i}+{\frac {m_{i}^{2}}{2p_{i}}}\sim E+{\frac {m_{i}^{2}}{2E}} 
\end{equation}
and equating time $t$ to baseline $L$ (since $v\sim \si{\clight}$), equation~\ref{eq:prop} can be approximated by
\begin{equation}
|\nu _{i}(L)\rangle =U^*_{\ell i}e^{-i\cdot m_{i}^{2}L/2E}|\nu _{i}(0)\rangle .
\end{equation}
The probability of flavour transition between two flavour eigenstates $\alpha$ and $\beta$ then is
\begin{equation}
{\displaystyle P(\nu_\alpha \rightarrow \nu_\beta)=\left|\left\langle \nu _{\beta }(L)|\nu _{\alpha }\right\rangle \right|^{2}=\left|U_{\alpha i}^{*}U_{\beta i}e^{-im_{i}^{2}L/2E}\right|^{2},}
\label{eq:trans}
\end{equation}
also commonly written as \cite{giuntibook}
\begin{equation}
{\begin{aligned}P(\nu_\alpha \rightarrow \nu_\beta)=\delta _{\alpha \beta }&{}-4\sum _{i>j}{\Re}(U_{\alpha i}^{*}U_{\beta i}U_{\alpha j}U_{\beta j}^{*})\sin ^{2}\left({\frac {\Delta m_{ij}^{2}L}{4E}}\right)\\&{}+2\sum _{i>j}{\Im}(U_{\alpha i}^{*}U_{\beta i}U_{\alpha j}U_{\beta j}^{*})\sin \left({\frac {\Delta m_{ij}^{2}L}{2E}}\right).\end{aligned}}
\label{eq:ImRe}
\end{equation}
Applying equation~\ref{eq:trans} to a system of only two neutrino flavours, $U$ in equation~\ref{eq:dirac} reduces to a simple rotation matrix
\begin{equation}
U={\begin{pmatrix}\cos \theta &\sin \theta \\-\sin \theta &\cos \theta \end{pmatrix}},
\end{equation}
leading to the probability function
\begin{equation}
{\displaystyle P(\nu_\alpha \rightarrow \nu_\beta)=\sin ^{2}(2\theta )\sin ^{2}\left({\frac {\Delta m^{2}L}{4E}}\right)}.
\end{equation}
(All equations from or derived from \cite{kopp}, if not indicated otherwise.)
\subsection{Neutrino Experiments}
\label{sec:neutrinoexperiments}
\begin{table}[htbp]
\centering
\begin{tabular}{llll}
\hline\hline
\textbf{Source} & \textbf{Type of $\nu$} & \textbf{${E} [\si{\mega\electronvolt}]$} & \textbf{$L [\si{\meter}]$}  \\ \hline
Sun & $\nu_e$  & $10^0$ & $1.5 \times 10^{11}$ \\
Atmospheric & $\nu_e, \nu_\mu, \bar{\nu}_e, \bar{\nu}_\mu$  & $10^3$ & $10^7$ \\
Accelerator & $\nu_\mu,\bar{\nu}_\mu$& $10^3$ & $10^3\ldots 10^6$ \\
Reactor & $\nue$ & $10^0$ & $10^1\ldots 10^5$ \\
\hline\hline
\end{tabular}
\caption{Properties of neutrinos by source, as orders of magnitude. \textit{Source:} \cite{Olive:2016xmw} .}
\label{tab:etypes}
\end{table}
Neutrino experiments can be divided according to neutrino source -- emitting neutrinos of different energy scales -- and baseline $L$ into \textit{solar}, \textit{atmospheric}, \textit{accelerator} and \textit{reactor} neutrino experiments. The mean energy and baseline of the respective sources are given in table~\ref{tab:etypes}.
When neutrino flavour oscillations exist, given a neutrino of a flavour $\alpha$ produced at the source, the probability to find it in another flavour state $\beta$ is non-zero, i.e.
\begin{equation}
P(\nu_\alpha\rightarrow\nu_\beta) > 0,
\end{equation}
which is called the oscillation or transition probability. This implies that the survival probability of flavour $\alpha$ is less than 1, i.e.
\begin{equation}
P(\nu_\alpha\rightarrow\nu_\alpha) < 1.
\end{equation}
One would therefore observe a \textit{disappearance} of neutrinos of flavour $\alpha$ at the detector, compared to the neutrinos of the same flavour at the source.

Neutrino disappearance has been observed for solar $\nu_e$, reactor $\nue$ and for atmospheric and accelerator $\nu_\mu$ on neutrino experiments (e.g. SAGE\cite{Abdurashitov:2009tn} and GNO \cite{Altmann:2005ix}), at the KamLAND reactor antineutrino experiment\cite{Eguchi:2002dm}, K2K \cite{Ahn:2006zza} and Super-Kamiokande \cite{Fukuda:1998mi} in the 1990s and 2000s. Following that, the MINOS \cite{Adamson:2013whj} and T2K \cite{Abe:2012gx} experiment found evidence for $\nu_\mu$ disappearance. $\nu_\mu \rightarrow \nu_\tau$ oscillations have been observed by the Super-Kamiokande \cite{Abe:2012jj} and OPERA \cite{Agafonova:2015jxn} experiments subsequently. As a result, the existence of neutrino oscillations, implying a non-zero mass of the neutrino, can be considered as proven. The discovery of neutrino oscillations led to Nobel prizes in 2015. \cite{Olive:2016xmw}
\begin{table}[htbp]
\centering
\begin{tabular}{lll}
\hline\hline
\textbf{Mass hierarchy} & \textbf{Parameter} & \textbf{Best fit value, $3\sigma$ limits} \\ \hline \hline
$m_1<m_2<m_3$ &&\\
$^*$&$\sin^2\theta_{12}$ & $0.297_{0.250}^{0.354}$\\
&$\sin^2\theta_{23}$ & $ 0.437_{0.379}^{0.616}$\\
&$\sin^2\theta_{13}$ & $ 0.0214_{0.0185 }^{0.0246}$\\
&$\Delta m^2$ &$(2.50_{2.37}^{2.63}\times 10^{-3}) \si{\square\electronvolt}$ \\
$^*$&$\Delta m_{21}^2$ &$(7.37_{6.93}^{7.97}\times 10^{-5}) \si{\square\electronvolt}$ \\ 
$^+$&$\delta_{CP}$ & $ (1.35_{0.92}^{1.99})\pi $\\
\hline
$m_3<m_1<m_2$&&\\
$^*$&$\sin^2\theta_{12}$ & $0.297_{0.250}^{0.354}$\\
&$\sin^2\theta_{23}$ & $  0.569_{ 0.383 }^{0.637}$\\
&$\sin^2\theta_{13}$ & $ 0.0218_{0.0186 }^{0.0248}$\\
&$\Delta m^2$ &$-(2.46_{2.33}^{2.60}\times 10^{-3}) \si{\square\electronvolt}$ \\
$^*$&$\Delta m_{21}^2$ &$(7.37_{6.93}^{7.97}\times 10^{-5}) \si{\square\electronvolt}$ \\
$^+$&$\delta_{CP}$ & $ (1.32_{0.83}^{1.99})\pi $\\
\hline\hline
\end{tabular}
\caption{Neutrino oscillation parameter fit values, derived from a global fit of current data of neutrino oscillation experiments. $^*$: Parameters are the same for both mass hierarchies. $^+$: For $\delta_{CP}$, the $2\sigma$ confidence level is given. 
Source: \cite{Olive:2016xmw}\cite{Capozzi2016218}.}
\label{tab:parameters}
\end{table}
\begin{figure}[htb]
	\centering
	  \includegraphics[trim={0 0.7cm 0 0},clip,width=0.5\textwidth]{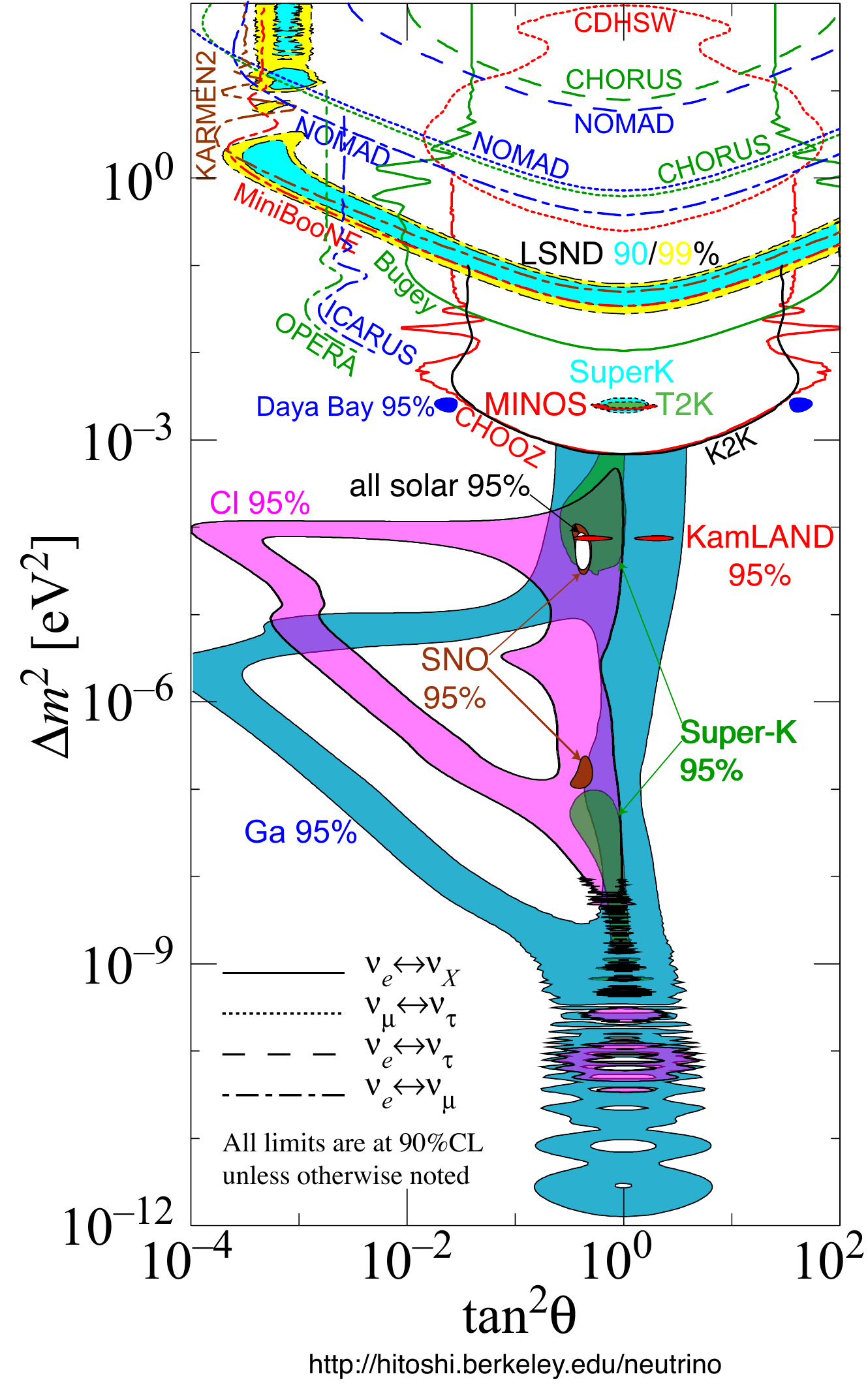}
	  \caption{$\Delta m^2$ and mixing angle $\theta$ favoured or excluded by various neutrino oscillation experiments on different oscillation types. \textit{Source}: \cite{Olive:2016xmw}.}
	\label{fig:globalplot}
\end{figure}
\begin{table}[htbp]
\centering
\begin{tabular}{lllllll}
\hline\hline
\textbf{Experiment} & \textbf{Location} & \textbf{$L [\si{\meter}]$} & \textbf{Scintillator} &\textbf{act. $[\si{\tonne}]$}& \textbf{$P_R [\si{\mega\watt}]$} & Ref\\ \hline
SoLid & Mol, Belgium & $5.5\ldots10$ & $\mathrm{PVT},\prescript{6}{}{\mathrm{Li}}$ & $1.6$ & $40\ldots80$ &\cite{Abreu:2017bpe} \cite{Arnold:2017lph}\\
Nucifer & Saclay, France& $7.2$ & $\mathrm{Gd}$ & $0.74$ & $67$ &\cite{Boireau:2015dda} \\
POSEIDON & Gatchina, Russia& $5$ & $\mathrm{Gd}$ & $\sim3$ & $100$ &\cite{Derbin:2012kf} \cite{Lhuillier}\\
STEREO & Grenoble, France& $8.8\ldots11.2$ & $\mathrm{Gd}$ & $1.75$ & $57$ &\cite{Haser:2016xlb} \cite{Lhuillier}\\
NEUTRINO-4 & Dimitrovgrad, Russia & $6\ldots11$ & $\mathrm{Gd}$ & $0.35$ & $16$ &\cite{Serebrov:2013}\cite{Serebrov:2016} \\
Hanaro & Daejon, Korea& $6$ & $\mathrm{Gd}$ & $\sim 1$ & $30$ &\cite{Yeo:2014spa} \cite{Lhuillier}\\
DANSS& Tver Oblast, Russia& $10\ldots12$ & $\mathrm{Gd}$ & $0.9$ & $3000$ &\cite{Alekseev:2016llm} \\
PROSPECT& United States& $7\ldots20$ & $\mathrm{Gd},\prescript{6}{}{\mathrm{Li}}$ & $11$ & $85$ &\cite{Ashenfelter:2013oaa} \\
NuLat & Idaho, United States& $3\ldots8$ & $\mathrm{PVT},\prescript{6}{}{\mathrm{Li}}$ & $1$ & $250\ldots1500$ &\cite{Lane:2015alq} \\
\hline\hline
\end{tabular}
\caption{Overview of very short-baseline reactor antineutrino experiments with baseline $L<10\si{\meter}$. \textit{act.} is an abbreviation for \textit{mass of active material}, $P_R$ denotes reactor power.}
\label{tab:shortbaseline}
\end{table}

Since then, efforts to investigate neutrino oscillations, particularly to determine the parameters described in section~\ref{sec:osc}, continue at an intense level. Strong evidence of $\nue$ disappearance at rather short baselines was reported by several reactor antineutrino experiments: by Double Chooz at $L= 1.1\si{\kilo\meter}$ \cite{Abe:2011fz}, by Daya Bay at $L= 1.65\si{\kilo\meter}$ \cite{Wang:2014nta} and by RENO at $L=1.38\si{\kilo\meter}$ \cite{Ahn:2012nd}. 

As efforts by neutrino experiments looking at all the sources to measure neutrino oscillations are ongoing, parameters are being measured with successively greater accuracy. A summary of parameter fit values as for 2016 combining different experiments is given on table~\ref{tab:parameters}. A global plot of $\Delta m^2$ versus mixing angle $\theta$ favoured or excluded by various neutrino oscillation experiments is shown in figure~\ref{fig:globalplot}.

As the three-flavour oscillation non-zero mass neutrino model has replaced the Standard Model's formulation of neutrinos, several anomalies discovered in the recent past in conjunction with reactor antineutrino experiments support the hypothesis that one or more additional neutrino flavours exist. Several experiments at a short baseline are in construction or have recently obtained data, including SoLid. An overview of experiments with baseline $L<10\si{\meter}$ is summarised on table~\ref{tab:shortbaseline}. Reactor antineutrino emission and the discovered anomalies are discussed in sections \ref{sec:rap} to \ref{sec:acceleratoranomaly}.

\subsection{Reactor Antineutrino Emission}
\label{sec:rap}
Nuclear fission reactors are an intense source of neutrinos\cite{Lasserre:2005qw}, emitting them isotropically \cite{Gratta:1999ny} as products of $\beta^-$-decay fission fragments\cite{Shirai:2005zz}. 

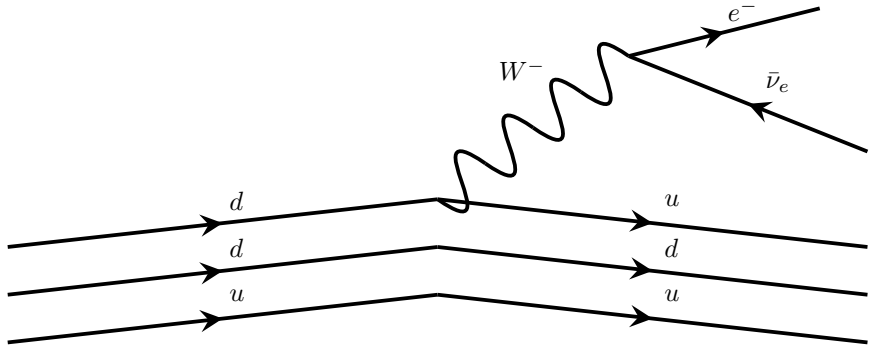
\begin{figure}[htb]
\centering
\begin{feynman}
\newcommand{\scl}{/4}
    \fermion[color=000000, endcaps=true, flip=false, label=d, labelDistance=0.12, labelLocation=0.50, showArrow=true, lineWidth=1.4285714285714286]{0\scl, 2\scl}{9\scl, 3\scl}
    \fermion[color=000000, endcaps=true, flip=false, label=d, labelDistance=0.12, labelLocation=0.50, showArrow=true, lineWidth=1.4285714285714286]{0\scl, 1\scl}{9\scl, 2\scl}
    \fermion[color=000000, endcaps=true, flip=false, label=u, labelDistance=0.12, labelLocation=0.50, showArrow=true, lineWidth=1.4285714285714286]{0\scl, 0\scl}{9\scl, 1\scl}
    \fermion[color=000000, endcaps=true, flip=false, label=u, labelDistance=0.12, labelLocation=0.50, showArrow=true, lineWidth=1.4285714285714286]{9\scl, 1\scl}{18\scl, 0\scl}
    \fermion[color=000000, endcaps=true, flip=false, label=d, labelDistance=0.12, labelLocation=0.50, showArrow=true, lineWidth=1.4285714285714286]{9\scl, 2\scl}{18\scl, 1\scl}
    \fermion[color=000000, endcaps=true, flip=false, label=u, labelDistance=0.12, labelLocation=0.50, showArrow=true, lineWidth=1.4285714285714286]{9\scl, 3\scl}{18\scl, 2\scl}
    \electroweak[color=000000, endcaps=true, flip=false, label=W^-, labelDistance=0.38, labelLocation=0.50, showArrow=false, lineWidth=1.4285714285714286]{9\scl, 3\scl}{13\scl, 6\scl}
    \fermion[color=000000, endcaps=true, flip=false, label=\nue, labelDistance=0.12, labelLocation=0.50, showArrow=true, lineWidth=1.4285714285714286]{18\scl, 4\scl}{13\scl, 6\scl}
    \fermion[color=000000, endcaps=true, flip=false, label=e^-, labelDistance=0.12, labelLocation=0.50, showArrow=true, lineWidth=1.4285714285714286]{13\scl, 6\scl}{17\scl, 7\scl}
\end{feynman}
\caption{Feynman diagram of the fundamental level of a $\beta^-$-decay producing $\nue$.}
\label{betafeyn}
\end{figure}

The most important processes in reactor $\nue$ production by $\prescript{235}{}{\mathrm{U}}$-fuelled $n$-induced nuclear fissions are of the type
\begin{equation}
\label{eq:betad}
n+\prescript{235}{}{\mathrm{U}}\rightarrow X_1 +X_2+2n+204\mathrm{MeV}.
\end{equation}
The fragments $X_1$ and $X_2$ will undergo several $\beta^-$-decays,
\begin{equation}
\prescript{A}{Z}{X}\rightarrow \prescript{A}{Z+1}{X'}+e^-+\nue,
\end{equation}
until they have reached a long-lived state, producing an average of $6 \nue$ per fission\cite{Lipari:2001is}. The Feynman diagram of the fundamental level of the $\beta^-$-decay,
\begin{equation}
n=udd\rightarrow udu+W^-\rightarrow udu+\nue+e^-=p+\nue+e^-,
\end{equation}
is shown in figure~\ref{betafeyn}.
The $\nue$'s energy is $<10\mathrm{MeV}$\cite{DiLella:1999ar}. Thus, reactor neutrinos  are not only produced at a very high rate, but their energies also lie within a limited spectrum.

The emission energy spectrum of reactor neutrinos depends on the composition of the fissioning actinides:
\begin{equation}
S(E_\nu)=\sum_i f_i \left( \frac{\mathrm{d}N_i}{\mathrm{d}E_\nu}\right),
\label{eq:em}
\end{equation}
$f_i$ being the number of fissions from actinide $i$ and $ \frac{\mathrm{d}N_i}{\mathrm{d}E_\nu}$ is the cumulative $\nue$ spectrum of $i$ normalised per fission. Thus, the parameters $f_i$ must be known, requiring detailed information about the reactor core, including the exact composition. The total reactor thermal energy $W_{th}$ is the sum of all the fission antinides' effective thermal energy $e_i$:
\begin{equation}
W_{th}=\sum_i f_i \cdot e_i.
\end{equation}

\begin{figure}[htb]
	\centering
	  \includegraphics[width=0.68\textwidth]{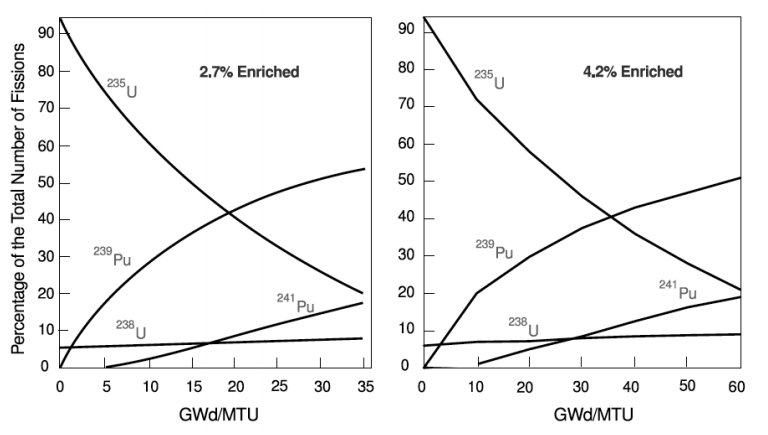}
	  \caption{Time-development of the composition of the nuclear core during a reactor cycle of a water reactor. Note the different time scales. \textit{Source:} \cite{Nieto:2003wd}.}
	\label{fig:fuelcycle}
\end{figure}
$99.9\%$ of the reactor power stems from the fission of the $\prescript{235}{}{\mathrm{U}}$, $\prescript{238}{}{\mathrm{U}}$, $\prescript{239}{}{\mathrm{Pu}}$ and $\prescript{241}{}{\mathrm{Pu}}$ isotopes, and only these are considered for $\nue$ energy spectrum calculations. The thermal energy of the reactor changes with time as the fuel gets burnt up. The time-development of fuel composition during a reactor cycle can be seen in figure~\ref{fig:fuelcycle}. For practical purposes, $f_i$ is expressed as the fraction of total number of fissions $F$. \cite{Hayes:2016qnu} So equation~\ref{eq:em} can be re-written as \cite{Cao:2011gb}
\begin{equation}
S(E_\nu)=\frac{W_{th}}{\sum_i\left(\frac{f_i}{F}\right)e_i}\cdot \frac{f_i}{F}\left(\frac{\mathrm{d}N_i}{\mathrm{d}E_\nu}\right).
\end{equation}

In order to obtain the Inverse $\beta$-decay spectrum -- or detection spectrum -- rather than the emission spectrum, the emission spectrum is folded with the Inverse $\beta$-decay cross section \cite{Mueller:2011nm} (Inverse $\beta$-decay is later explained in section~\ref{sec:ibd}).

The calculations of reactor antineutrino spectra are quite complex in nature as they rely on parameters that are difficult to determine, such as the core composition and the individual emission spectra of the fragments' energy dissipation, and an inaccurate estimation of these easily propagates to the estimate of total emission spectrum. This leads to final relative uncertainties in the $10\%\ldots20\%$ range \cite{Mueller:2011nm}. As neutrino oscillation experiments rely on an exact estimate of neutrino emission for disappearance measurements (see section~\ref{sec:osc}), emission calculations have a direct impact on uncertainty or error of these types of experiments.

\subsection{Reactor Anomaly}
\begin{figure}[htb]
	\centering
	  \includegraphics[width=0.53\textwidth]{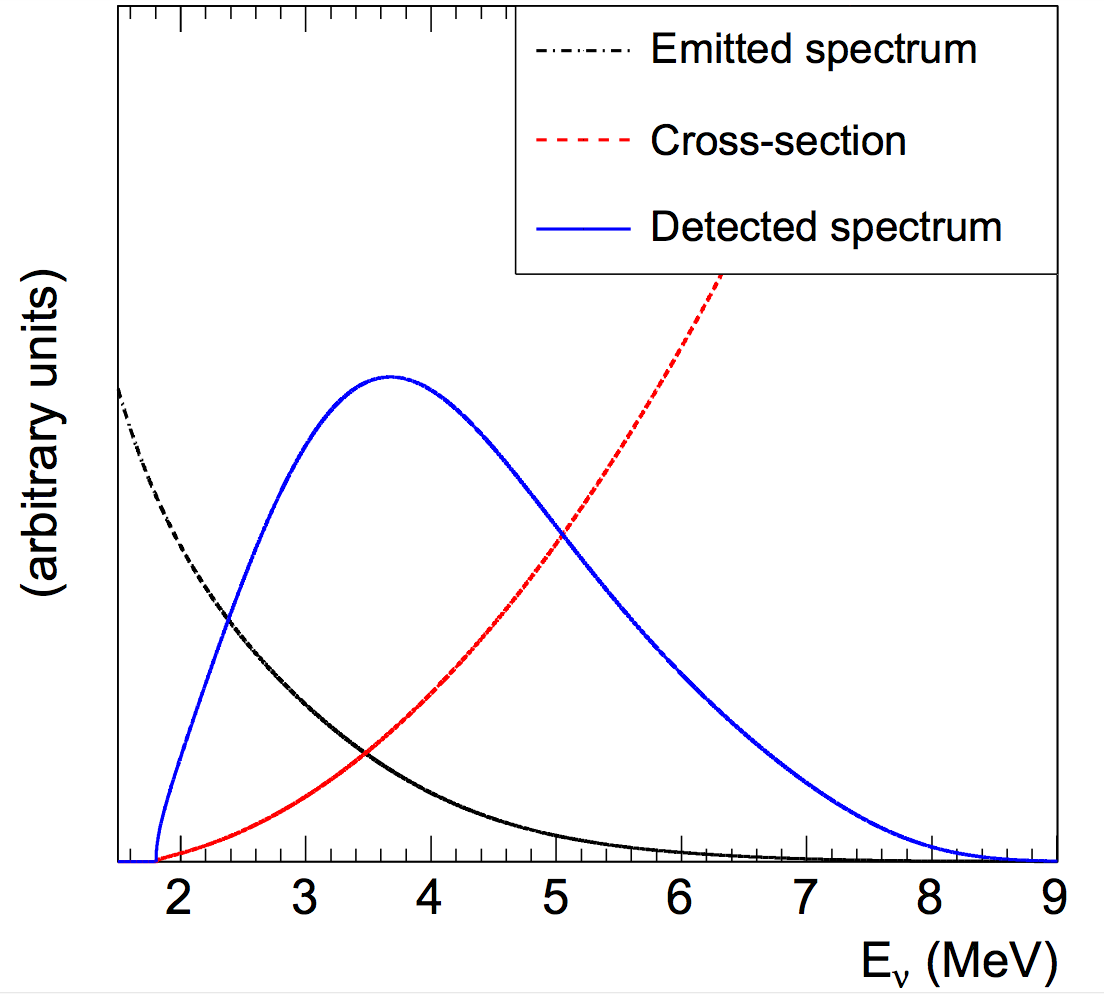}
	  \caption{$\nue$ detection spectrum  for the $\prescript{235}{}{\mathrm{U}}$ isotope. The shape results from folding the emitted spectrum (black dashed-dotted curve), parameterisation and IBD cross section (red dashed curve). \textit{Source}: \cite{Mueller:2011nm}.}
	\label{fig:spectrumnew}
\end{figure}
A more accurate calculation for $\prescript{235}{}{\mathrm{U}}$ and $\prescript{239}{}{\mathrm{Pu}}$ antineutrino detection spectra has been performed in 2011 \cite{Mueller:2011nm} by combining data from nuclear databases and electron energy spectra measured in the 1980s at the ILL reactor facility for the dominant $\prescript{235}{}{\mathrm{U}}$, $\prescript{239}{}{\mathrm{Pu}}$ and $\prescript{241}{}{\mathrm{Pu}}$ isotopes (see figure~\ref{fig:spectrumnew}). 
\begin{figure}[htb]
	\centering
	  \includegraphics[width=0.98\textwidth]{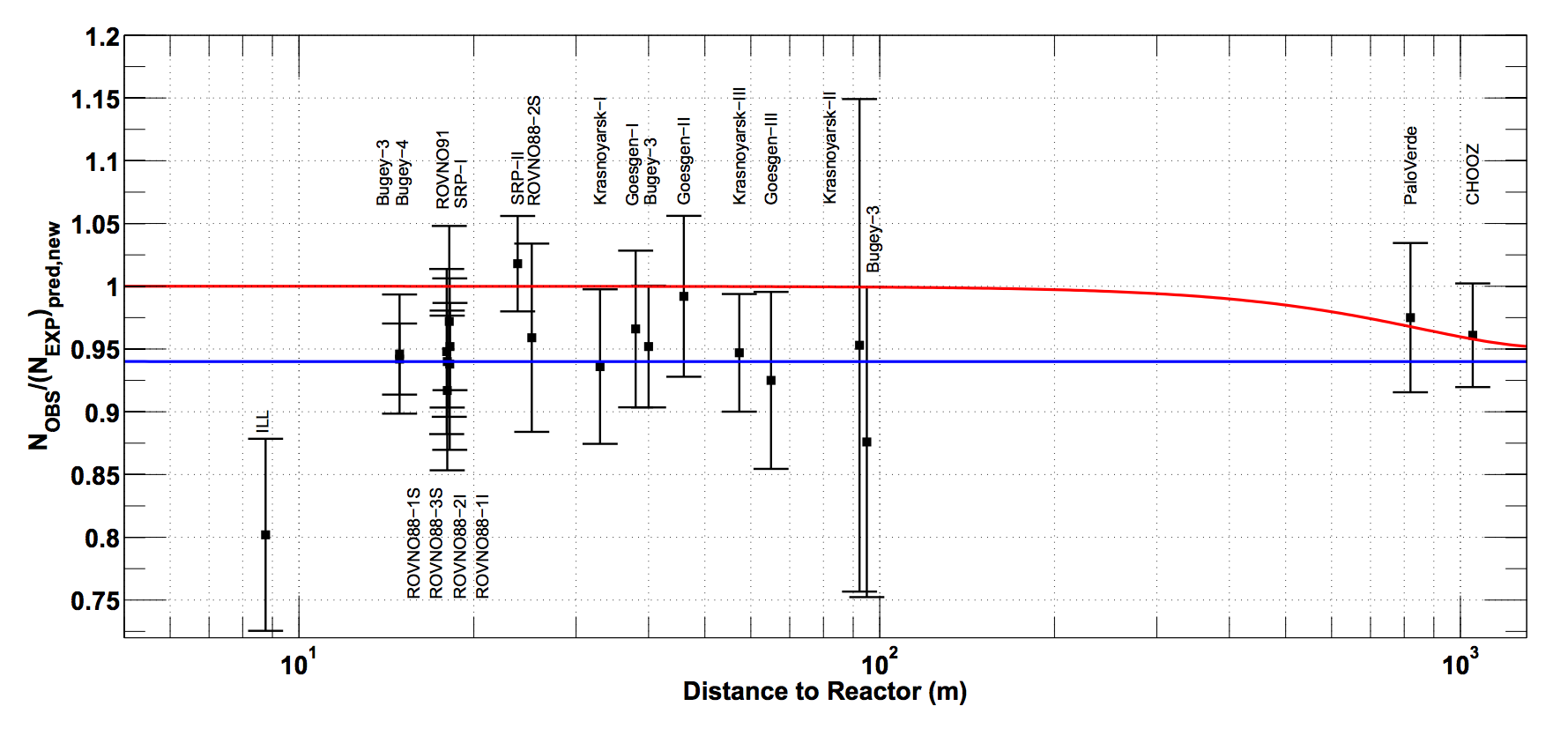} 
	  \caption{The ratio of observed $\nue$ events to expected events in case of the new calculations of the three neutrino flavour mixing model (red line) and the case of the existence of a sterile neutrino (blue line). The reactor antineutrino experiments seem to favour the sterile neutrino model. \textit{Source}: \cite{Mention:2011}.}
	\label{fig:reactoranomaly}
\end{figure}

These calculations shifted the ratio of predicted to observed $\nue$ events combining various reactor antineutrino experiments at a baseline $L<100\si{\meter}$ from $0.976\pm0.024$ -- prediction and observation being consistent in this case -- to $0.943\pm0.023$ -- leading to a deficit at the $98.6\%$ C.L. This is referred to as the \textit{reactor antineutrino anomaly}. \cite{Mention:2011}

The deficit favours the existence of at least one additional flavour eigenstate, the  sterile flavour $\nu_s$,  over the three-flavour mixing model (see figure~\ref{fig:reactoranomaly}). If one additional sterile neutrino is assumed, it would have a new mass eigenstate $m_4$ with $\Delta m_{41}^2 \sim 1 \si{\square\electronvolt}$ \cite{Gariazzo:2016lsd}. The sterile neutrino hypothesis is discussed in section~\ref{sec:sterileneutrino}.
A deficit of neutrinos compared to expectation was also observed by the reactor experiments Double Chooz in 2014 \cite{Abe:2014bwa}, RENO in 2015 \cite{Kim:2014rfa} and Daya Bay in 2016 \cite{PhysRevLett.116.061801}.
Despite the re-calculations of reactor neutrino spectra seeming to favour the sterile neutrino hypothesis, this might be as well be a result of a systematic error, affecting the results of all previous reactor antineutrino experiments simultaneously. Systematic errors cannot be excluded, as currently unknown, unconsidered or underestimated effect might have led to an incorrect estimation of the predicted reactor detection spectrum. However, an unknown background is unlikely to be the explanation, as the measured deficit is time-independent and correlated to reactor power \cite{Olive:2016xmw}.

\subsection{Gallium Anomaly}
\begin{figure}[htb]
	\centering
	  \includegraphics[width=0.72\textwidth]{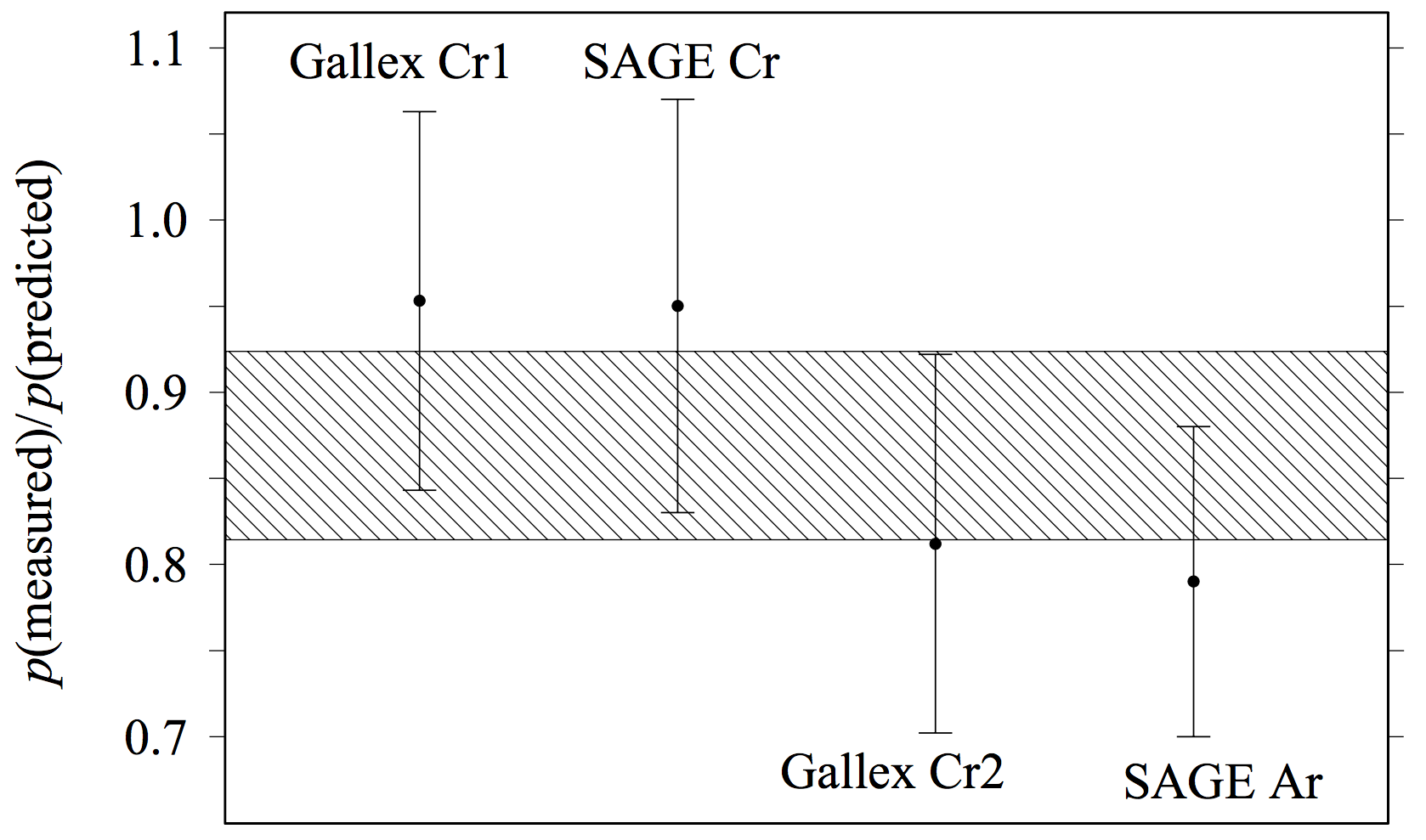} 
	  \caption{The ratio of observed $\nue$ events to expected events for the two GALLEX experiments with the $\prescript{51}{}{\mathrm{Cr}}$ source and the SAGE experiment with both $\prescript{51}{}{\mathrm{Cr}}$ and $\prescript{37}{}{\mathrm{Ar}}$ sources. The error bar shows the $1\sigma$-limit. A deficit up to $20\%$ of the expected rate was observed. \textit{Source}: \cite{Abdurashitov:2009tn}.}
	\label{fig:galliumanomaly}
\end{figure}
A second observation supporting the sterile neutrino hypothesis is the \textit{Gallium anomaly}. The GALLEX and SAGE detectors have been tested with intense $\prescript{51}{}{\mathrm{Cr}}$ and $\prescript{37}{}{\mathrm{Ar}}$ sources placed inside the detectors. Ratios $R$ of expected to observed $\nu_e$ events,
\begin{equation}
 \begin{aligned}
R_{\text{GALLEX-1},\prescript{51}{}{\mathrm{Cr}}}&=0.953\pm 0.11, \\
R_{\text{GALLEX-2},\prescript{51}{}{\mathrm{Cr}}}&=0.812^{+0.10}_{-0.11}, \\
R_{\text{SAGE},\prescript{51}{}{\mathrm{Cr}}}&=0.95\pm0.12, \\
R_{\text{SAGE},\prescript{37}{}{\mathrm{Ar}}}&=0.791^{+0.084}_{-0.078},
 \end{aligned}
\end{equation}
subscripts denoting detector and source, have been measured;  the quoted uncertainties are at the $1\sigma$ level. A clear deficit of neutrinos up to $20\%$ of the expected number of events was observed (see figure~\ref{fig:galliumanomaly}).\cite{Abdurashitov:2009tn} \cite{Kaether:2010ag} However, these numbers do not take into account cross section uncertainty.

Both GALLEX and SAGE use the reaction
\begin{equation}
\nu_e + \prescript{71}{}{\mathrm{Ga}} \rightarrow \prescript{71}{}{\mathrm{Ge}}+e^-
\end{equation}
for detection, a reaction whose cross section has a large uncertainty \cite{Giunti:2010zu}. Re-calculation with consideration of the cross section uncertainty modified the results and the $1\sigma$ level to  new ratios of 
\begin{equation}
 \begin{aligned}
R_{\text{GALLEX-1},\prescript{51}{}{\mathrm{Cr}}}&=0.84^{+0.13}_{-0.12}, \\
R_{\text{GALLEX-2},\prescript{51}{}{\mathrm{Cr}}}&=0.71^{+0.12}_{-0.11}, \\
R_{\text{SAGE},\prescript{51}{}{\mathrm{Cr}}}&=0.84^{+0.14}_{-0.13}, \\
R_{\text{SAGE},\prescript{37}{}{\mathrm{Ar}}}&=0.70^{+0.10}_{-0.09}.
 \end{aligned}
\end{equation}
These have slightly higher uncertainty, but significant lower result values have been obtained, leading to the confirmation of the Gallium anomaly at the $3.0\sigma$ level. \cite{Giunti:2010zu} Combined analysis of the Gallium anomaly experiments and the reactor anomaly experiment disfavour the standard three-flavour mixing model even at a C.L. of $99.9\%$, or $3.3\sigma$ \cite{Kopp:2013vaa}.

\subsection{Accelerator Anomaly}
\label{sec:acceleratoranomaly}
The \textit{accelerator anomaly}, or \textit{LSND anomaly}, stems from measurements at the LSND accelerator neutrino experiment in Los Alamos \cite{PhysRevLett.102.101802} that observed $\nue$ in a $\bar{\nu}_\mu$-beam at a baseline of $L=30\si{\meter}$. LSND measured an excess of neutrinos which would be consistent with the existence of a sterile neutrino with $\Delta m^2_{41} \sim 1 \si{\square\electronvolt}$. Further support of this excess was given by MiniBooNE at Fermilab at the $2.8\sigma$ level \cite{PhysRevLett.110.161801}. However, the KARMEN experiment could not reproduce the excess, despite having measured a similar phase space to LSND \cite{PhysRevD.65.112001}. In addition, other accelerator neutrino experiments also disfavour the sterile neutrino hypothesis, such as CDHSW \cite{Dydak:1983zq}, MINOS \cite{Adamson:2011ku}, OPERA \cite{Agafonova2013} and ICARUS \cite{Antonello:2012pq}. However, not all regions of the relevant parameter space have been measured by these experiments \cite{Lane:2015alq}. 

In conclusion, the high confidence put on the sterile neutrino hypothesis by reactor experiments was not reproduced by accelerator neutrino experiments.

\subsection{The Sterile Neutrino Hypothesis}
\label{sec:sterileneutrino}
As neutrinos have non-zero mass, they are a superposition of negative and positive chirality states. Only the negative chirality component interacts weakly. For neutrinos, the negative chirality state corresponds essentially exclusively to negative helicity, and for antineutrios to positive helicity (with a small correction, that would vanish in the limit of zero neutrino mass). Neutrinos therefore interact essentially always as helicity left-handed particles, and anti-neutrinos as helicity right-handed particles. 

There is no evidence so far that predominantly right-handed neutrinos or predominantly left-handed antineutrinos exist, nevertheless, if they exist, they should be \textit{sterile} neutrinos, not coupling neither to $W^\pm$ nor to $Z^0$ bosons, as obtained experimental data is compatible only with 3 flavour neutrinos that are coupled to the $Z$ boson \cite{PhysRevD.86.010001}. They could contribute significantly to open fundamental questions in physics such as the generation of the observed matter-antimatter asymmetry, the generation of neutrino masses and their disparity. \cite{Olive:2016xmw}

\subsubsection{Sterile Neutrino Mass Mixing}
The sterile neutrino hypothesis suggests one or more additional neutrinos to the three active flavour neutrinos which are already known ($\nu_e$, $\nu_\mu$, $\nu_\tau$), and is called the 3+$n$ model depending on the number $n$ of how many new flavour eigenstates are suggested. These $n$ new \textit{sterile} flavours would correspond to $n$ corresponding additional mass eigenstates. \cite{Olive:2016xmw} Taking the case of the 3+1 model, i.e. of the model suggesting one additional sterile neutrino, the mass mixing equations~\ref{eq:massmixing} and~\ref{eq:massmixing2} would be expressed as
\begin{equation}
\begin{pmatrix}\nu_e \\\nu_\mu \\ \nu_\tau \\ \nu_s \end{pmatrix}= U^* \begin{pmatrix}\nu_1 \\\nu_2 \\ \nu_3 \\ \nu_4 \end{pmatrix},
\end{equation}
with $\nu_s$ denoting the suggested sterile neutrino, increasing the number of linearly independent mass splitting factors $\Delta m_{ij}$ from 3 to 4. In case of more than one sterile neutrino, the equation would simply contain more vector elements for the additional neutrino states. For $n$ additional sterile neutrinos, the dimensionality of the PMNS matrix (equation~\ref{eq:pmns} and~\ref{eq:dirac}) together with the number of mass splittings would also have to increase by $n$, thus the $3\times3$ matrix for the three-flavour model would become a $4\times4$ matrix together with 4 mass splitting factors in the 3+1 model and to a $5\times5$ matrix with 5 mass splitting factors in the 3+2 model \cite{Ke:2015xka}. 
In the 3+1 model, the new $4\times4$ mixing matrix $\widetilde{U}$ is expressed as
\begin{equation}
    \begin{aligned}
    \widetilde{U} &=&
        \left(\begin{matrix}    
        c_{14} & 0 & 0 & s_{14} \\
                -s_{14}s_{24} & c_{24} & 0 & c_{14}s_{24} \\
                -c_{24}s_{14}s_{34} & -s_{24}s_{34} & c_{34} & c_{14}c_{24}s_{34} \\
                -c_{24}c_{34}s_{14} & -s_{24}c_{34} & -s_{34} & c_{14}c_{24}c_{34}
            \end{matrix}\right)
    \left( \begin{matrix}
    U_{e1} & U_{e2} & U_{e3} & 0 \\
    U_{\mu1} & U_{\mu2} & U_{\mu3} & 0 \\
    U_{\tau1} & U_{\tau2} & U_{\tau3} & 0 \\
    0 & 0 & 0 & 1
        \end{matrix} \right)
         \\
    &=& \left( \begin{matrix}
    c_{14}U_{e1} & c_{14}U_{e2} & c_{14}U_{e3} & s_{14} \\
    \cdots & \cdots & \cdots & c_{14}s_{24} \\
    \cdots & \cdots & \cdots & c_{14}c_{24}s_{34} \\
    \cdots & \cdots & \cdots & c_{14}c_{24}c_{34}
        \end{matrix} \right)
         \\
    &=& \left( \begin{matrix}
    \widetilde{U}_{e1} & \widetilde{U}_{e2} & \widetilde{U}_{e3} & \widetilde{U}_{e4} \\
    \widetilde{U}_{\mu1}& \cdots & \cdots & \cdots \\
    \widetilde{U}_{\tau1}& \cdots & \cdots & \cdots \\
    \widetilde{U}_{s1}& \cdots & \cdots &\cdots 
        \end{matrix} \right).
\end{aligned}
\end{equation}
A written-out representation of $\widetilde{U}$ is given in the appendix in equation~\ref{eq:fullu}.
The elements of the $3\times3$ PMNS matrix $U_{\alpha i}$ from equation~\ref{eq:pmns} are included in this parameterisation. The CP phases are omitted as they do not affect survival probability for reactor neutrino oscillation.\cite{Kang:2013gpa}

The sterile neutrino is called \textit{sterile} as -- unlike the confirmed 3 flavour eigenstates -- it does not couple to the $W^\pm$ nor to the $Z^0$ boson, thus does not interact weakly \cite{Lipari:2001is}. Not to mention, that -- as for any other neutrino -- it neither does have charge nor colour, hence it is only affected by the gravitational force. Therefore oscillations into the sterile neutrino state would simply lead to disappearance of neutrinos in observations. The additional disappearance caused by the existence of sterile neutrinos would be able to explain the measured neutrino deficit of the reactor and Gallium anomaly.
\subsubsection{Sterile Neutrino Oscillations}
\begin{figure}[htb]
	\centering
	  \includegraphics[width=0.7\textwidth]{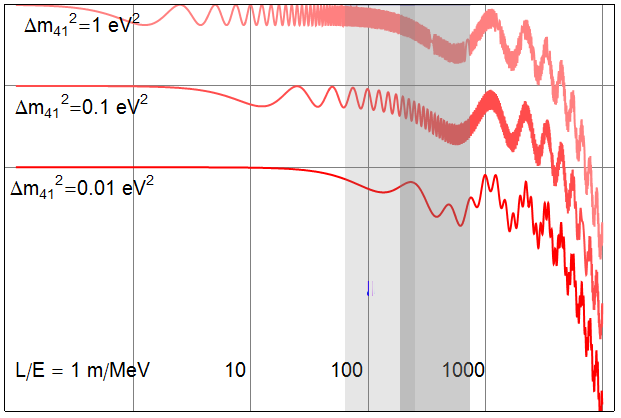}
	  \caption{Survival probability for reactor antineutrinos for different sterile neutrino mass splitting factors $\Delta m_{41}^2$ as a function of baseline-to-energy ratio $L/E$. \textit{Source:} \cite{Kang:2013gpa} (edited).}
	\label{fig:LtoE}
\end{figure}
\begin{figure}[htb]
	\centering
	  \includegraphics[width=0.7\textwidth]{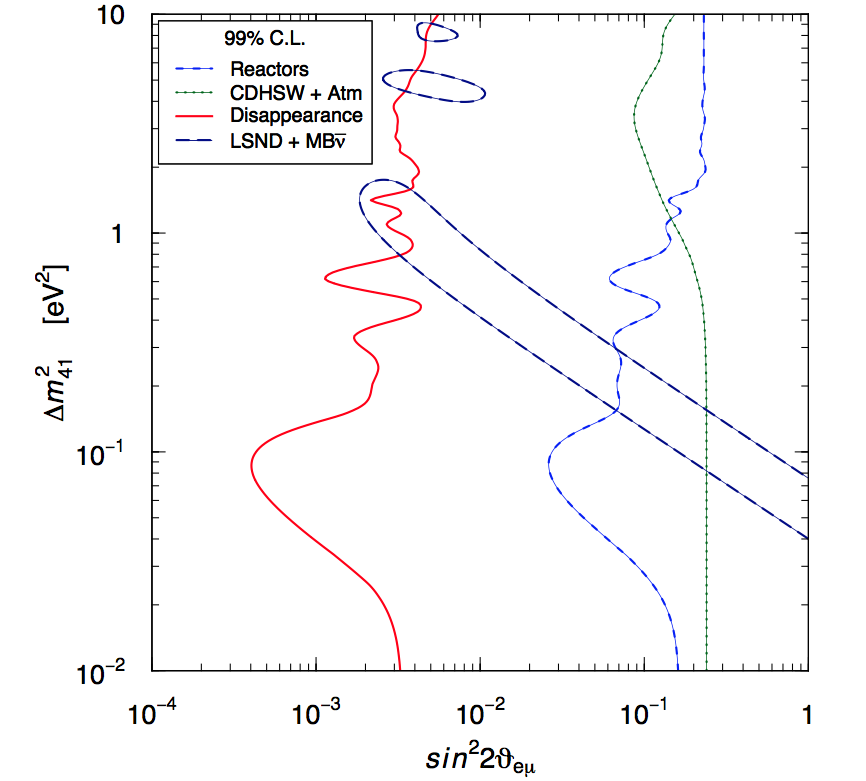}
	  \caption{Exclusion curves for $\sin^22\theta_{12}$ and $\Delta m^2_{41}$ for the 3+1 sterile neutrino model for reactor (blue), atmospheric (green) and accelerator (dark blue) experiments and their combined fit (red) at a $99\%$ C.L. \textit{Source:} \cite{Giunti:2011gz}.}
	\label{fig:excl}
\end{figure}

In the case of the 3+1 model, equation~\ref{eq:ImRe} applied to the case $\bar{\nu}_e\rightarrow\bar{\nu}_e$, where it becomes the survival probability of reactor antineutrinos, will become \cite{Kang:2013gpa}
    \begin{equation}
\begin{aligned}
    P(\bar{\nu}_e\rightarrow\bar{\nu}_e) &=& \left|\sum_{j=1}^4\left|\widetilde{U}_{ei}\right|^2 \exp{i\frac{\Delta m_{j1}^2L}{2E_\nu}}\right|^2 \\
        &=& 1- \sum_{i<j}4\left|\widetilde{U}_{ei}\right|^2\left|\widetilde{U}_{ej}\right|^2\sin^2\left(\frac{\Delta m_{ij}^2L}{4E_\nu}\right),
\end{aligned}
    \end{equation}
approximated by \cite{Kang:2013gpa}
\begin{equation}
\begin{aligned}
    P(\bar{\nu}_e\rightarrow\bar{\nu}_e) = 1&-&c_{14}^4c_{13}^4\sin^2 2\theta_{12}\sin^2\left(\Delta m_{21}^2\frac{L}{E}\right) \\
    &-& c_{14}^4\sin^2 2\theta_{13}\sin^2\left(\Delta m_{31}^2\frac{L}{E}\right)   \\
    &-& \sin^2 2\theta_{14}\sin^2\left(\Delta m_{41}^2\frac{L}{E}\right). 
\end{aligned}
\label{eq:Kang}
\end{equation}

The reactor antineutrino survival probability is plotted in figure~\ref{fig:LtoE} for sterile neutrino mass splitting factors $\Delta m_{41}=0.01\si{\square\electronvolt},0.1\si{\square\electronvolt},1 \si{\square\electronvolt}$ as a function of baseline-to-energy ratio $L/E$. 
As can be seen, sterile neutrinos of the range of $\Delta m_{41}^2 \sim 1\si{\electronvolt}$ would agree with the deficit of neutrinos in the reactor antineutrino anomaly. Indeed, in the 3+1 and 3+2 models that are commonly considered\cite{Olive:2016xmw}, the discussed anomalies would be consistent with fits to $\Delta m_{41}^2\sim1\si{\square\electronvolt}$ in the 3+1 model or $\Delta m_{41}^2,\Delta m_{51}^2\sim1\si{\square\electronvolt}$ in the 3+2 model (see figure~\ref{fig:excl})\cite{Gariazzo:2016lsd}.

\begin{figure}[htb]
	\centering
	  \includegraphics[width=0.97\textwidth]{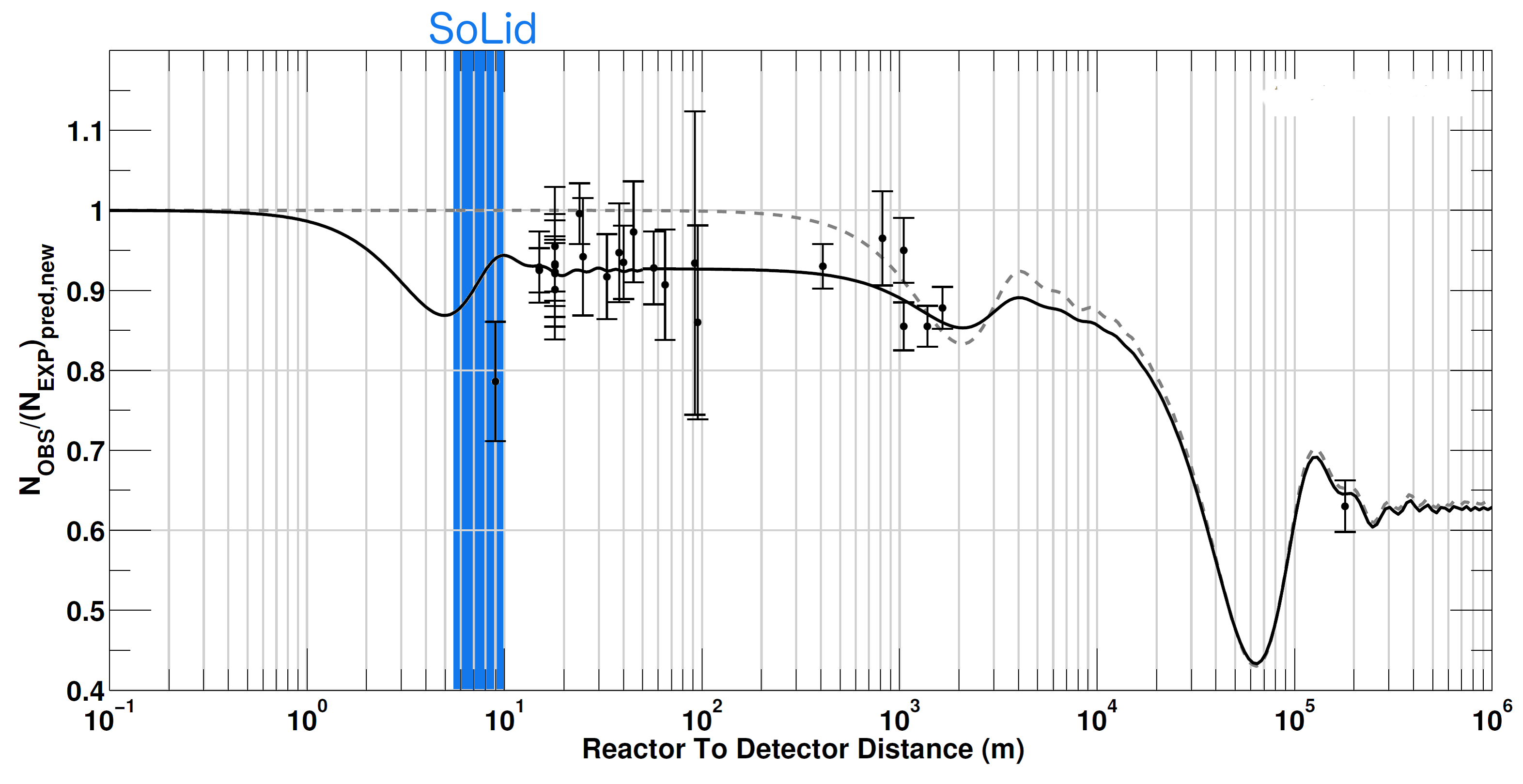}
	  \caption{Ratio of observed to expected neutrino rate is shown for different reactor antineutrino experiments. The dashed line represents the three-flavour mixing model, while the solid line corresponds to the 3+1 mode. SoLid will search for sterile neutrinos at a baseline from $L=5.5\ldots10\si{\meter}$ (blue area).  \textit{Source:} \cite{Gando:2013zla} (edited).}
	\label{fig:range}
\end{figure}

Experimental data suggest that the mass splittings $\Delta m_{21}^2$ ($\mathcal{O}(10^{-5})\si{\square\electronvolt}$) and $\Delta m_{31}^2$ ($\mathcal{O}(10^{-3})\si{\square\electronvolt}$) are very small compared to $\Delta m_{41}^2$ (or $\Delta m_{51}^2$) $\sim 1\si{\square\electronvolt}$. In the short-baseline range, $\nu_\alpha\rightarrow\nu_s$ oscillations would be dominant to the level where oscillations between the three standard neutrinos are negligible, with these neutrinos becoming more dominant as baseline increases. Therefore $\Delta m_{21}$ and $\Delta m_{31}$ can be set to $0$, or to \textit{degenerate} state, for approximation in the short-baseline range in the sterile neutrino models. 
In the case of the 3+1, equation~\ref{eq:Kang} can be approximated for very short-baseline reactor antineutrino experiments such as SoLid by\cite{Collin:2016rao} \cite{wouter}:
\begin{equation}
{\begin{aligned}P(\nu_\alpha \rightarrow \nu_\beta)=\delta _{\alpha \beta }&{}-4\left(\delta_{\alpha\beta}-U_{\alpha 4}U_{\beta 4}^*\right) U_{\alpha 4}^*U_{\beta 4}\sin ^{2}\left({\frac {\Delta m_{41}^{2}L}{4E}}\right) \end{aligned}}
\end{equation}
for any oscillation $\nu_\alpha \rightarrow \nu_\beta$ which is in the particular case of $\bar{\nu}_e\rightarrow\bar{\nu}_e$ survival probability would become
\begin{equation}
\begin{aligned}
    P(\bar{\nu}_e\rightarrow\bar{\nu}_e) &=& 1- \sin^2 2\theta_{14}\sin^2\left(\Delta m_{41}^2\frac{L}{E}\right),
\end{aligned}
\end{equation}
the function that will be used for fitting the sterile neutrino parameters $\theta_{14}$ and $\Delta m_{41}^2$ in short-baseline reactor antineutrino experiments. SoLid will measure these neutrino oscillation parameters at a well-defined range from $L=5.5\si{\meter}$ to $L=10\si{\meter}$, as shown in figure~\ref{fig:range}, currently uncovered by reactor antineutrino experiments.

\subsection{Other Applications of SoLid}
\label{sec:otherapplications}
Although the hunt for sterile neutrinos is what SoLid mainly aims for, SoLid also allows the investigation of the $5\si{\mega\electronvolt}$ excess, and its technology might be used for reactor monitoring for nuclear non-proliferation applications.
\subsubsection{$5 \si{\mega\electronvolt}$ Excess}
\begin{figure}[htb]
	\centering
	  \includegraphics[width=0.685\textwidth]{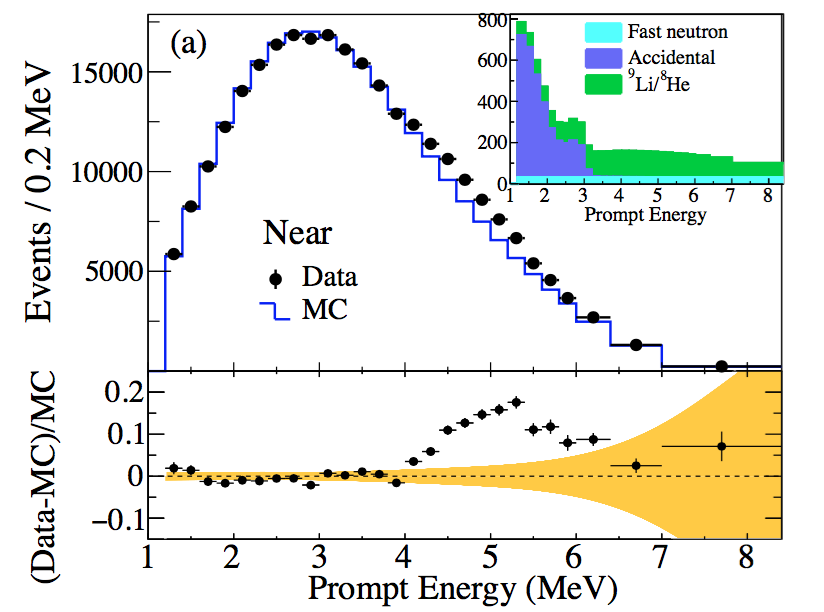}
	  \caption{Comparison of neutrino events  expected by simulation (blue) and observed (black dots) by the RENO near detector at a baseline of $L=293\si{\meter}$. An excess of $\nue$ occurs at the region surrounding $5\si{\mega\electronvolt}$ at a level of $3.6\sigma$. \textit{Source}: \cite{Seo:2016uom}.}
	\label{fig:renobump}
\end{figure}

The precise measurement of reactor antineutrino flux of $\prescript{235}{}{\mathrm{U}}$-based reactors  is also of interest due to an anomaly recently discovered by reactor antineutrino experiments: A discrepancy, or excess, at the region of about $5\si{\mega\electronvolt}$ has been discovered by Double Chooz in 2011 \cite{Abe:2011fz} and Daya Bay in 2012 \cite{An:2012eh}, and later confirmed at a $3.6\sigma$ level by RENO in 2015 (see figure~\ref{fig:renobump}) \cite{Seo:2016uom}.

The discrepancy at $5\si{\mega\electronvolt}$ is not understood yet. Possible causes of the excess might be unaccounted background of the reactor flux, or even effects due to new physics \cite{Lane:2015alq}.

Double Chooz is located at a baseline of $L=1.1\si{\kilo\meter}$ from the reactor, Daya Bay at a baseline of $L=1.64\si{\kilo\meter}$ and RENO at $L=1.38\si{\kilo\meter}$ for the far detector and $L=294\si{\meter}$ for the near detector \cite{Kim:2014rfa} (see section~\ref{sec:neutrinoexperiments}). Hence, the baseline range of $L<290\si{\meter}$ is uncovered so far. SoLid, along with other very short-baseline reactor antineutrino experiments, will provide evidence whether the $5 \si{\mega\electronvolt}$ excess can be reproduced at the very near distance from the reactor.

\subsubsection{Non-Proliferation}
SoLid is a particle physics experiment aiming at the search on sterile neutrinos. However, despite not being a main target of the experiment, the SoLid detector also could explore potential applications of its technology in nuclear non-proliferation.

Monitoring the $\mathrm{Pu}$ content of nuclear reactors forms an essential measure in non-proliferation efforts. Usually, $\gamma$-ray spectroscopy is applied for this purpose, however short-baseline $\nue$ flux  monitoring has been suggested in the past with increasing recent interest as an appropriate tool of cross-validation in determining the composition of nuclear fuel \cite{Ashenfelter:2013oaa} \cite{Cao:2011gb}.

\begin{figure}[htb]
	\centering
	  \includegraphics[width=0.685\textwidth]{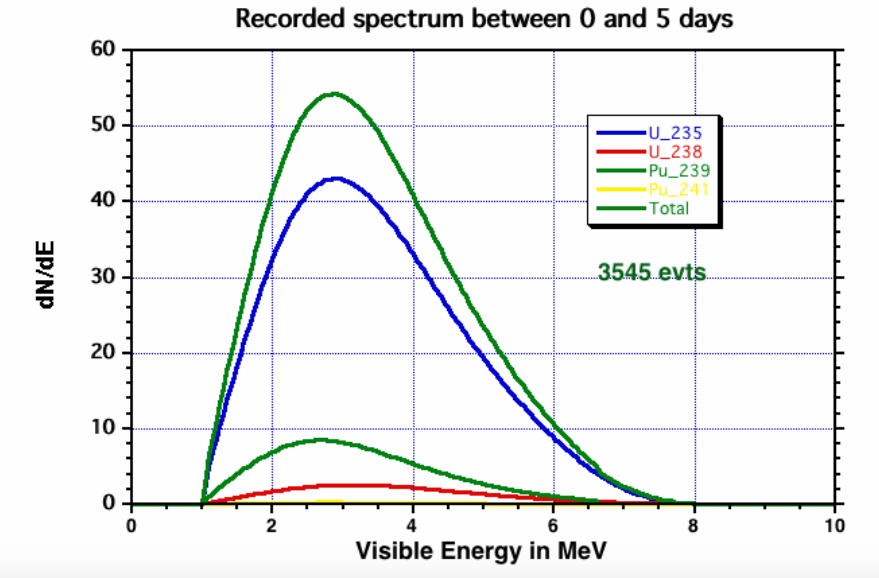}
	  \caption{Inverse $\beta$-decay $e^+$ energy spectra of different isotopes at a $150\si{\meter}$ baseline from a $1\si{\giga\watt}$-reactor (simulation). \textit{Source}: \cite{Cribier:2007zh}.}
	\label{fig:rech}
\end{figure}

Reactors emit a characteristic $\nue$ flux and energy spectrum according to their thermal power and to the  isotopic composition of their fuel, namely $\prescript{235}{}{\mathrm{U}}$, $\prescript{238}{}{\mathrm{U}}$, $\prescript{239}{}{\mathrm{Pu}}$ and $\prescript{241}{}{\mathrm{Pu}}$ (see figure~\ref{fig:rech}) \cite{Christensen:2014pva}. This allows detection of $\mathrm{Pu}$ content in nuclear reactors by obtaining their $\nue$ spectrum and flux. Because of the high penetration depth of neutrinos, it is possible to perform the measurements remotely. \cite{Cribier:2007zh} 

Due to the compactness of the technology used for SoLid, as well as its usage of solid material (PVT and \LiZnS), SoLid's novel neutrino detection technology is a potential candidate for employment in non-proliferation monitoring applications.
\clearpage

\section{The SoLid Detector}
\label{sec:detector}
The SoLid detector is unique not only in terms of its very short baseline to the reactor core, but also in terms of the employed scintillators -- Polyvinyl toluene (PVT) and \LiZnS\phantom{a}--  and in providing a high level of segmentation and granularity of the detector modules.

\subsection{BR2 Reactor}
\begin{figure}[htb]
	\centering
	  \includegraphics[width=0.755\textwidth]{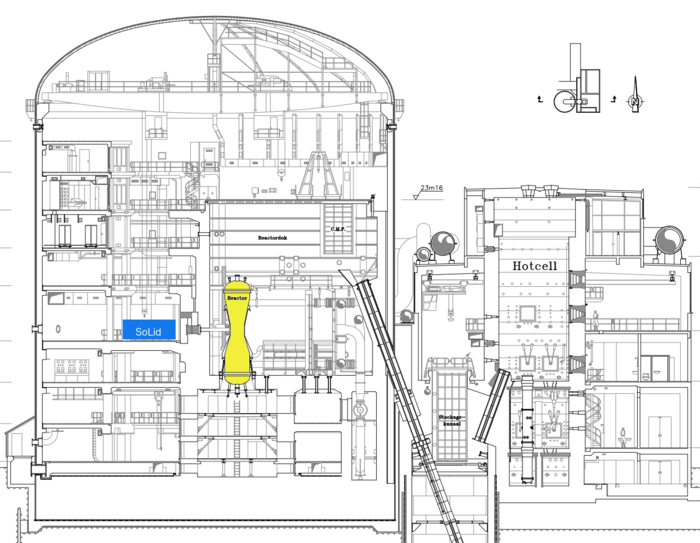}
	  \caption{The BR2 reactor facility at SCK\textbullet CEN in Mol, Belgium. The reactor is shown in yellow and the SoLid detector module is shown in blue. \textit{Source}: \cite{joppen} (edited).}
	\label{br2}
\end{figure}

 SoLid's neutrino source is the BR2 nuclear fission reactor located at the Belgian nuclear research facility SCK\textbullet CEN in Mol. BR2 is designed and operated as a tank-type material research reactor, with a power range up to $100\si{\mega\watt}$ \cite{joppen}. A picture of BR2's architecture can be seen in figure~\ref{br2}.

\begin{figure}[htb]
	\centering
	  \includegraphics[width=0.55\textwidth]{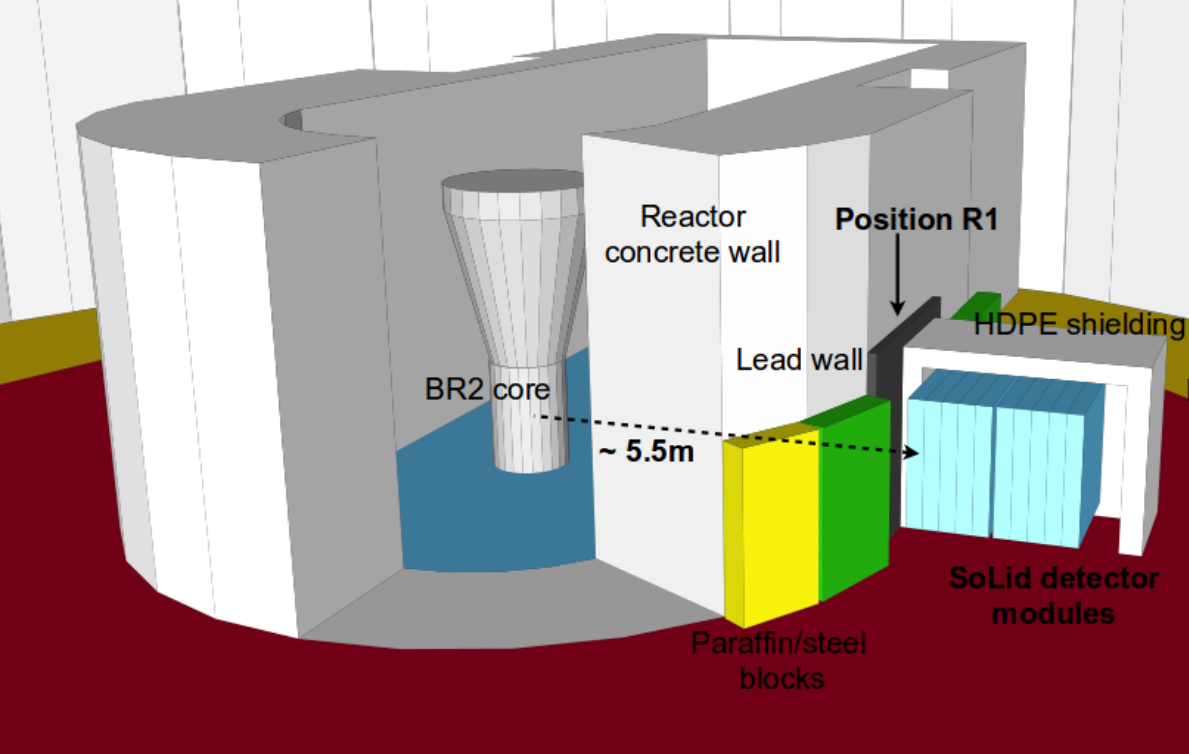}
	  \caption{A SoLid module placed in the BR2 reactor hall. \textit{Source}: \cite{Michiels:2016qui} (edited).}
	\label{michsolid}
\end{figure}

Its core measures only $50\si{\centi\meter}$ in diameter \cite{vsi}, which makes it an almost point-like $\nue$ source.
The architecture allows the SoLid detector modules to be placed at a very short distance of $5.5\ldots10\si{\meter}$ from the reactor core, as shown in figure~\ref{michsolid}. SoLid is intended to be the only experiment running at BR2, contributing to the stability of background conditions as other experiments might contribute to spallation neutron background \cite{Michiels:2016qui}.


BR2 is operated with $\prescript{235}{}{\mathrm{U}}$ as fuel, usually in cycles of three to four weeks \cite{wouter} at $40\ldots80\si{\mega\watt}$ thermal power \cite{vsi}.

\begin{figure}[htb]
	\centering
	  \includegraphics[width=0.7\textwidth]{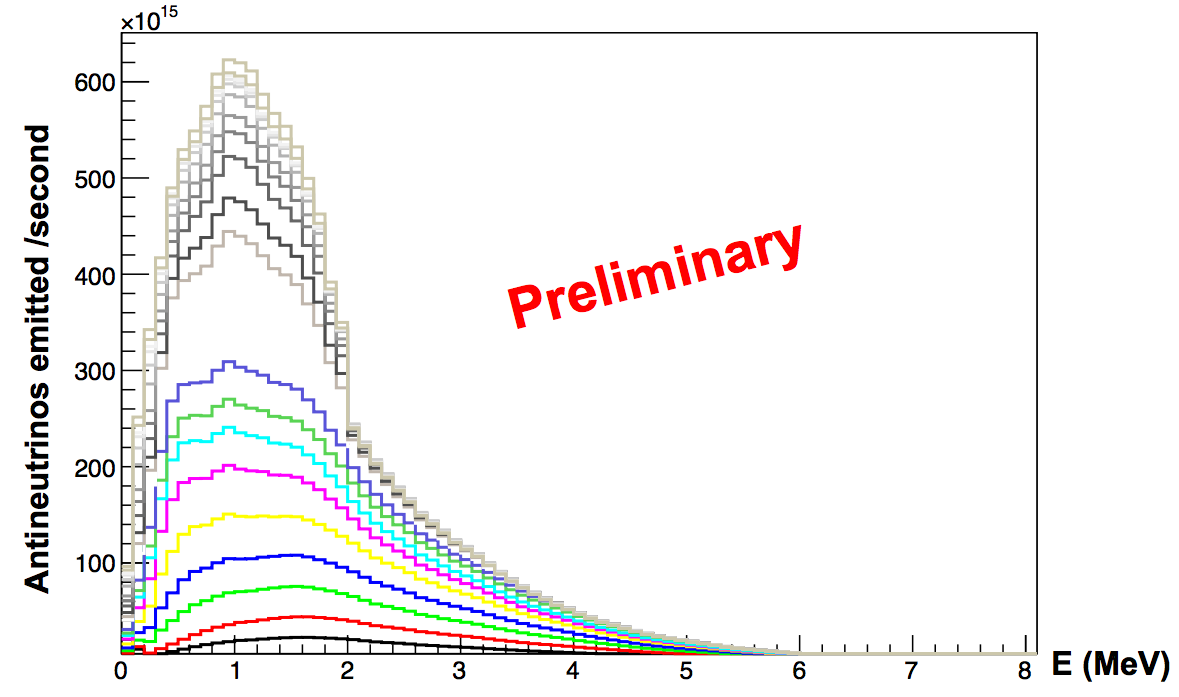}
	  \caption{Preliminary calculations of the $\nue$ energy spectrum of the BR2 reactor. Different colours indicate different measurement periods. \textit{Source}: \cite{kalcheva}.}
	\label{fig:kalchevaf}
\end{figure}

A very intense  neutrino flux of $\sim 10^{10} \nue \si{\per\second}$ is expected \cite{vsi} with the $\nue$ energy spectrum peaking at $\sim 1\si{\mega\electronvolt}$ and ranging to $\sim 6\si{\mega\electronvolt}$, as shown in figure~\ref{fig:kalchevaf} \cite{kalcheva}.

\subsection{IBD Capture}
\label{sec:ibd}
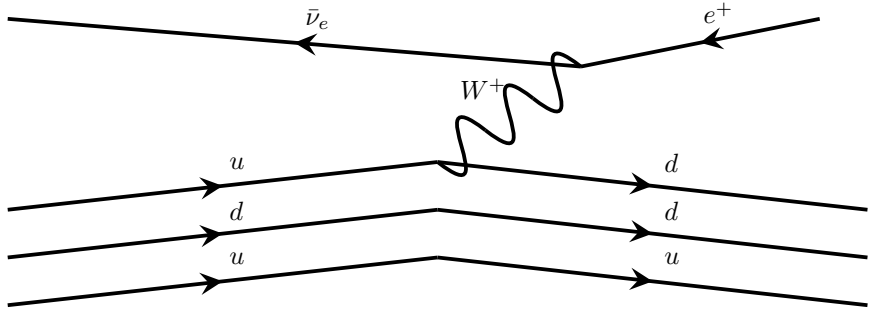
\begin{figure}[htb]
\centering
\begin{feynman}
\newcommand{\scl}{/4}
    \fermion[color=000000, endcaps=true, flip=false, label=u, labelDistance=0.12, labelLocation=0.50, showArrow=true, lineWidth=1.4285714285714286]{0\scl, 2\scl}{9\scl, 3\scl}
    \fermion[color=000000, endcaps=true, flip=false, label=d, labelDistance=0.12, labelLocation=0.50, showArrow=true, lineWidth=1.4285714285714286]{0\scl, 1\scl}{9\scl, 2\scl}
    \fermion[color=000000, endcaps=true, flip=false, label=u, labelDistance=0.12, labelLocation=0.50, showArrow=true, lineWidth=1.4285714285714286]{0\scl, 0\scl}{9\scl, 1\scl}
    \fermion[color=000000, endcaps=true, flip=false, label=u, labelDistance=0.12, labelLocation=0.50, showArrow=true, lineWidth=1.4285714285714286]{9\scl, 1\scl}{18\scl, 0\scl}
    \fermion[color=000000, endcaps=true, flip=false, label=d, labelDistance=0.12, labelLocation=0.50, showArrow=true, lineWidth=1.4285714285714286]{9\scl, 2\scl}{18\scl, 1\scl}
    \fermion[color=000000, endcaps=true, flip=false, label=d, labelDistance=0.12, labelLocation=0.50, showArrow=true, lineWidth=1.4285714285714286]{9\scl, 3\scl}{18\scl, 2\scl}
    \electroweak[color=000000, endcaps=true, flip=false, label=W^+, labelDistance=0.28, labelLocation=0.30, showArrow=false, lineWidth=1.4285714285714286]{9\scl, 3\scl}{12\scl, 5\scl}
    \fermion[color=000000, endcaps=true, flip=false, label=\nue, labelDistance=0.12, labelLocation=0.50, showArrow=true, lineWidth=1.4285714285714286]{12\scl, 5\scl}{0\scl, 6\scl}
    \fermion[color=000000, endcaps=true, flip=false, label=e^+, labelDistance=0.16, labelLocation=0.50, showArrow=true, lineWidth=1.4285714285714286]{17\scl, 6\scl}{12\scl, 5\scl}
\end{feynman}
\caption{Feynman diagram of Inverse $\beta$-decay $\nue+p\rightarrow e^+ + n$.}
\label{fig:ibetafeyn}
\end{figure}

Reactor neutrino detection is based on Inverse $\beta$-Decay (IBD),
\begin{equation}
\label{eq:IBD}
\nue+p\rightarrow e^+ + n,
\end{equation}
with an energy threshold of $1.8\si{\mega\electronvolt}$ \cite{Boehm:2000va}.
Equation~\ref{eq:IBD} refers to the same process as
\begin{equation}
\nue+udu \rightarrow W^+ + \nue + udd \rightarrow e^+ + udd
\end{equation}
shown in figure~\ref{fig:ibetafeyn}, the quark triplet $udu$ equating to a $p$, and the triplet $udd$ equating to a $n$. It has a well-known cross section of \cite{Bemporad:2001qy}\cite{Vidyakin:1987ue}

\begin{equation}
 \sigma_{\text{IBD}}=0.952\left(\frac{E_e p_e}{1\si{\square\mega\electronvolt}}\right)\times 10^{-43}\si{\square\centi\meter}.
\end{equation}

The products of the IBD process are a positron ($e^+$) and a neutron ($n$). These product are easily detectable. At SoLid, the products are measured using two different scintillating materials for conversion into visible photons: Polyvinyl toluene -- or PVT -- for $e^+$ capture and Lithium-6 Silver-doped Zinc Sulphide (\LiZnS) layers for $n$ capture. 
\begin{figure}[htb]
	\centering
	  \includegraphics[width=0.4\textwidth]{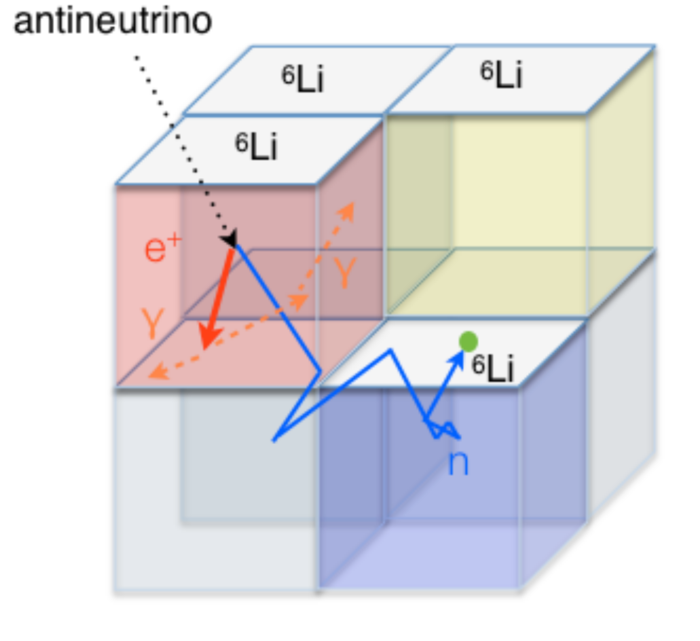}
	  \caption{An array of PVT cubes with \LiZnS-layers. The PVT is used for positron capture, while the \LiZnS-sheets capture the neutrons. \textit{Source:} \cite{cubesy}.}
	\label{fig:cubes}
\end{figure}
\subsubsection{Detector Components}
\begin{figure}[htb]
	\centering
	  \includegraphics[width=0.7\textwidth]{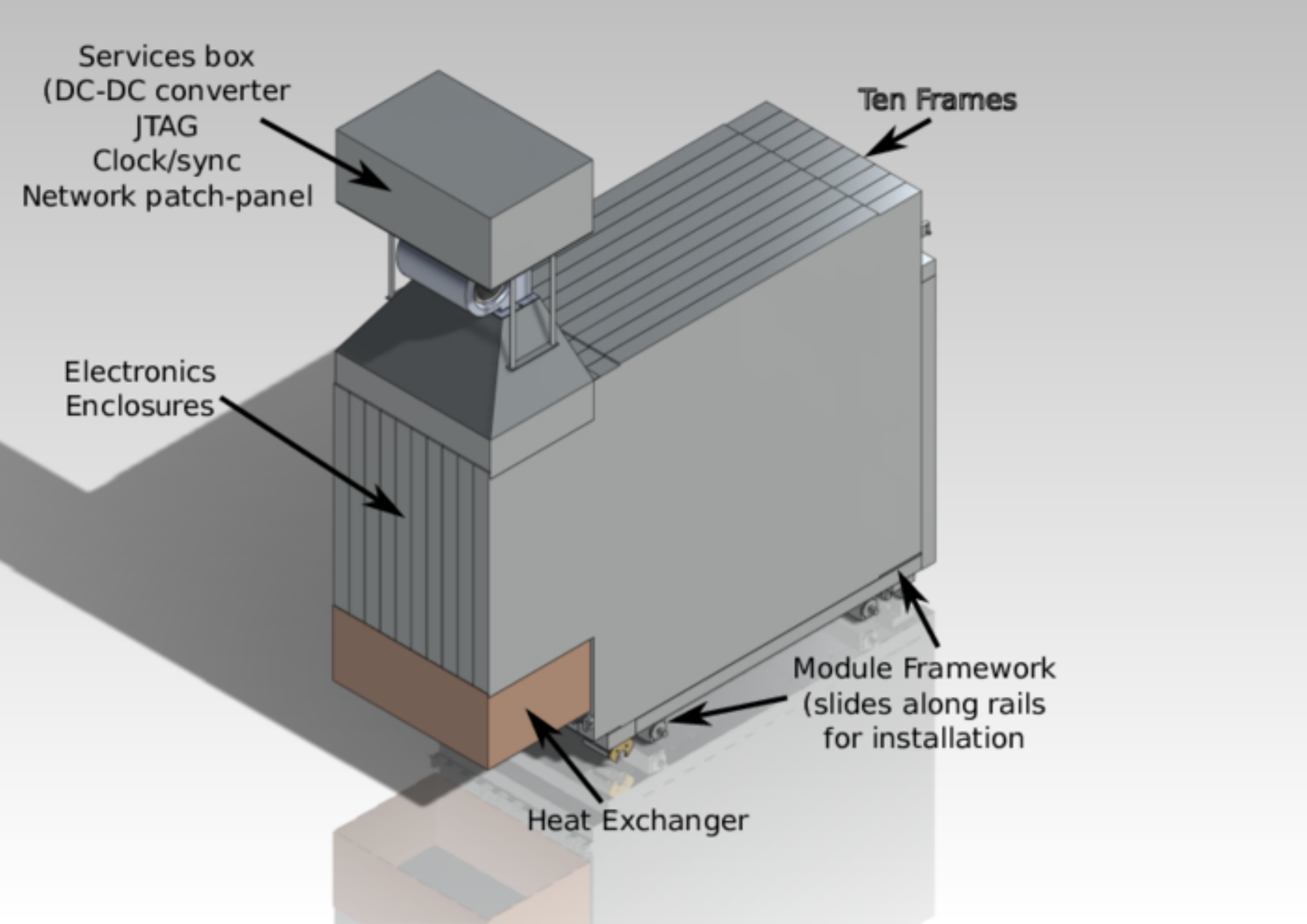}
	  \caption{Three-dimensional rendering of a ten-plane SoLid module. \textit{Source:} \cite{Arnold:2017lph}.}
	\label{fig:module}
\end{figure}

The SoLid module is built out of 10 planes, which each contain 256 cubes containing the scintillation materials. The scintillators capture the IBD products and convert them into light to be guided to the photon sensors. A sketch of a full detector module is shown in figure~\ref{fig:module}.

\begin{figure}[htb]
	\centering
	  \includegraphics[width=0.4\textwidth]{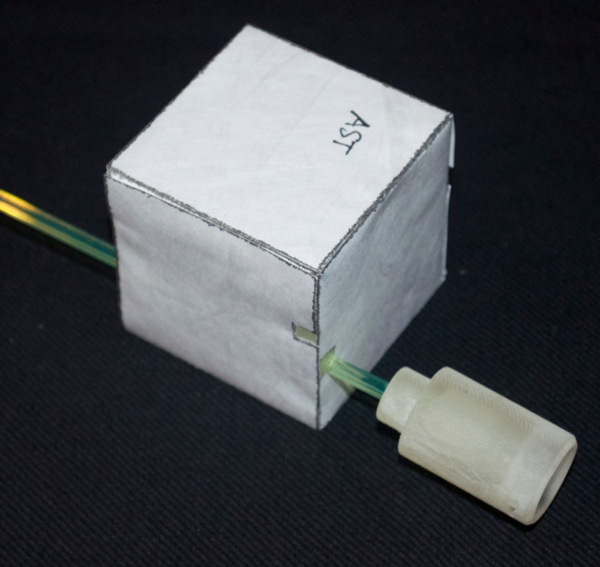}
	  \caption{A PVT cube wrapped in \textsc{Tyvek} sheets as optical shielding and one WLS fibre attached to it. \textit{Source:} \cite{michielsmaster}.}
	\label{fig:tyvec}
\end{figure}

\paragraph{The Cube} consists of PVT and measures $5\si{\centi\meter}\times 5\si{\centi\meter}\times5\si{\centi\meter}$. A layer of \LiZnS\phantom{a}is attached to one side of the cube. Several cubes are arranged discretely as an array (see figure~\ref{fig:cubes}). Each cube contains four grooves for placement of Wavelength-Shifting (WLS) fibres which guide the light to the photosensors.
The single cubes are optically shielded from each other by being wrapped in \textsc{Tyvek} paper. \cite{Moortgat:2015bwg} The assembly of a cube with one WLS fibre can be seen in figure~\ref{fig:tyvec}.

\paragraph{The Plane} consists of 256 cubes, arranged as a $16\times16$ array, with the photosensors being placed at the frame and the read-out electronic modules being attached at one side. Since two Wavelength-Shifting Fibres cross each row both in the horizontal and the vertical direction, 64 photosensors are used per plane, also meaning that the electronics board is designed to read out 64 channels.

\paragraph{The Module} which can be seen in figure~\ref{fig:module} is built upon 10 planes. The modules are independent of each other mechanically as well as in terms of power supply, clock and control distribution, heat-exchanger and cooling-air blower, which means they can be operated separately for commissioning \cite{Arnold:2017lph}.

\subsubsection{Positron Capture}
\label{sec:positroncapture}
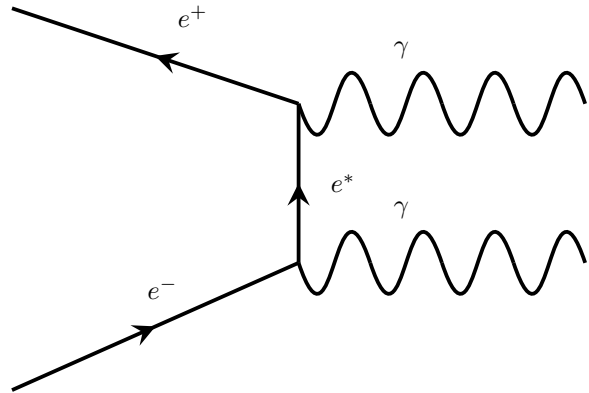
\begin{figure}[htb]
\centering
\begin{feynman}
\newcommand{\scl}{/6}
    \fermion[color=000000, endcaps=true, flip=false, label=e^-, labelDistance=0.22, labelLocation=0.50, showArrow=true, lineWidth=1.4285714285714286]{0\scl, 0\scl}{9\scl, 4\scl}
    \fermion[color=000000, endcaps=true, flip=false, label=e^+, labelDistance=0.22, labelLocation=0.50, showArrow=true, lineWidth=1.4285714285714286]{9\scl, 9\scl}{0\scl, 12\scl}
    \fermion[color=000000, endcaps=true, flip=false, label=e^*, labelDistance=0.12, labelLocation=0.50, showArrow=true, lineWidth=1.4285714285714286]{9\scl, 4\scl}{9\scl, 9\scl}
    \electroweak[color=000000, endcaps=true, flip=false, label=\gamma, labelDistance=0.28, labelLocation=0.30, showArrow=false, lineWidth=1.4285714285714286]{9\scl, 4\scl}{18\scl, 4\scl}
    \electroweak[color=000000, endcaps=true, flip=false, label=\gamma, labelDistance=0.28, labelLocation=0.30, showArrow=false, lineWidth=1.4285714285714286]{9\scl, 9\scl}{18\scl, 9\scl}
\end{feynman}
\caption{Feynman diagram of electron-positron annihilation $e^+ + e^- \rightarrow \gamma + \gamma$.}
\label{fig:electronpositronannihilation}
\end{figure}

PVT is used as the scintillator material which catches the positron. The positron is absorbed within $\mathcal{O}\left({10^{-8}}\right)\si{\second}$ by the PVT and is emitted as a short ($\mathcal{O}\left({10^{-7}}\right)\si{\second}$), intense light pulse \cite{Michiels:2016qui} \cite{Ryder:2015sma}. Thereafter, the positron annihilates with an electron coming from the detector via the process 

\begin{equation}
e^+ + e^- \rightarrow \gamma + \gamma 
\end{equation}
as depicted in figure~\ref{fig:electronpositronannihilation}. The $\gamma$s each have an energy of $511\si{\kilo\electronvolt}$ \cite{Moortgat:2015bwg}.

The light pulse is captured by the SiPMs. The total energy deposited in the cube can be used to reconstruct the $\nue$ energy, when correcting for the small fraction of annihilation $\gamma$ energy on average deposited in the interaction cube \cite{Ryder:2015sma}.

\subsubsection{Neutron Capture}
\label{sec:neutroncapture}
Whilst positrons are captured by the PVT scintillator, neutrons are caught by the \LiZnS\phantom{a}sheets. And while positrons cause a prompt response from the scintillator, neutrons do not; they undergo thermalisation while being scattered through the material before they are captured by $\prescript{6}{}{\mathrm{Li}}$ via the process 
\begin{equation}
n+\prescript{6}{}{\mathrm{Li}}\rightarrow \prescript{3}{}{\mathrm{H}}+\alpha+4.78\si{\mega\electronvolt}=\prescript{3}{}{\mathrm{H}}+\prescript{4}{}{\mathrm{He}^{2+}}+4.78\si{\mega\electronvolt}.
\end{equation}

\begin{figure}[htb]
	\centering
	  \includegraphics[width=1.0\textwidth]{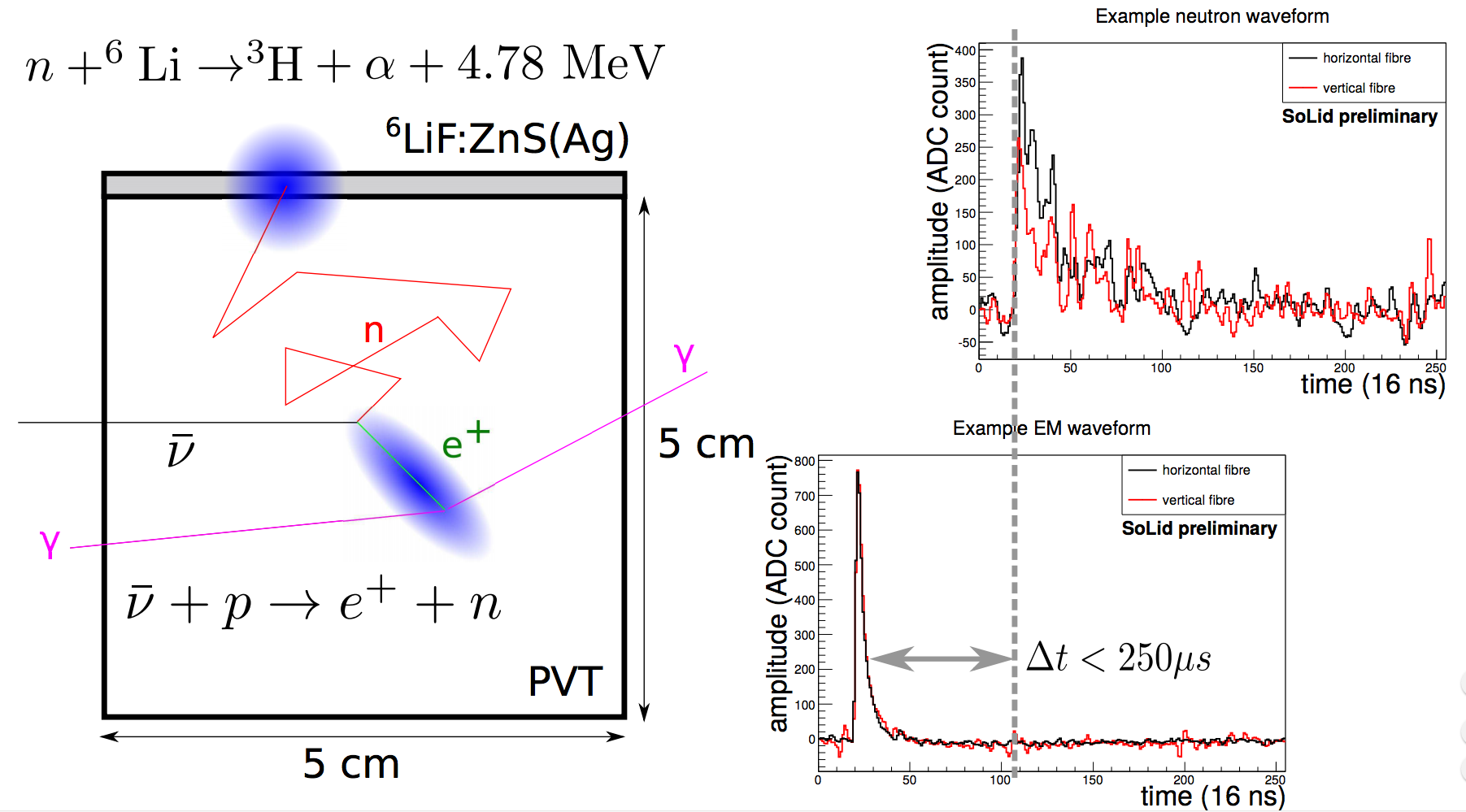}
	  \caption{Right: Interaction of $n$ and $e^+$ in the detector cube. Left: Comparison of the signals of $n$ capture (upper) in \LiZnS, and $e^+$ capture in PVT (lower). \textit{Source:} \cite{Ryder:2015sma}.}
	\label{fig:scintillator}
\end{figure}

Both $\prescript{3}{}{\mathrm{H}}$ and the $\alpha$-particle ($=\prescript{4}{}{\mathrm{He}^{2+}}$) contain sufficient energy to excite the electrons in the $\mathrm{ZnS}$ crystals. Scintillation light is emitted by de-excitation of these electrons over a longer period of $\mathcal{O}\left(10^{-6}\right)\si{\second}$, thus one order of magnitude higher than the positron light pulse -- this leads to a shape for the neutron light emission which is distinct from the short, high-amplitude pulse generated by positron scintillation. 
The neutron \LiZnS\phantom{a}scintillation is delayed by $\mathcal{O}\left(10^{-4}\right)\si{\second}$ when compared to the prompt response of the PVT scintillator to the positron. The difference of $n$ and $e^+$ signal is shown in figure~\ref{fig:scintillator}, with a sketch of the underlying interaction within the cube. \cite{michielsmaster} \cite{Ryder:2015sma}\cite{Sweany:2014ena}

\subsubsection{Light Capture}
\begin{figure}[htb]
	\centering
	  \includegraphics[width=0.55\textwidth]{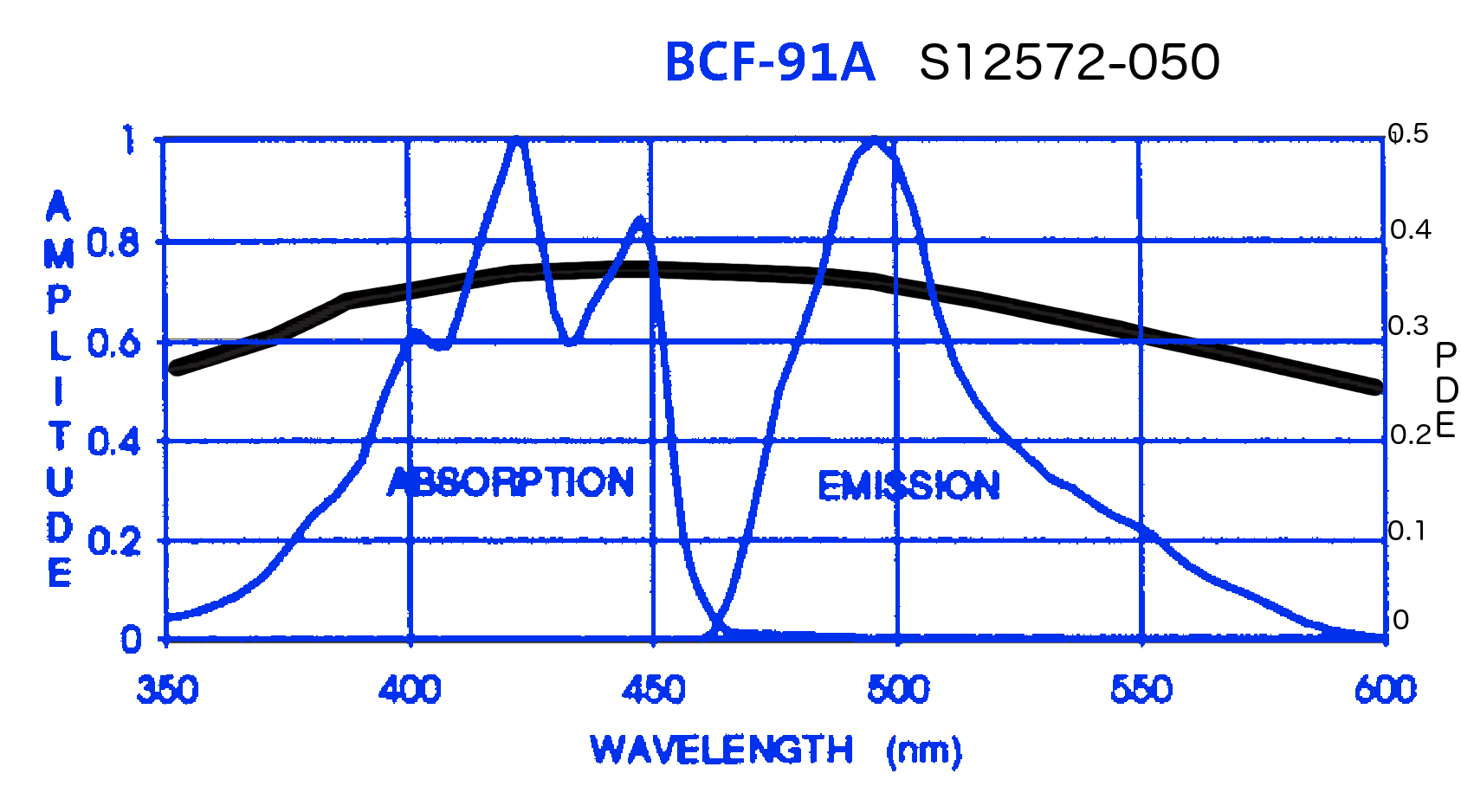}
	  \caption{Absorption and emission spectra of the WLS fibre is shown in blue, Photon Detection Efficiency (PDE) of the SiPM is shown in black. \textit{Source:} \cite{saintgobain} \cite{hamamatsudata} (edited).}
	\label{fig:wlsvssipm}
\end{figure}

\paragraph{Wavelength-Shifting Fibres} collect and convert the light signals produced by $n$ and $e^+$ capture within the cubes in order to transport them to the photosensors \cite{Schluter:2011rv}. They absorb photons at a given wavelength and subsequently emit them isotropically at a longer wavelength, thus at a lower energy (see figure~\ref{fig:wlsvssipm}). 
The purpose of using WLS fibre is to randomise the direction of the emitted light, in order that a fraction of the light can be captured by total internal reflection. 
The WLS fibre used at SoLid -- \textsc{Saint-Gobain bcf-91a} -- consists of a polystyrene-based core and two layers of polymethyl methacrylate (PMMA), a highly transparent thermoplastic material, and shifts the light from blue to green spectrum. \cite{verstraten} \cite{Oliveira20102098} It is rectangular in shape and has a cross section of $3\si{\milli\meter}\times3\si{\milli\meter}$, corresponding to the active area of the SiPM sensor (see \cite{saintgobain} \cite{hamamatsudata} for technical information). 

\paragraph{Silicon Photomultipliers} (SiPMs) -- also referred to as Multi-Pixel Photon Counters (MPPCs) -- are used as photosensors. SiPMs are pixel arrays of avalanche photodiodes operated $10\ldots20\%$ above breakdown voltage, in so-called Geiger mode \cite{Buzhan:2001xq}. Their advantages lie in their robustness, high photon-detection efficiency, single photon detection capability as well as their resolution, their low operational voltage and inexpensiveness. However, their high Dark-Count Rate (DCR) at room temperature, their excess noise due to inter-pixel cross-talk and their generation of after-pulses are disadvantages of the devices \cite{Chmill:2016ghf} \cite{Otte:2006gm}. DCR in SoLid is decreased by cooling down the detector module to $5\si{\degreeCelsius}$.
On SoLid, \textsc{Hamamatsu S12572-050} devices are employed, containing 3600 pixels each, peak sensitivity wavelength of $450\si{\nano\meter}$ and a Photon Detection Efficiency (also called \textit{quantum efficiency}) of $0.35$ and nominal DCR of typically $1\si{\kilo\hertz}$, decreased by one order of magnitude of by cooling down the detector \cite{hamamatsudata} (see \cite{wouter} for DCR--temperature dependence).

Figure~\ref{fig:wlsvssipm} shows the absorption and emission spectra of the WLS fibre, as well as Photon Detection Efficiency (PDE) of the SiPM.
\subsubsection{Background}
As SoLid is an overground experiment in proximity to the reactor core, a significant amount of background is faced in comparison to underground experiments \cite{wouter}. These backgrounds include:
\paragraph{Cosmic Muons} cause PVT scintillation when passing through the cube, thereby causing significant contribution to the background \cite{michielsmaster}.
\paragraph{Cosmogenic Background} is background caused by spallation reactions, such as decays of isotopes produced by muon-induced spallation reactions, spallation reactions caused by muon-induced neutrons entering from outside the detector, or nuclear recoils initiated by such neutrons \cite{Galbiati:2005ft}.
 Fast neutrons from cosmic ray showers might cause a similar signal to a positron signal by inducing a $p$ recoil signal and being captured shortly after, mimicking an IBD signal. In order to reduce this effect, a water shield is built around the SoLid detector module.  \cite{verstraten}
\paragraph{Intrinsic Background} can occur by contamination of the \LiZnS-layer by $\prescript{214}{}{\mathrm{Bi}}$ in the fabrication process. The $\prescript{214}{}{\mathrm{Bi}}$ isotope decays via $\beta^-$-decay to $\prescript{214}{}{\mathrm{Po}}$, emitting an  $e^-$. The $\prescript{214}{}{\mathrm{Po}}$ itself decays by emitting an $\alpha$-particle, looking signal-wise like a neutron. The $e^-$ and $\alpha$-particle can fake an IBD event \cite{verstraten} \cite{verreyken}.
\begin{figure}[!htb]
	\centering
	  \includegraphics[width=0.3\textwidth]{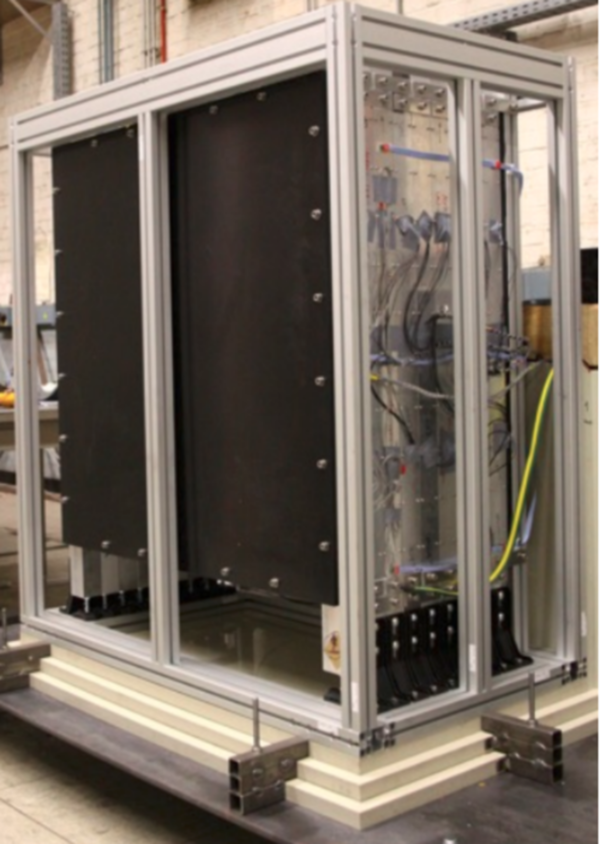}
	  \caption{The SM1 prototype module. \textit{Source:} \cite{Michiels:2016qui}.}
	\label{fig:sm1}
\end{figure}
\begin{figure}[!htb]
	\centering
	  \includegraphics[width=0.9\textwidth]{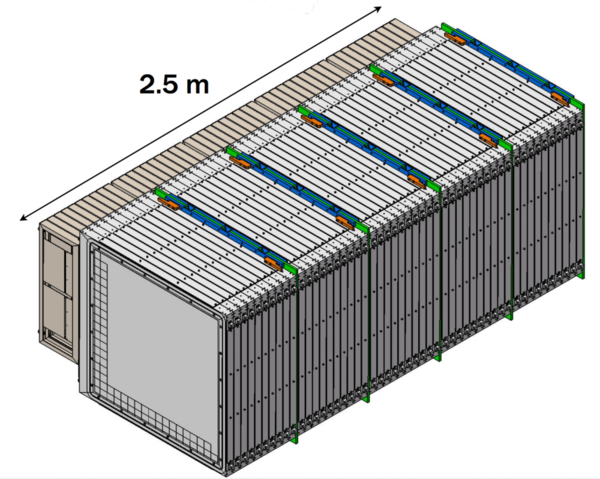}
	  \caption{Three-dimensional rendering of the $1.6\si{\tonne}$-detector, consisting of 50 planes, used in Phase 1. \textit{Source:} \cite{neutrinovacheret}.}
	\label{fig:phase}
\end{figure}
\begin{figure}[!htb]
	\centering
	  \includegraphics[width=0.9\textwidth]{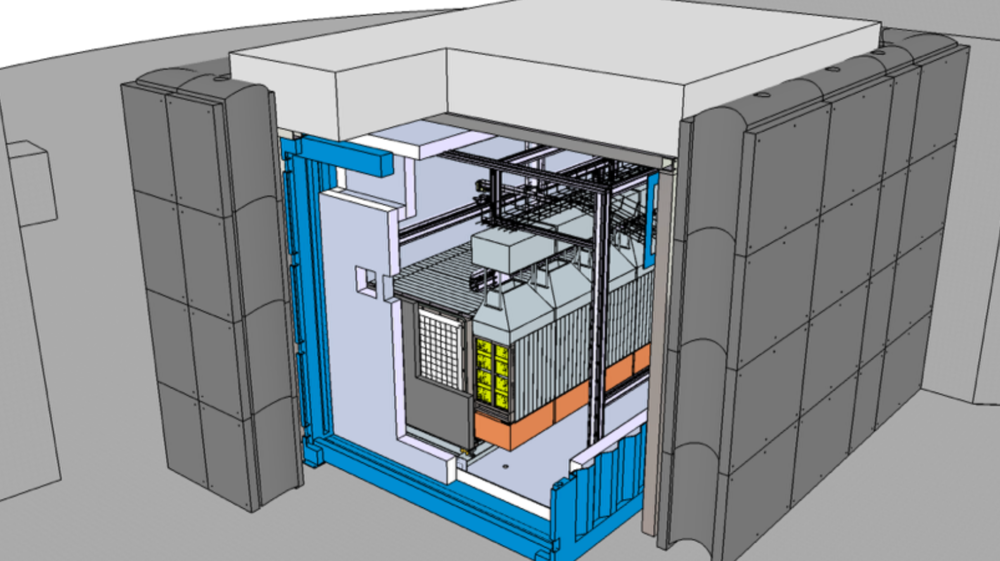}
	  \caption{Three-dimensional rendering of Phase 1-detector placed in a container with surrounding water shielding. \textit{Source:} \cite{appryder}.}
	\label{fig:withshielding}
\end{figure}

\subsection{Status of the experiment}
SoLid is planned to enter its first large-scale data-taking phase in 2017. Two prototype phases have run earlier in order to demonstrate feasibility: NEMENIX and SM1.

\subsubsection{NEMENIX}
The first prototype, NEMENIX, was built in 2013. The $6\si{\kilogram}$-detector consists of 64 PVT cubes and 32 read-out channels, forming a detecting volume of approximately $20\si{\centi\meter}\times 20\si{\centi\meter} \times 20\si{\centi\meter}$. It was designed to provide proof of concept as well as acquire estimates of background and signal coincidence rates on the BR2 reactor site. \cite{michielsmaster} \cite{verreyken}

\subsubsection{SM1}
The SM1 prototype detector, shown in figure~\ref{fig:sm1},  contains $288\si{\kilogram}$ of active scintillation material, arranged in 9 planes with 256 cubes each. A total of 100 channels is used for read-out. It took data in 2015 both when the BR2 reactor was running and when it was turned off. During the turn-off period, various radioactive sources have been used to investigate background conditions: $\prescript{137}{}{\mathrm{Cs}}$, $\prescript{60}{}{\mathrm{Co}}$, $\mathrm{AmBe}$ and $\prescript{252}{}{\mathrm{Cf}}$. The latter two have been used for neutron studies, while $\prescript{137}{}{\mathrm{Cs}}$ and $\prescript{60}{}{\mathrm{Co}}$ were used for the calibration of the electromagnetic signals. \cite{Saunders:2016gcc} \cite{verreyken} An energy threshold trigger has been used for read-out \cite{Arnold:2017lph}.

\subsubsection{Phase 1}
The first main run of SoLid -- Phase 1 -- will start in 2017. $1.6\si{\tonne}$ of material is being used as scintillation material. 5 modules, thus 50 planes, are employed, and 3200 SiPM channels are used for read-out. \cite{Arnold:2017lph} A rendering of the bare detector can be seen in figure~\ref{fig:phase} and the rendering of the detector placed in the container is shown in figure~\ref{fig:withshielding}. Among upgrades in online software, storage, controls and detector technology -- an additional \LiZnS-layer will be added -- a new trigger system will be implemented which significantly increases purity of the data and decreases data rate.

\subsubsection{Phase 2}
\begin{figure}[!htb]
	\centering
	  \includegraphics[width=0.7\textwidth]{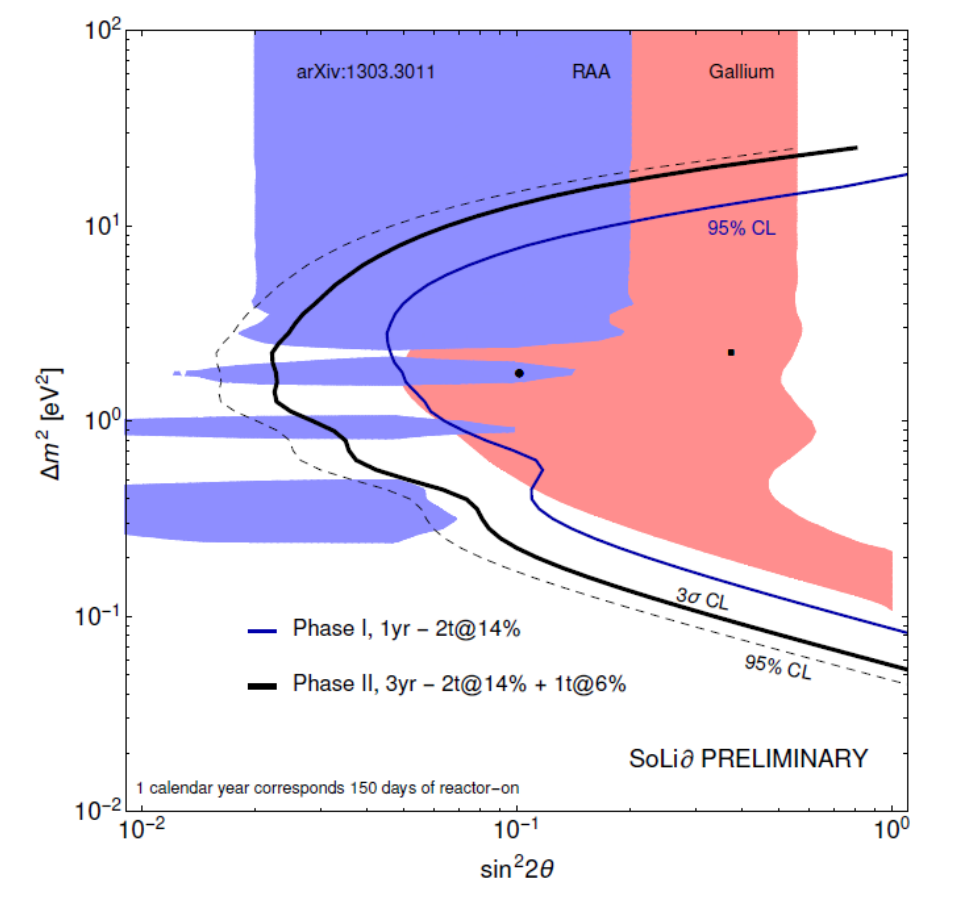}
	  \caption{The estimated sensitivity of the SoLid experiment for Phase 1 (blue line) and Phase 2 (black line) to the reactor antineutrino anomaly (RAA) region (blue area) and the Gallium anomaly region (red area) as a function of $\Delta m^2$ and $\sin^22\theta$. The $95\%$ C.L. are shown. \textit{Source:} \cite{Ryder:2015sma}.}
	\label{fig:sensitivity}
\end{figure}

The detector can be upgraded by adding more modules or by constructing an additional detector station with higher energy resolution. A potential upgrade which should be run as Phase 2 will have total sensitive mass up to $3\si{\tonne}$. \cite{Abreu2016} A sensitivity plot for both Phase 1 and Phase 2 can be seen on figure~\ref{fig:sensitivity}.

\clearpage

\section{The SoLid Read-out System}		
\label{sec:readout}
\begin{figure}[htb]
	\centering
	  \includegraphics[width=0.85\textwidth]{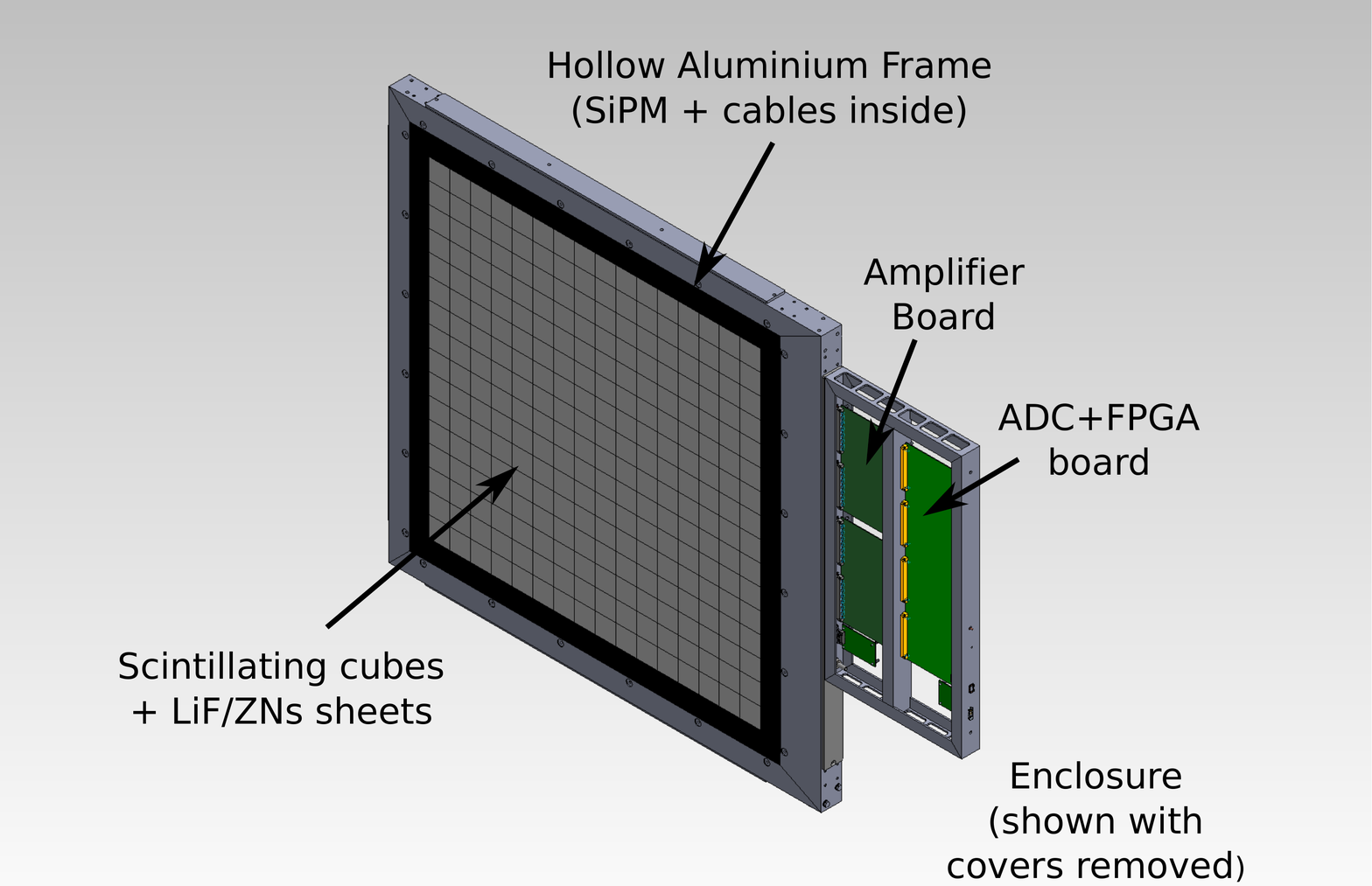}
	  \caption{Three-dimensional rendering of a plane containing the cubes, the frame and the read-out electronics. \textit{Source:} \cite{Arnold:2017lph}.}
	\label{fig:frame}
\end{figure}

The read-out system for acquiring the SiPM signals is composed out of two parts: the analogue front-end and the digital front-end, that also includes the trigger on a Field-Programmable Gate Array (FPGA) chip, as well as the power- and the I$^2$C-board. The read-out electronics is placed on side of the frame containing the cubes, as shown in figure~\ref{fig:frame}.

The boards are custom-made to serve the needs of the SoLid experiment.

\begin{figure}[htb]
	\centering
	  \includegraphics[width=0.85\textwidth]{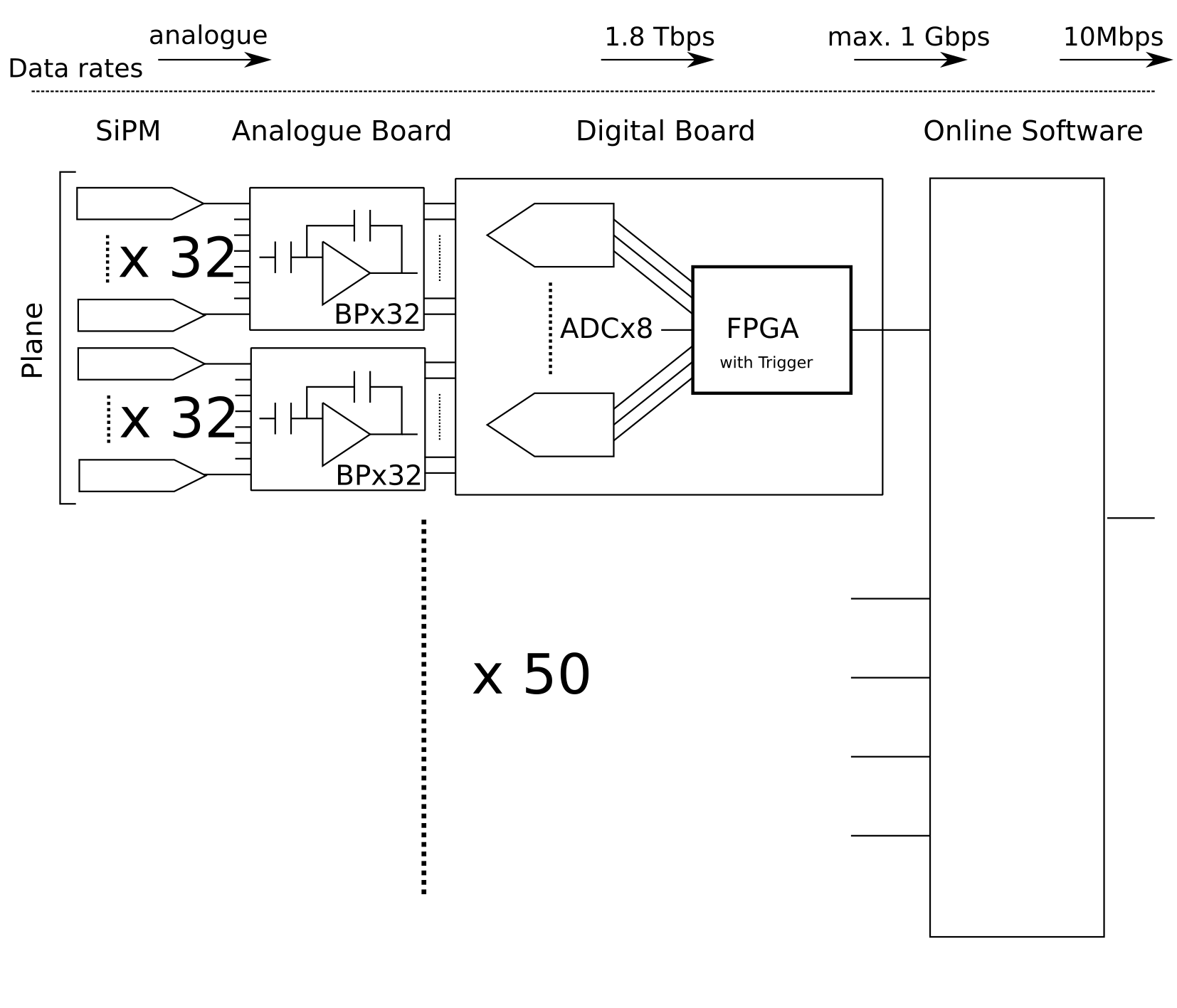}
	  \caption{Flow of the signal: The SiPM input signal is shaped by the analogue board, then sampled and processed by the digital board before being fed into the online software. The expected data rate at each stage is outlined.}
	\label{fig:dataflow}
\end{figure}

As can be seen in the data flow diagram in figure~\ref{fig:dataflow}, the SiPM sensor signals are shaped and amplified by a band-pass filter on the analogue front-end board before being sampled and processed by the digital board's components. The data then is sent out via Ethernet to the central online software.

\subsection{Analogue Front-end}
Two analogue front-end boards are used per plane, each serving 32 channels. Each board provides the bias voltage to the SiPMs, according to the value set by the digital board. On the input side, it receives, amplifies, band-pass filters and shapes the signals. Accordingly, it is connected to the digital board on one end and to the SiPMs on the other, using differential signalling for the signals. \cite{Arnold:2017lph}
\subsection{Digital Front-end}
\begin{figure}[htb]
	\centering
	  \includegraphics[width=0.85\textwidth]{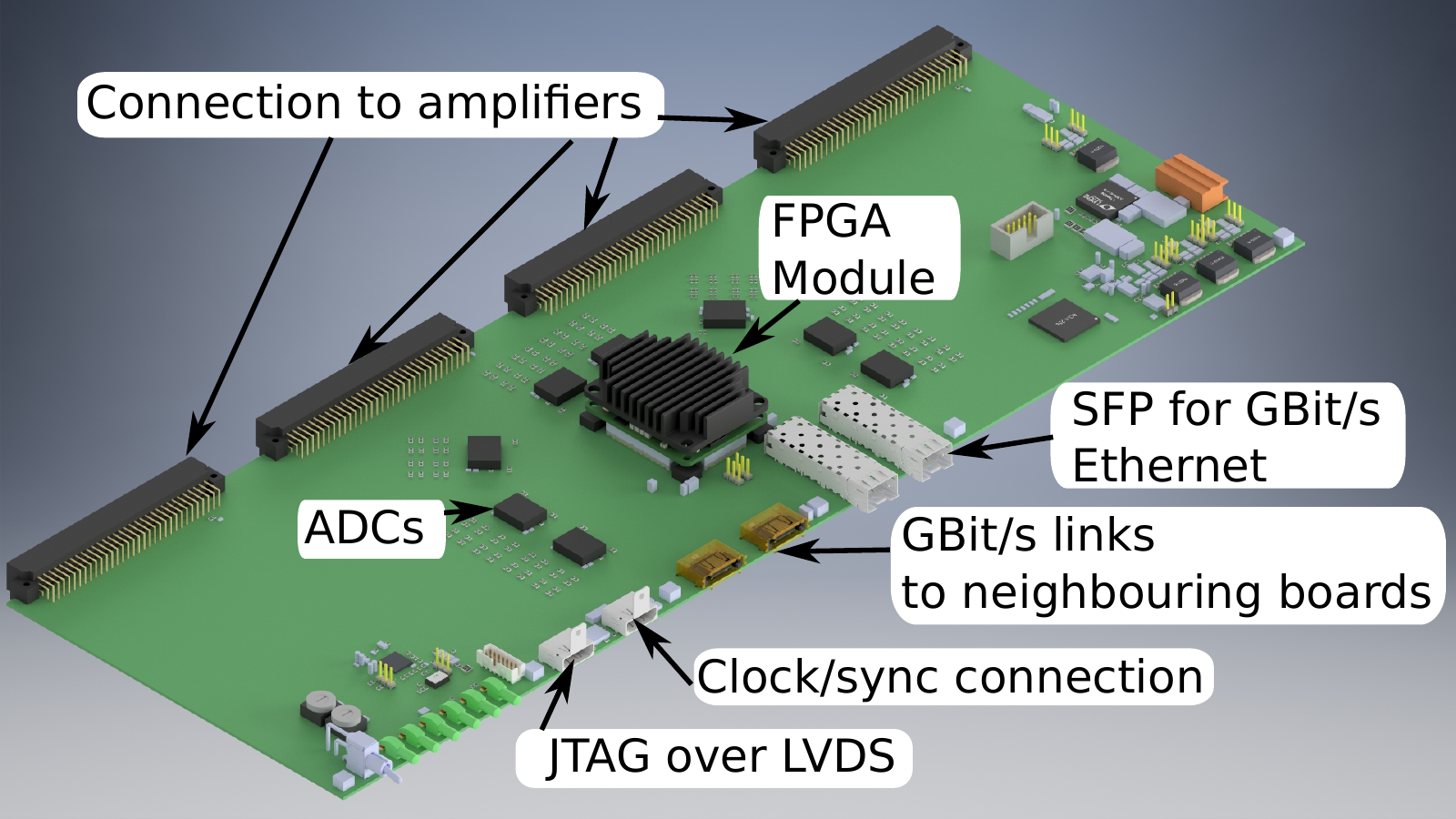}
	  \caption{Three-dimensional rendering of a digital front-end board. \textit{Source:} \cite{Arnold:2017lph}.}
	\label{fig:digitalboard}
\end{figure}

The digital front-end board receives the signals from the analogue board. 8 Analogue-Digital Converters (ADCs) sample 8 channels each at a rate of $40\si{\mega\hertz}$ and a resolution of $14\si{\bit}$. The digitised signals are fed into the \textsc{Trenz} FPGA module containing a \textsc{Xilinx Artix-7} FPGA chip. The digital board connects to JTAG over LVDS, contains a $\si{\giga\bit\per\second}$-Ethernet interface using Small Form-factor Pluggable transceivers (SFPs), direct $2.5\si{\giga\bit\per\second}$ links to neighbouring boards and is connected to the per-module clock and control signal distribution. A three-dimensional rendering can be seen in figure~\ref{fig:digitalboard}. \cite{Arnold:2017lph}
\subsection{Other Boards}
The read-out electronics assembly also contains two smaller boards, each designated for a specific function. The power board, located under the digital front-end board, provides $5\si{\volt}$ outputs for digital and analogue use each, $-3.3\si{\volt}$ for the FPGA, and $70\si{\volt}$ for the SiPM bias voltage. The I$^2$C board, which is positioned beneath the analogue board, contains an I$^2$C communication interface.

\subsection{Shielding}
\begin{figure}[htb]
	\centering
	  \includegraphics[width=0.8\textwidth]{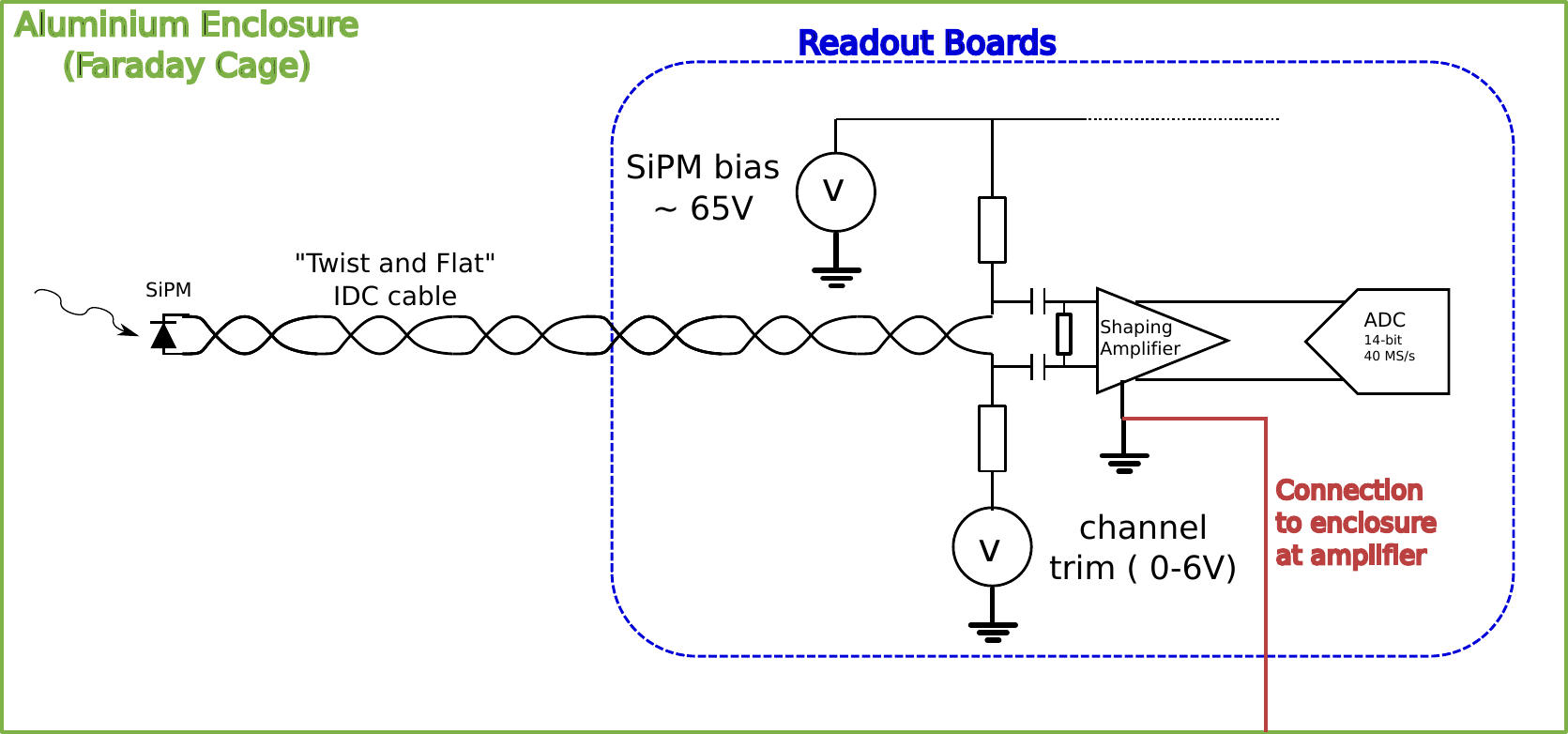}
	  \caption{The shielding and grounding scheme of the read-out electronics. \textit{Source:} \cite{Arnold:2017lph}.}
	\label{fig:shielding}
\end{figure}

As the plane's frame is made of hollow extruded aluminium sections that are connected together and to the read-out electronics boards, it is a Faraday cage blocking electrical fields. A sketch of the shielding scheme can be seen in figure~\ref{fig:shielding}.
\subsection{Data Generation and Rates}
\begin{table}[]
\centering
\begin{tabular}{llll}
\hline\hline
\textbf{Stage}  & \textbf{Data rate $R [\si{\per\second}]$} & \textbf{Data rate $R [\si{\per\day}]$} & \textbf{Reduction factor $r$} \\ \hline
Digital board   &    $=1.8\si{\tera\bit}$                &             $=19\si{\peta\byte}$                           &                         --      \\
Online software (maximum) &   $\sim1.0\si{\giga\bit}$                 & $\sim11\si{\tera\byte}$                   &                          $\sim1800$     \\
Data storage &  $\sim10\si{\mega\bit}$                                        &$\sim100\si{\giga\byte}$                                &     $\sim100$ \\
\hline\hline
\end{tabular}
\caption{Data rates at the different stages of the data acquisition chain per second and per day. Reduction factor is given to the previous stage.}
\label{tab:datarates}
\end{table}

SoLid faces a large amount of background signals due to it being an overground experiment in proximity to the reactor core. This requires the read-out system -- and the trigger part -- to reject as many background signals as possible. 
The extent to which this rejection has to take place is dependent on data generation on one side and on the data handling capacity of the online software on the other.

\paragraph{Data Generation} is determined by sampling rate and resolution. Since a $14\si{\bit}$-ADC is used at a sampling frequency of $40\si{\mega\hertz}$, with 64 channels per plane. The data rate -- or bit rate -- per plane amounts to
\begin{equation}
R_{f}=\text{\#Channel}\times\text{Resolution}\times\text{Frequency}=64\times14\si{\bit}\times40\si{\mega\hertz}=36\si{\giga\bit\per\second}.
\end{equation}
Since 50 planes are employed on Phase 1, the data rate for the whole detector is 
\begin{equation}
R_d=1.8 \si{\tera\bit\per\second}.
\end{equation}

\paragraph{Data Handling Capacity} is set by the capacity of the online software to process data. The software is estimated to be able to handle data up to a rate of $R_s\sim1\si{\giga\bit\per\second}$ \cite{whitepaper}. The minimum data reduction factor therefore is
\begin{equation}
r_d=\frac{R_d}{R_s}\sim 1800.
\end{equation}
As reduction should be above the minimum limit in order to prevent data pile-up or even data loss, a reduction factor of $\mathcal{O}(10^4)$ is aimed for. The online software reduces the amount of data by additional two order of magnitudes, leading to a data storage rate 
\begin{equation}
R_s\sim 10\si{\mega\bit\per\second}\sim 100\si{\giga\byte\per\day},
\end{equation}

limited by the storage capacity at BR2.
A summary of data rates and reduction factors is given in table~\ref{tab:datarates}. Obviously, the total reduction of the data rate by a factor of one million forms a challenge, in particular due to the fact that high efficiency is required and that a large amount of background data is faced. It is the task of the trigger to provide efficient, yet pure data reduction.

\subsection{The SoLid Firmware}
\label{sec:firmware}
\begin{sidewaysfigure}[htb]
	\centering
	  \includegraphics[width=1.0\textwidth]{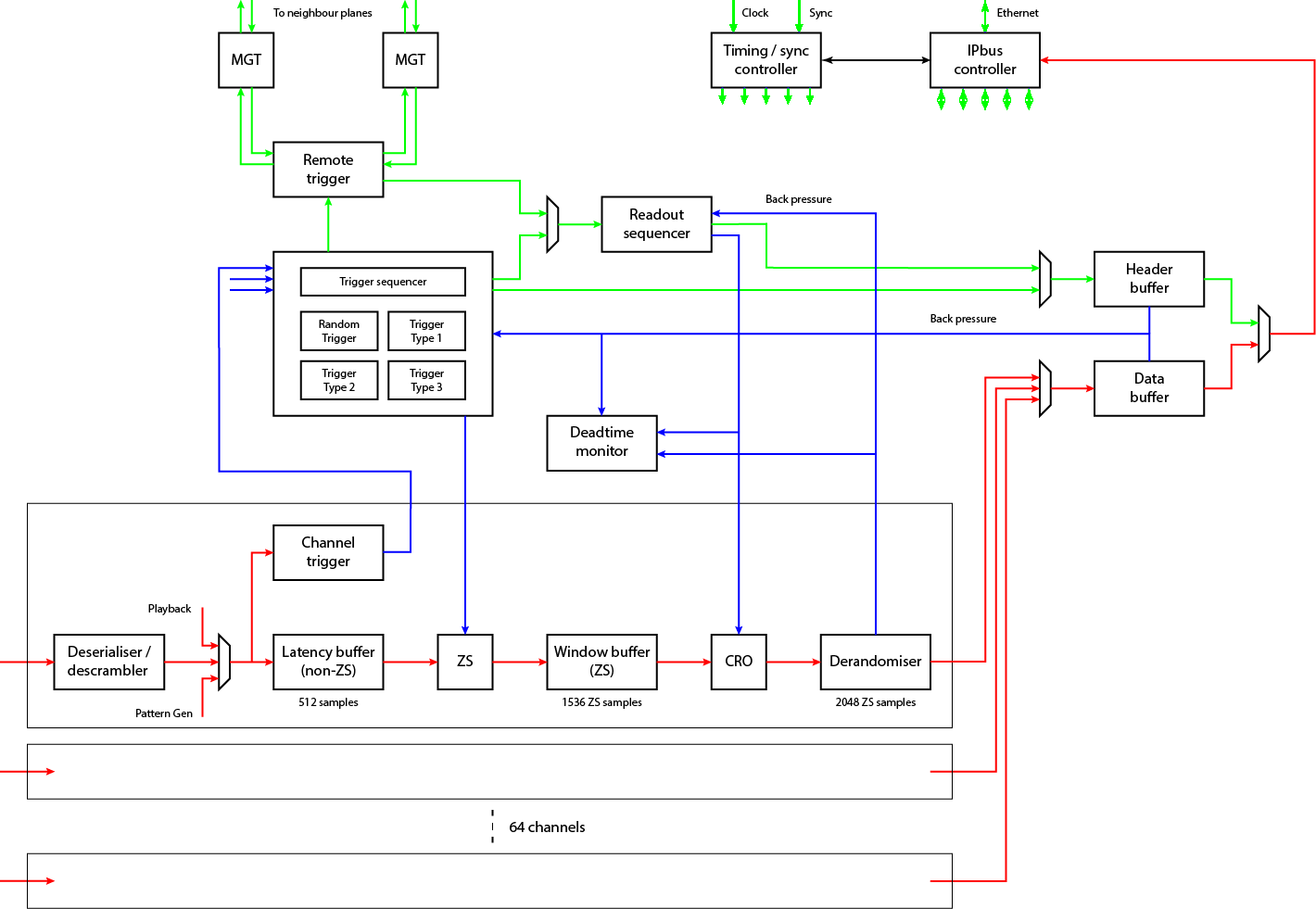}
	  \caption{Block diagram of the firmware. \textit{Source:} \cite{Arnold:2017lph} (edited).}
	\label{fig:solidreadout}
\end{sidewaysfigure}

The SoLid FPGA firmware is responsible for buffering the data, triggering on it, and for communication with other planes as well as with the data acquisition device. Also, it has the slow-control for the SiPMs integrated in its functionality. It forms therefore a crucial part of the experiment's read-out chain. A block diagram with the firmware modules is shown in figure~\ref{fig:solidreadout}. 

The firmware is based on the \textit{IPbus} protocol, a gigabit Ethernet-based reliable high-performance protocol designed for particle physics experiments\cite{Larrea:2015wra} \cite{Dasgupta:2015ohj}. 
Three tasks are carried out by the firmware: \cite{Arnold:2017tms}
\begin{enumerate}
\item The \textit{buffer}, that segments incoming data into chunks and applies zero-suppression (red in figure~\ref{fig:solidreadout}).
\item The \textit{trigger}, that decides which part of the data is sent out (blue in figure~\ref{fig:solidreadout}).
\item The \textit{control part}, that consists of the slow-control for SiPM and the communication link to the DAQ online system and to other planes of the detector (green in figure~\ref{fig:solidreadout}).
\end{enumerate}
\subsubsection{Buffer}
\begin{figure}[htb]
	\centering
	  \includegraphics[width=0.95\textwidth]{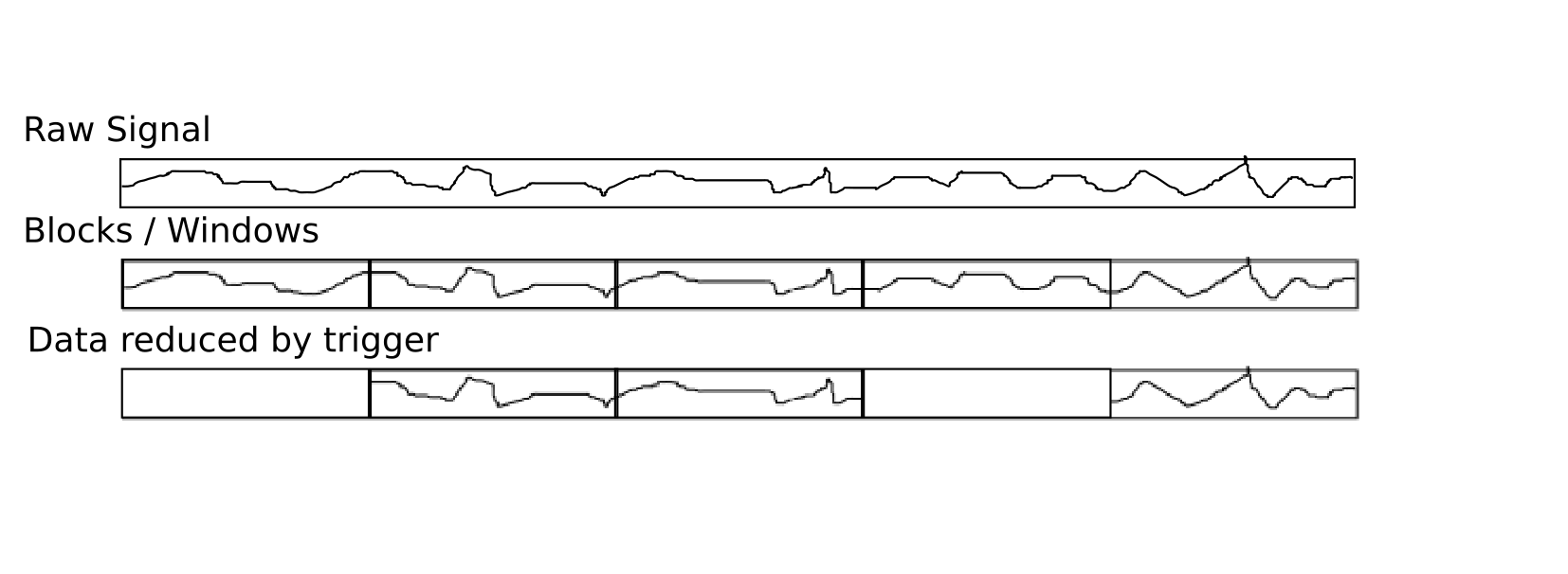}
	  \caption{Raw data is segmented into defined blocks or time windows, usually consisting of 256 samples. The trigger decides on a block-by-block basis whether data should be read out.}
	\label{fig:timewindow}
\end{figure}
The buffer stores the data coming from the ADC and prepares them to be sent out as soon as an event is triggered.

For each channel, data is fed through a channel pipeline, consisting of the deserialiser, the latency buffer, the zero suppression, the window buffer, channel read-out and the derandomiser. Each channel pipeline is connected to the main data buffer.

The \textbf{deserialiser} converts the serial data received from the ADC and converts them into $14\si{\bit}$ words. Alternatively to the deserialiser, data can be fed into the buffer from a \textbf{pattern generator} or as \textbf{playback} of existing data. It hands the data to the trigger and the latency buffer.

Data is divided into different segments of equal size by the buffer, usually 256 samples, as shown in figure~\ref{fig:timewindow}. A \textbf{latency buffer} holds the data streams for the time corresponding to two blocks in order to allow the trigger to make a decision.

The \textbf{zero suppression} suppresses data below a certain threshold. Hence it sets all data points below threshold $\theta$ to 0,
 \begin{equation}
 X(t)=\begin{cases}
         0, &\text{if } X(t)<{\theta}, \\
         X(t), &\text{otherwise},
 \end{cases}
 \end{equation}
with expected data reduction by the factor of $\mathcal{O}(10)$.
The samples put to 0 are removed and replaced by a marker. The marker contains information of how many sequential samples have been removed.  The threshold $\theta$ can be varied depending on which type of trigger has been activated.

After zero suppression has been applied where appropriate, the \textbf{window buffer} holds the data for a certain time period to allow the trigger to initiate read-out of data recorded earlier. An overflow of the window buffer is critical as this would disrupt the pipeline in such a way that no more data may be read out until re-synchronisation. 

The \textbf{channel read-out} hands single blocks to the derandomiser following a read-out request.

The \textbf{derandomiser} stores blocks for transfer into the main readout buffer. An overflow will result in ignorance of read-out requests, in which  the deadtime monitor will account for the missed blocks.

All the 64 channel's derandomisers are fed into the main data buffer. \cite{newb}

\subsubsection{Trigger}
The \textbf{channel trigger} performs feature extraction and triggering on the buffer as data come in. It decides real-time on a block-by-block basis and forwards the decisions to the local trigger. Its parameters are programmable. The channel trigger is described in section~\ref{sec:trigger}.

The \textbf{local trigger} contains one or more trigger generator units. The local trigger is responsible for making the trigger decision for one plane.  Trigger generator units form the local trigger based on information provided by the channel trigger values. Trigger decisions are passed to the readout sequencer, but also can be sent to neighbouring planes. Information is sent to the read-out controller on how many triggers are active and which features have caused them. At the current stage, the local trigger uses \textsc{or}-logic, i.e. logical disjunction, to combine the channel trigger decisions, but can be extended to contain more sophisticated algorithms, as will be discussed in section~\ref{sec:prospects}.

The \textbf{deadtime monitor} records deadtime caused by overflow or other malfunction of the firmware for each channel and each trigger. Deadtime can occur both by suppression of individual blocks or triggers. \cite{newb}

\subsubsection{Controls}
\begin{figure}[htb]
	\centering
	  \includegraphics[width=0.4\textwidth]{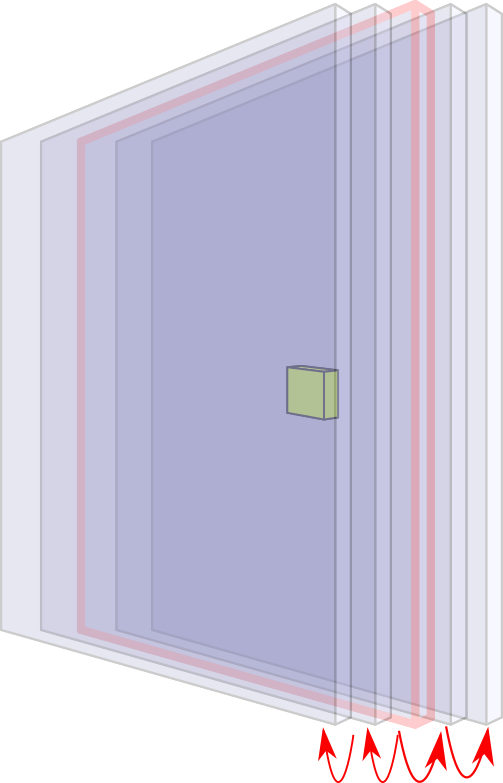}
	  \caption{Following a trigger decision on one plane, the trigger can be spread along neighbouring planes.}
	\label{fig:remotetrigger}
\end{figure}
The control part is responsible both for communication, including sending data out, and slow-control of the SiPMs.

The \textbf{read-out sequencer} translates all kinds of trigger decisions into the sequence of data blocks to be sent out. Each trigger type is linked to an offset and a block-count in a table that is used for the translation. Blocks tagged by multiple triggers are merged by the read-out sequencer. The read-out provides the read-out controller with information about the channels to be read out and whether read-out is skipped due to deadtime.

The read-out controller linked to the \textbf{header buffer} collects data from the channel buffers to the main data buffer. It formats the data and appends header and trailer information in order that the data sent out can be decoded.

The \textbf{\textit{IPbus} controller} is responsible for the \textit{IPbus} register and encodes and decodes communication from and to these registers.

The \textbf{timing/sync controller} controls clock synchronisation and generation from a distributed clock input.

The \textbf{remote trigger} is able to spread trigger decision onto neighbouring planes via serial link (see figure~\ref{fig:remotetrigger}). It also receives trigger decisions coming from remote planes and forwards them to the read-out sequencer and, where required, to other planes.

The \textbf{slow control} (not shown on the block diagram in figure~\ref{fig:solidreadout}) sets the bias voltages for the SiPM sensors.

\subsubsection{Hardware}
\begin{figure}[htb]
	\centering
	  \includegraphics[width=0.5\textwidth]{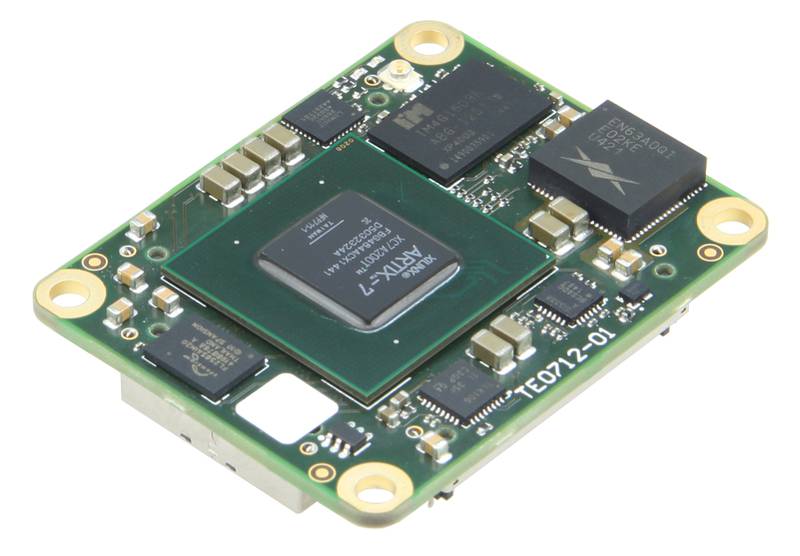}
	  \caption{\textsc{Trenz} module containing the \textsc{Xilinx} Artix 7 FPGA. \textit{Source:} \cite{trenz}.} 
	\label{fig:trenz}
\end{figure}

\begin{table}[]
\centering
\begin{tabular}{llllll}
\hline\hline
&\textbf{Logic cells}  & \textbf{CLBs} & \textbf{DSPs} & \textbf{Block RAM $[\si{\kilo\bit}]$} & \textbf{I/Os}\\ \hline
\textbf{Total} & $215360$ & $33650$ & $740$ & $13140$ & $500$ \\
\textbf{Per Channel} & $3365$ & $526$ & $12$ & $205$ & $8$ \\
\hline\hline
\end{tabular}
\caption{Available resources for the \textsc{Xilinx} Artix 7-200T FPGA for the whole device and divided by the 64 read-out channels. \textit{Source:} \cite{xilinx}.}
\label{tab:artix}
\end{table}
As buffer and trigger logic for 64 channels have to fit simultaneously on a single FPGA, a medium-density device -- the \textsc{Xilinx} Artix 7-200T FPGA -- is used. The FPGA is placed on a commercial module by \textsc{Trenz Electronics} shown in figure~\ref{fig:trenz}. The \textsc{Trenz} board contains -- beside the FPGA -- a clock chip (\textsc{Silicon Labs} Si5338), 4 Multi-Gigabit Transceivers (MGT), $256\si{\mega\bit}$ Quad-SPI Flash memory, $1 \si{\giga\byte}$ DDR3 SDRAM, DC-DC converters, 75 LVDS pairs and 4 single-ended I/O pins. \cite{trenz} \cite{trenz2}

The \textsc{Xilinx} 7-Series, to which the employed Artix 7-200T belongs, rely on $28\si{\nano\meter}$ technology. The Artix family is optimised for high logic throughput and low-power applications using serial links. A summary of key parameters can be seen in table~\ref{tab:artix}.  \cite{xilinx}

Irradiation can cause damage to the FPGA and other electronic parts \cite{Arnold:2015ylm}. However, the ionising radiation being faced is about the same as normal background levels. The increase in neutron and gamma flux by the reactor is compensated for by water and containment shielding of SoLid. Even if a Single Event Upset (SEU) occurs, the FPGA simply can be re-flashed due to it being connected to a JTAG link.

\clearpage
\section{The SoLid Trigger}
\label{sec:trigger}
As data rate would be too high for sending out all incoming singals, a decision has to be made which data are probably meaningful in terms of their content and thereby sent for further processing. This decision is made by the \textit{trigger}. A trigger thus classifies data on a rolling base into probably useful data -- to be sent out -- and probably meaningless data -- to be discarded. Hence it can be seen as a \textit{real-time classifier}, and statistical classification theory -- as part of pattern recognition and machine learning theory -- can be applied. 

This sections forms the part of the research carried out by me. A theoretical introduction in classification theory is given in section~\ref{sec:classificationtheory}. Methodology and designs of feature extraction and machine learning algorithms are discussed in section~\ref{sec:featureextraction}. Discussion of evaluation results are shown in section~\ref{sec:results}. Finally, an implementation is presented in section~\ref{sec:implementation} and prospect for the plane-level trigger is given in section~\ref{sec:prospects}. Conventions used in the mathematical representation of the trigger are given in the appendix~\ref{app:convention}.

\subsection{Trigger Types}
As described in section~\ref{sec:ibd}, the IBD process produces a positron and a neutron. While the positron signals appear with no significant latency, the neutron signal is delayed by $\mathcal{O}\left(10^{-4}\right)\si{\second}$. 
And while the positron signal forms a short light pulse, the neutron appears as a long signal of $\mathcal{O}\left(10^{-6}\right)\si{\second}$. 
Therefore, different triggers are used for the two kinds of signal. The positron signal -- due to it discriminability and briefness -- is fairly easy to distinguish. A threshold trigger is used. However, designing the neutron trigger is more challenging, as its signals usually do not reach high amplitudes. This prevents the usage of a threshold trigger due to the length of the signal, as the number of non-neutrons mis-classified as neutrons must be limited. This section deals with theory, methods and results that were applied to evaluating the neutron trigger.

\subsection{Classification Theory}
\label{sec:classificationtheory}
\subsubsection{Signal Space}
Preliminary, the concept of \textit{signal space} shall be introduced, often being credited to Wozencraft and Jacobs \cite{9780881335545}, as the data to be classified actually can be seen as a representation of its underlying -- analogue -- signal. 

The idea is to view signal waveforms as vectors in a vector space called \textit{signal space}. These vectors represent analogue, continuous signals. The constituent vectors are time-dependent finite-energy functions
\begin{equation}
u(t): \bm{R}\rightarrow \bm{C},
\end{equation}
where the condition of being finite-energy within an interval $[a,b]$ equates to them being square-integrable, i.e.
\begin{equation}
\label{eq:one}
\|u\|^2=\int_a^b \left|u(t)\right|^2\,dt=\int_a^b u(t)\overline{u(t)}\,dt \neq -\infty,\infty
\end{equation}

This derives from the definition of their norm $\|u\|$ as the square root of the inner product with itself,
\begin{equation}
\label{eq:two}
\|u\|={\sqrt {\langle x,x\rangle }},
\end{equation}
and the definition of the inner product of any two vectors within the interval $[a,b]$ 
\begin{equation}
\label{eq:three}
\langle v,w\rangle =\int _{a}^{b}v(t)\overline{w(t)}\,dt.
\end{equation}

The signal space is a $\bm{L}_2$ Lebesgue Pre-Hilbert space. In the case the interval $[a,b]$ is set to $[-\infty,\infty]$, it even is a complete $\bm{L}_2$ Lebesgue Hilbert space.\cite{Gallager} However, for the purposes of the trigger, only the signal within a time-interval will be used. Also, as will be discussed later in section~\ref{sec:algorithms}, due to considerations with regards to the FPGA resources, only time-domain features of the signals are used, leaving out features in complex frequency domain, hence all signals can be considered \textit{real} for our purposes. 


\subsubsection{Discrete Time Domain}
\begin{figure}[htb]
	\centering
	  \includegraphics[width=0.85\textwidth]{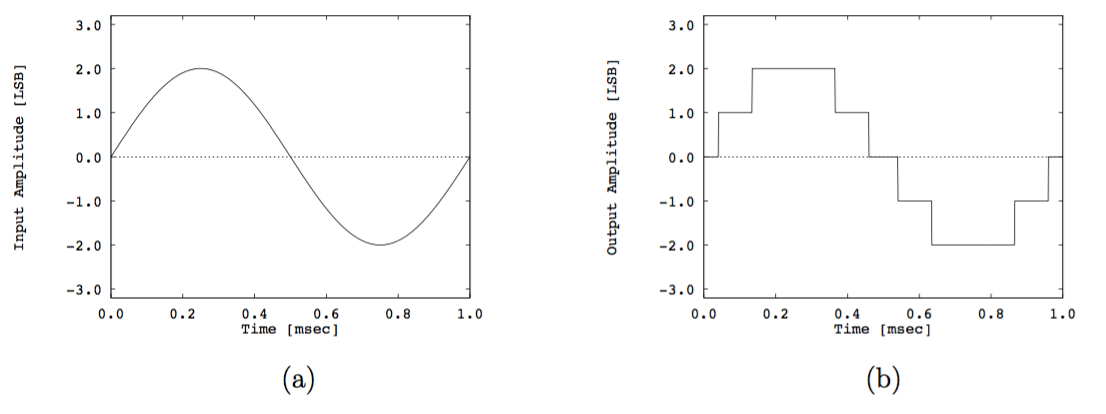}
	  \caption{(a) Continuous input signal. (b) Signal after time-discretisation (Dirac deltas connected by lines) and quantisation. \textit{Source:} \cite{wannamaker}.}
	\label{fig:discretisation}
\end{figure}
An $\bm{L}_2$ Hilbert space can be used as a representation for analogue signals, but the reality on an FPGA looks quite differently, as one can imagine. FPGAs, like any other digital device, handle signals $u(t)$ discretely both on the $u$ (signal) and $t$ (time) axis. Physically, this corresponds to the process of converting the SiPM signal input to a digital signal, as  done by the Analogue-Digital Converter (ADC).

Discretisation in time is achieved by multiplying the signal $u$ with the sampling function -- or Dirac comb -- $\operatorname{III}_T(t)$ that is a periodic series with period $T$ of Dirac delta functions $\delta$ \cite{cordoba1989},
\begin{equation}
\label{eq:diraccombs}
\operatorname{III}(u)=u(t)\cdot\operatorname{III}_T(t) =\ u(t)\cdot\sum_{k=-\infty}^{\infty} \delta(t - k T).
\end{equation}

Discretisation on the signal axis, called \textit{quantisation}, or digitisation, can be formalised by \cite{lipshitz1992quantization}
\begin{equation}
Q(u)=\Delta \cdot \left\lfloor {\frac {u}{\Delta }}+{\frac {1}{2}}\right\rfloor,
\end{equation}
for quantiser step size $\Delta$, with $\lfloor . \rfloor$ denoting the floor function. Combined time-discretisation and quantisation is a composition of the respective functions:
\begin{equation}
\label{eq:comp}
x(t) = Q\circ\operatorname{III}\circ u(t),
\end{equation}
with an example shown in figure~\ref{fig:discretisation}. Equation~\ref{eq:comp} corresponds to an \textit{ideal} ADC, while in the real case, channel noise $\widetilde{u}$ and effects caused by ADC non-linearity $\widetilde{Q}$ have to be considered:
\begin{equation}
x(t) = \left(Q+\widetilde{Q}\right)\circ\operatorname{III}\circ \left(u+\widetilde{u}\right)(t).
\end{equation}

Two values with connection to time- and value-discretisation are of characteristic importance for the process: \textit{Nyquist frequency} and \textit{quantisation noise}.

According to the Nyquist-Shannon sampling theorem \cite{shannon1949}, all parts of a signal have to be below Nyquist frequency in order that they can be accurately reconstructed. Information for parts of the signal with higher frequency might be lost. Nyquist frequency $f_n$ is half the sampling frequency $f_s$ \cite{hufschmid}:
\begin{equation}
{\displaystyle f_{\text{n}}={\frac {1}{2}}\cdot f_{\text{s}}}.
\end{equation} 

\begin{table}[]
\centering
\begin{tabular}{llll}
\hline\hline
\textbf{Period $T [\si{\nano\second}]$}  & \textbf{Quantisation step size $\Delta [\si{\micro\volt\per\bit}]$} & \textbf{Nyquist frequency $f_n [\si{\mega\hertz}]$} & \textbf{$SQNR_{max} [\si{\decibel}]$} \\ \hline
$25$&  $201.4$                &            $20$                           &                         $88.8$      \\
\hline\hline
\end{tabular}
\caption{Characteristic values for time-discretisation and quantisation.}
\label{tab:charval}
\end{table}
Flooring values in the quantisation process leads to quantisation noise. Given the general formula of signal-to-noise ratio ($\mathrm{SNR}$) \cite{hufschmid},
\begin{equation}
\mathrm{SNR} = \frac{\sigma^2_\mathrm{signal}}{\sigma^2_\mathrm{noise}}
\end{equation}
where $\sigma^2$ is the variance of signal and noise, respectively, signal-to-quantisation-noise can be approximated by \cite{li}:
\begin{equation}
\label{eq:SQNR}
{\mathrm  {SQNR}}={\frac  {3\cdot 4^{n\si{\bit} }\cdot P_{{x^{n\si{\bit} }}}}{x_{{max}}^{2}}}\sim\left( \log_{10}\left(\frac{P_{{x^{n\si{\bit}}}}}{x_{max}^2}\right)+6n\si{\bit} +4.8\right)[\si{\decibel}],
\end{equation}
with $n\si{\bit}$ being the number of bits used and $P_{{x^{n\si{\bit}}}}$ being calculated by the probability distribution function $pdf(x)$:
\begin{equation}
P_{{x^{n\si{\bit} }}}=\int _{{}}^{{}}x^{2}pdf(x)\,dx =E[x^{2}].
\end{equation}
As the exact $pdf$ is not known, the maximum SQNR is more feasible to calculate and might be sufficient, simplifying equation~\ref{eq:SQNR} to:
\begin{equation}
\label{eq:sqnrmax}
{\mathrm  {SQNR_{max}}}={{3\cdot 4^{n\si{\bit}_{max} } }}\sim \left(6n\si{\bit}_{max} +4.8\right)[\si{\decibel}].
\end{equation}
A table with the values indicated in equations~\ref{eq:diraccombs} to~\ref{eq:sqnrmax} is shown in table~\ref{tab:charval}. The short positron light pulse is expected to have a bandwidth of $\sim 10\si{\mega\hertz}$, thus lies below Nyquist frequency.

\subsubsection{Data Representation}
\begin{figure}[htb]
	\centering
	  \includegraphics[angle=270,width=0.55\textwidth]{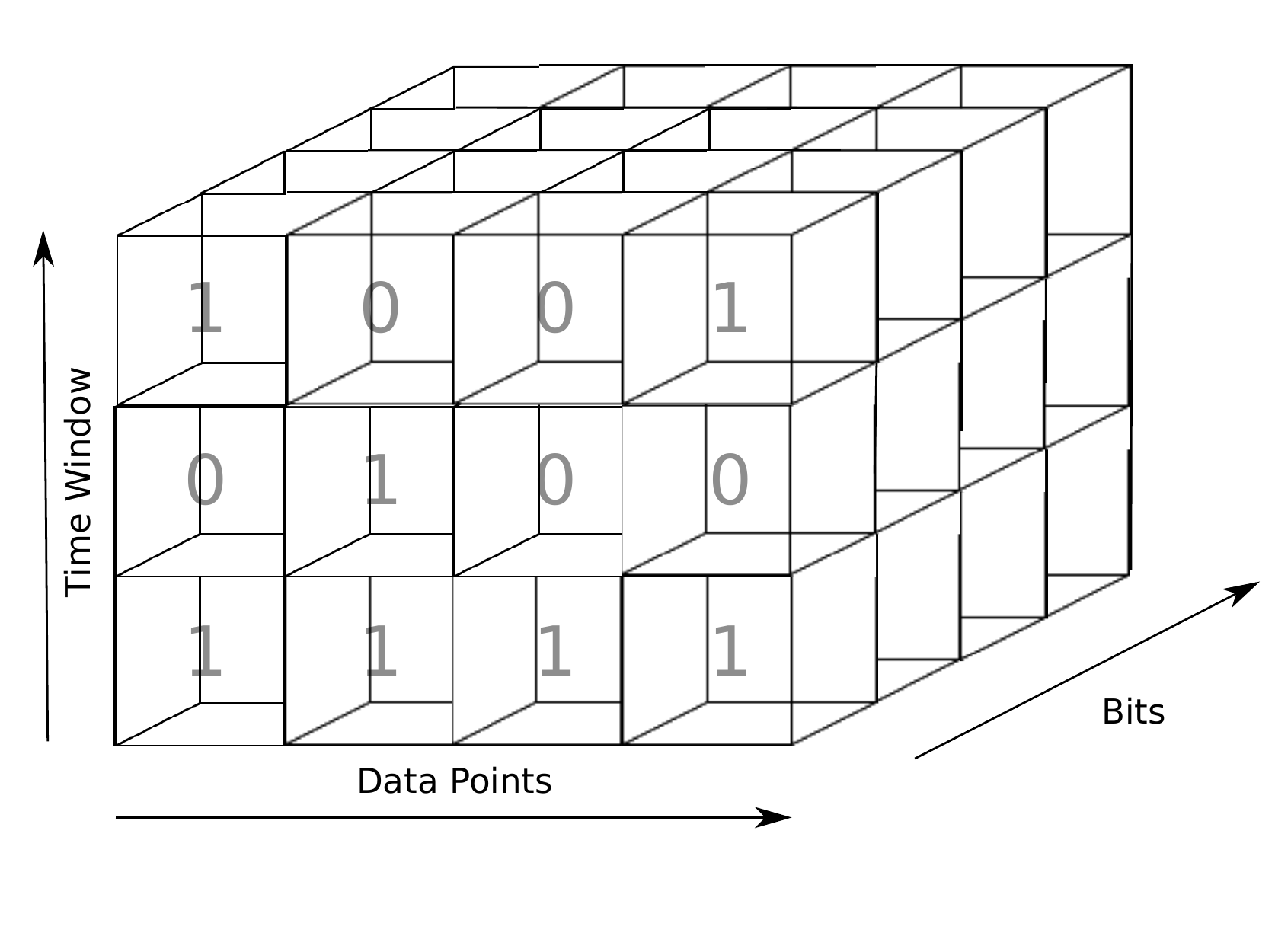}
	  \caption{Data cube (or tensor) containing signal information. Corresponds to equation~\ref{eq:tensorrepresentation}.}
	\label{fig:datacube}
\end{figure}
For usability, discrete signals are often represented as vectors, with the Dirac combs being mapped to the integer vector items:
\begin{equation}
\label{eq:realnotation}
\mathbf{X}^{\bm{R}^n}_k {{=}} x\left(T\cdot k+C\right),
\end{equation}
with the time offset $C$ chosen so that the Dirac combs are located at time $(T\cdot k+C)$ and dimensionality $n$ equate to the number of entries. One could take a step closer to reality, and map this vector into Boolean space $\bm{B}^n$ or $\bm{B}^{m\times n}$,
\begin{equation}
\label{eq:representation}
\mathbf{X}^{\bm{B}^n}_{((ij-1)+j)}=\mathbf{X}^{\bm{B}^{m\times n}}_{ij} = j\text{th position in binary representation of}(\mathbf{X}^{\bm{R}^n}_i),
\end{equation}
where rows $i$ represent data points corresponding to $\mathbf{X}^{\bm{R}^n}_i$ and columns $j$ represented bit value in the case of a $\mathbf{X}^{\bm{B}^{m\times n}}$ matrix, or position $i+j$ for the $\mathbf{X}^{\bm{B}^n}$ vector. 

Data points can be further segmented into $l$ time windows of length $m$, giving a multi-dimensional array, or tensor, or data cube $\in\bm{B^{l\times m \times n}}$.
This is the case for SoLid where data is segmented into packages. The new tensor will then be:
\begin{equation}
\label{eq:tensorrepresentation}
\begin{cases}
\mathbf{X}^{\bm{B}^{l \times m\times n}}_{kij} = \mathbf{X}^{\bm{B}^{m\times n}}_{uv}\\
u=i+\lfloor\frac{k}{l}\rfloor\cdot i\\
v=j
\end{cases}
,
\end{equation}
represented in figure~\ref{fig:datacube}. Note that the different notations of the discrete signal are isomorphic ($x\cong \mathbf{X}^{\bm{R}^n}\cong\mathbf{X}^{\bm{B}^n}\cong\mathbf{X}^{\bm{B}^{m\times n}}\cong\mathbf{X}^{\bm{B}^{l \times m\times n}}$).

As a convention within this thesis, if data vector $\mathbf{X}$ appears without a superscript, it means the representation in $\bm{B}^{l \times m\times n}$ is used.

\subsubsection{Trigger Formalisation}
The purpose of the trigger is data reduction. The data to be sent out should be much smaller than the acquired data on the trigger level, or
\begin{equation}
\dim \mathbf{X}_s \ll \dim \mathbf{X},
\end{equation}
achieved by discarding data points $i$ in the $\mathbf{X}^{\bm{B}^{m\times n}}_{ij}$ representation of the data vector. Note that dimensionality for an $l\times m \times n$ structure is
\begin{equation}
\dim \mathbf{X}=l\times m \times n.
\end{equation}
The size of each data point  and the number of data points cannot be reduced. For SoLid, these are $n=14$ and $m=256$. It is therefore the number of windows $l$ that is reduced.
 The decision whether a window $l$ can be discarded or not is a binary classification problem.

The formal definition of a classifier $f(\mathbf{X})$ is \cite{jameswitten} \cite{Bishop}
\begin{equation}
\label{eq:classificationgeneral}
\mathbf{\hat{{Y}}} = f(\mathbf{X}),
\end{equation}
where $\mathbf{\hat{Y}}$ denotes the \textit{class} and $\mathbf{X}$ is the data cube, or tensor, representing the data set. Each time window $\mathbf{X}_{k..}$ is mapped to a class $\mathbf{\hat{Y}}_i$, taking a value of a finite set of classes $\mathbb{K}$. 
$\mathbb{K}$ in the present case is $\{\text{discard},\text{not discard}\}$ and can be represented as $\{0,1\}$ in $\bm{B}$. 

There is a problem in doing so: The dimensionality of each row $k$ in Boolean space $\bm{B}^{l\times m\times n}$ can be high, and in fact too high for a computation device to practically handle. In the case of a time-wise segmented signal, such as for SoLid, the dimensionality is
\begin{equation}
\dim \mathbf{X}_k = m\times n =\frac{\text{length of time window}}{T}\cdot {{n\si{\bit}}}_{\text{data point}},
\end{equation}
that for SoLid is $\dim X_k = 3584$.  The classification in the FPGA has to be made as data come in, i.e. in real-time, not requiring more than a few clock cycles. Estimating the number of parallel operations $o_p$ using
\begin{equation}
o_p\approx\frac{\dim \mathbf{X}_k}{{I_{CLB}}\cdot c_{max}}
\end{equation}
with $I_{CLB}$ being the number of Look-up Table (LUT) inputs and $c_{max}$ being the maximum number of clock cycles, which can be put to $I_{CLB}=6$ and $c_{max}\sim5$, it is 120 parallel operations that would have to be processed per clock cycle.  This is too high for a 64-channel FPGA to handle in real-time. The dimensionality therefore has to be \textit{reduced}. This is commonly achieved by splitting classification into two parts: Feature extraction and application of the machine learning algorithm.

\subsection{Trigger Design}
\label{sec:featureextraction}
The trigger classification function $f(\mathbf{X}_{l..})$ -- operating on the matrix $\mathbf{X}_{l..}$ referring to a single time window $l$ of the tensor $\mathrm{X}$ -- actually is a composite of two functions,
\begin{equation}
\label{eq:composition}
h\circ g(X):
\end{equation}
the mapping of 

\begin{equation}
\label{eq:mapping}
g(\mathbf{X}_{l..}):\mathbf{X}_{l..}\in \bm{B}^{l\times m \times n} \mapsto \mathbf{F} \in \bm{B}^{l \times p \times n},
\end{equation}
with $p\ll m$, thus resulting in a feature vector $\mathbf{F}$ having much lower dimensionality than the input data, and the actual classification $h(\mathbf{F})$, which is the application of the machine learning algorithm, giving the flow
\begin{equation}
\left(\bm{L}_2\bm{H} \stackrel{\operatorname{III}} \rightarrow \right)\bm{B}^{l\times m \times n} \stackrel g \rightarrow \bm{B}^{l\times (p \ll m) \times n}  \stackrel h \rightarrow \bm{B}^l.
\end{equation}
In terms of computation, the algorithm $g(\mathbf{X})$ should reduce the number of parallel processes by several orders of magnitude, while the splitting of the classification into two separate processes should increase computation time by just the factor of two. 

For usability, the representation of $\mathbf{F}$ as a real matrix $\bm{R}^{l \times p}$ shall be used, rather than the representation in Boolean space (that is closer to reality). As will be seen later in section~\ref{sec:perceptron}, this representation -- being a manifold on the original Hilbert space -- contains a property very useful for classification

The function $g(\mathbf{X})$ is called \textit{feature extraction}, with an allocated \textit{feature space} $\mathbf{B}^{l \times p \times n}$. These should meet several requirements, namely
\begin{enumerate}
\item being a good practical representation of signal space,
\item having low dimensionality $p$ ($p$ indicates also the number of features),
\item allowing high-accuracy classification, and
\item being resource-efficient. 
\end{enumerate}

\subsubsection{Metaheuristics}
\label{sec:metaheuristics}
\begin{figure}[htb]
	\centering
	  \includegraphics[width=0.5\textwidth]{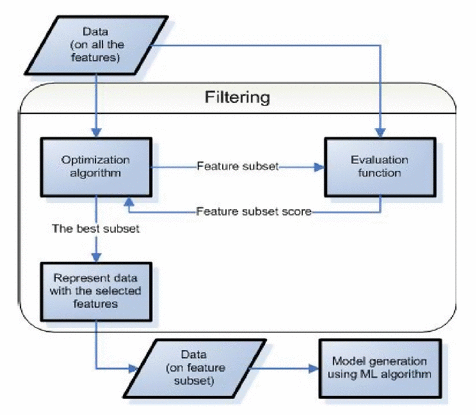}
	  \caption{In the filter method, the quality of a feature space is evaluated using a score independent from the machine learning algorithm. \textit{Source:} \cite{mladenic}.}
	\label{fig:filter}
\end{figure}
\begin{figure}[htb]
	\centering
	  \includegraphics[width=0.5\textwidth]{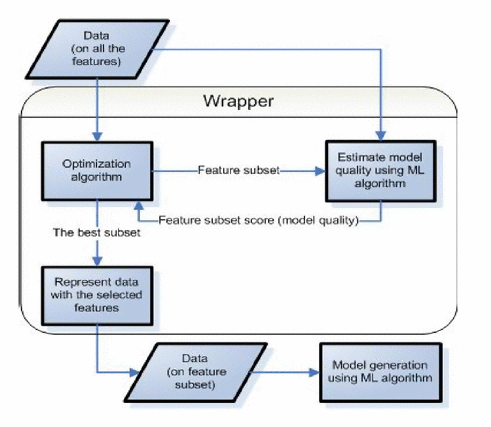}
	  \caption{In the wrapper method, the quality of a feature space is evaluated using the machine learning algorithm. \textit{Source:} \cite{mladenic}.}
	\label{fig:wrapper}
\end{figure}
Finding a suitable feature space is a crucial, but also a complex task \cite{liumotoda}. It consists of pre-processing the data, extracting features from the pre-processed data (called \textit{feature extraction} or \textit{feature engineering}), forming a primary feature space, and then finding a suitable subset of that space, usually of low dimensionality, forming the secondary feature space, a process that is called \textit{feature selection}.

Feature selection is a well-investigated field -- called \textit{metaheuristics}.  Two main approaches -- or metaheuristic methods -- exist for feature selection: The wrapper method, and the filter method. The wrapper method selects the feature set using evaluation by the classifier (see figure~\ref{fig:wrapper}), while in the filter method, the selection of feature is independent from the used classifier (see figure~\ref{fig:filter}). Also, hybrid methods exist.  \cite{Sanchez-Marono:2007:FMF:1777942.1777962}

As the machine learning algorithm is not determined beforehand, the filter method is used for finding the SoLid trigger feature space. It will be explained later in section~\ref{sec:thresholdfunction} that in the particular case of using a threshold function as the machine learning algorithm, the method is the same as the wrapper method. Individual algorithms are determined and evaluated on an individual basis. They will be ranked according to their performance, the correlation between them is determined and the most suitable chosen as the appropriate feature subset.

\subsubsection{Feature Evaluation}
\label{sec:evaluation}
\begin{figure}[htb]
	\centering
	  \includegraphics[width=0.85\textwidth]{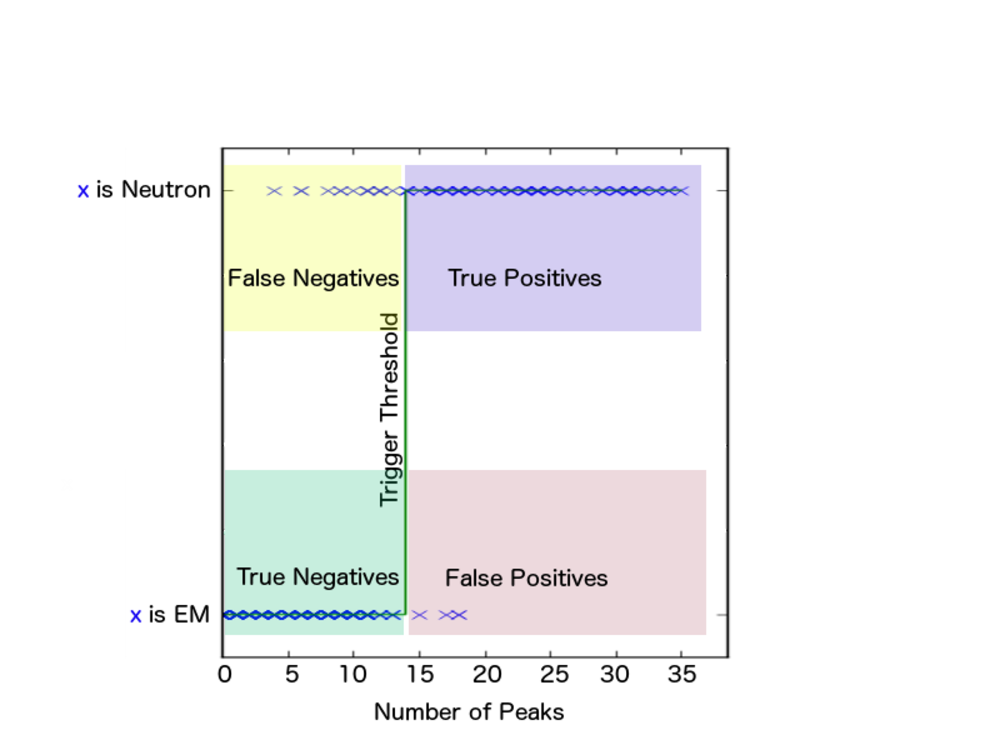}
	  \caption{Example of categorisation for the Number-of-Peaks feature corresponding to $i$. The threshold $\theta_i$ is shown in green, the values $\mathbf{F}_{li}$ in blue, and the value $\mathbf{E}_i$ on the $y$-axis. The differently coloured areas mark the four different categories $TP$, $TF$, $FP$ and $FN$.}
	\label{fig:solexample}
\end{figure}
Evaluation is based on binary classifiers. The different algorithms are assessed individually, using a threshold classification function on each of the features,
\begin{equation}
\label{eq:trigger}
h_i(\mathbf{F}_{li},\theta_i)= 
\begin{cases}
    \text{1},& \text{if } x\geq \theta_i\\
    \text{0},              & \text{otherwise}.
\end{cases}
\end{equation}
The simple threshold trigger algorithm is one such fuction.
For each feature $\mathbf{F}_{li}$, the threshold $\theta_i$ is swept from the minimum and maximum values within $\mathbf{F}_{li}$, leading to a Boolean binary classification vector $\hat{\mathbf{Y}} \in \bm{B}^l$. 
Assuming that a corresponding vector $\mathbf{E} \in \bm{B}^l$ exists representing whether the value $\mathbf{F}_{li}$ is actually linked to a physical IBD event, four categories are defined: \cite{fawcett2006introduction}

True positive (TP), correct hit:
\begin{equation}
h_i(\mathbf{F}_{li})=\mathbf{E}_l=1.
\end{equation}
True negative (TN), correct pass:
\begin{equation}
h_i(\mathbf{F}_{li})=\mathbf{E}_l=0.
\end{equation}
False positive (FP), false alarm:
\begin{equation}
h_i(\mathbf{F}_{li})\neq\mathbf{E}_l=0.
\end{equation}
False negative (FN), miss:
\begin{equation}
h_i(\mathbf{F}_{li})\neq\mathbf{E}_l=1.
\end{equation}
An example for SoLid signals is shown in figure~\ref{fig:solexample}.

These abbreviations simultaneously stand for the number of how many times one of these categories has occurred within a given sample, i.e.
\begin{equation}
TP=\left|\{\mathbf{F}_{li}|\forall \mathbf{F}_{li}: h_i(\mathbf{F}_{li})=\mathbf{E}_l=1\}\right|,
\end{equation}
\begin{equation}
TN=\left|\{\mathbf{F}_{li}|\forall \mathbf{F}_{li}: h_i(\mathbf{F}_{li})=\mathbf{E}_l=0\}\right|,
\end{equation}
\begin{equation}
FP=\left|\{\mathbf{F}_{li}|\forall \mathbf{F}_{li}: h_i(\mathbf{F}_{li})\neq\mathbf{E}_l=0\}\right|,
\end{equation}
and
\begin{equation}
FN=\left|\{\mathbf{F}_{li}|\forall \mathbf{F}_{li}: h_i(\mathbf{F}_{li})\neq\mathbf{E}_l=0\}\right|.
\end{equation}
In addition, the positives $P$ and the number of negatives $N$ simply give the total numbers of these values in $\mathbf{E}$.

Having these definitions, the following indicators for classification performance and accuracy can be used: \cite{citeulike:12882259}

\textbf{Sensitivity} or true positive rate (TPR) or trigger \textbf{efficiency}:
\begin{equation}
\label{eq:efficiency}
\mathit{TPR} = \frac{\mathit{TP}}{P} = \frac{\mathit{TP}}{(\mathit{TP}+\mathit{FN})}
\end{equation}
\textbf{Precision} or positive predictive value (PPV) or trigger \textbf{purity}:
\begin{equation}
\label{eq:purity}
\mathit{PPV} = \frac{\mathit{TP}}{(\mathit{TP} + \mathit{FP})}
\end{equation}
\textbf{Fallout} or false positive rate (FPR) or \textbf{fake rate} of the trigger:

\begin{equation}
\label{eq:fakerate}
\mathit{FPR} = \frac{\mathit{FP}}{N}=\frac{ \mathit{FP}}{(\mathit{FP} + \mathit{TN})}
\end{equation}

Two characteristic plots can be obtained from those values: The receiver operating characteristic (ROC) curve, that shows efficiency versus fake rate, and a curve showing efficiency versus purity.

\subsubsection{Feature Extraction Algorithms}
\label{sec:algorithms}
The set of algorithms to be evaluated should contain high variety, as in this phase as many possible algorithms should be assessed in order to increase the likelihood to find well-performing algorithms. However, reasoning on which features will lead to effective classification is essential, bearing in mind that adding randomness to the creation of algorithms could cover the part that pure thought cannot. 
In other words, if $\mathbb{G}$ is the set of all functions $g_i$ mapping $\bm{B}^{m\times n}\rightarrow \bm{R}$, increasing the number of functions will cover more elements of that set $\mathbb{G}$.

The creation of some algorithms seem comprehensible, whereas others do not. As a high number of algorithms will be reduced after assessment on a later stage, that however cannot be considered to be a disadvantage.
Fourier transformations on FPGAs usually use a lot of resources \cite{uzun2005fpga}, often above 1000 LUTs \cite{fft}. As the FPGA usage of hardware resources has strict limits (see table~\ref{tab:artix}), frequency-domain features hence will not be considered. Therefore only time-domain features are subsequently presented.

The single algorithms will be -- in reference to equation~\ref{eq:mapping} -- be denoted as $g_i(\mathrm{X}_{l..})$,
 mapping the signal from a given time window to one specific dimension of the feature vector. All the selected functions will then give $g$, i.e.
\begin{equation}
\mathbf{F}=\sum_i g_i(\mathrm{X}_{l..}),
\end{equation}
considering that all feature extraction functions $g_i$ project to spaces orthogonal to each other.

The values $X[t]$ will be notated as elements of an array isomorphic to the discrete real array $\bm{R}^n$, as given in equation~\ref{eq:realnotation}, with the notation  rule
\begin{equation}
X[t]=\mathbf{X}^{\bm{R}^n}_t,
\end{equation}
and the whole set $X$ shall only consist of elements within a given time window $l$, i.e.
\begin{equation}
X=\{X[t-l],X[t-l+1],\ldots,X[t]\}.
\end{equation}

In addition to the function operating in the discrete signal, also a representation operating on the continuous signal $g_c(u):\bm{L}_2\bm{H}\rightarrow \bm{R}$ is given where appropiate.

\paragraph{Maximum Amplitude}
Maximum amplitude is the feature used to trigger on the SM1 phase of SoLid. It simply uses the maximum value within a time window for triggering, i.e.
\begin{equation}
\label{eq:maxa}
g_c\left(u\right)=\max{u}
\end{equation}
for the continuous value and
\begin{equation}
\label{eq:maxa2}
g_i\left(X\right)=\max{X}
\end{equation}
for the discrete value.

\paragraph{Number-of-Peaks}
\begin{figure}
\centering
\begin{subfigure}{.5\textwidth}
  \centering
  \includegraphics[width= \linewidth]{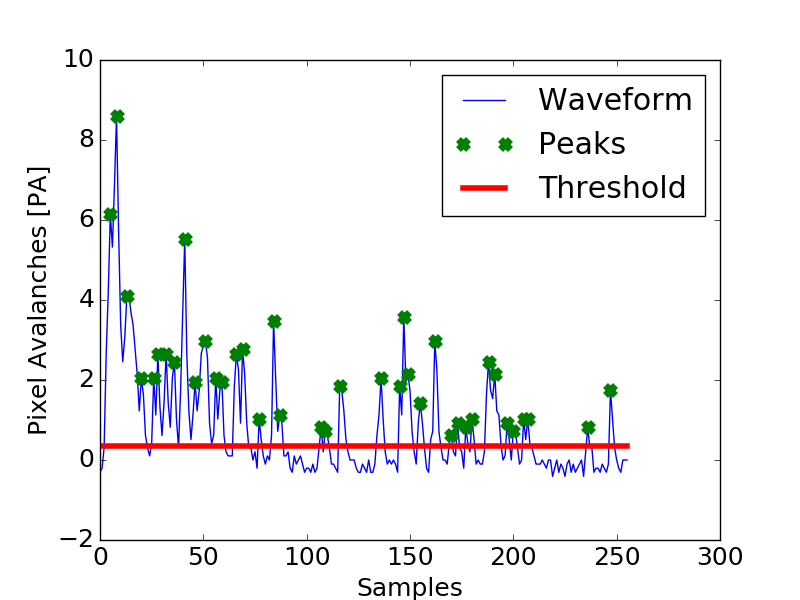}
  \caption{Neutron signal.}
  \label{fig:sub10}
\end{subfigure}%
\begin{subfigure}{.5\textwidth}
  \centering
  \includegraphics[width= \linewidth]{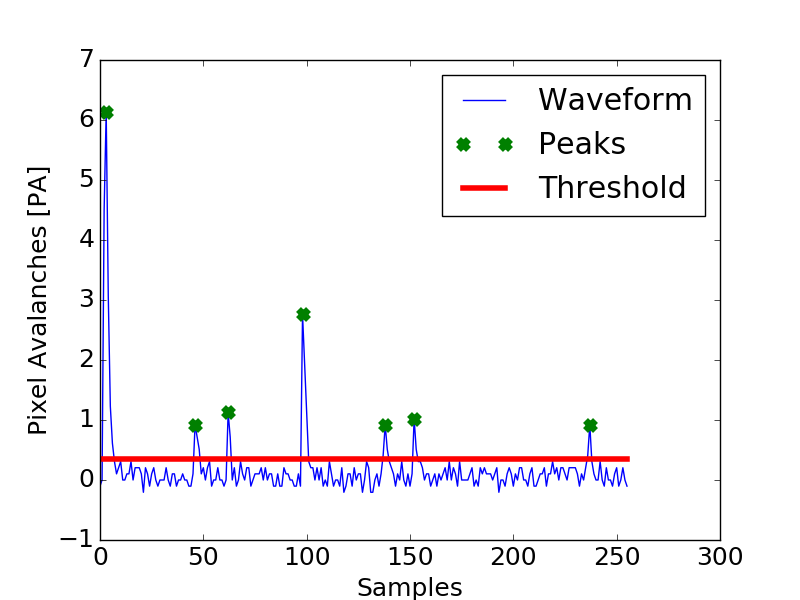}
  \caption{Photon/dark count (EM) signal.}
  \label{fig:sub20}
\end{subfigure}
\caption{Peak finding algorithm. Green crosses represent found maxima above threshold.}
\label{fig:npeak}
\end{figure}
The Number-of-Peaks feature counts the number of maxima above a certain threshold $\theta$ within a time window. An example is shown in figure~\ref{fig:npeak}. The threshold is set in order that pure noise-induced maxima are not considered. The reasoning behind this algorithm is that each time the \LiZnS-layer is emitting a photon, a peak on the input signal occurs.

It may be expressed in continuous space as
\begin{equation}
\label{eq:npeaks}
g_c\left(u\right)=\left\vert \left\{ t|\forall t: {u(t)>\theta} \land {\frac{du(t)}{dt}=0 \land \frac{d^2 u(t)}{dt^2}<0} \right\}\right\vert
\end{equation}
and in discrete space either as
\begin{equation}
\label{eq:npeaksd}
g_i\left(X\right)=\left\vert \left\{ t| \forall t:{X[t]>\theta} \land{ \Delta X[t-1]\geq0 \land \Delta X[t]<0}\right\}\right\vert
\end{equation}
with the definition
\begin{equation}
\Delta X[a] = X[a] - X[a-1],
\end{equation}
or
\begin{equation}
\label{eq:npeaksd2}
g_i\left(X\right)=\left\vert \left\{ t| \forall t:{X[t]>\theta} \land{ X[t-1]\geq X[t-2] \land X[t]<X[t-1]}\right\}\right\vert.
\end{equation}
In other words, the cardinality -- i.e. the number of elements -- $\left|.\right|$ of the set $\{.\}$ of points that are a maximum above a certain threshold forms the feature.

One variation of the algorithm that also has been assessed is to add the condition that after a peak has found, following peaks will be ignored during a certain interval (\textit{time veto}).

\paragraph{Weighted Peaks/Number of Photon Avalanches}
Another variation of the Number-of-Peaks feature is to weight the peaks by their amplitude. This equals to the number of pixel photon avalanches (PA). In continuous space this is expressed as
\begin{equation}
g_c\left(u\right)=\left\vert \left\{ u(t)\cdot t|\forall t: {u(t)>\theta} \land {\frac{du(t)}{dt}=0 \land \frac{d^2 u(t)}{dt^2}<0} \right\}\right\vert
\end{equation}
and in discrete space as
\begin{equation}
g_i\left(X\right)=\left\vert \left\{ X[t]\cdot t| \forall t:{X[t]>\theta} \land{ \Delta X[t-1]\geq0 \land \Delta X[t]<0}\right\}\right\vert.
\end{equation}

\paragraph{Time-over-Threshold}
\begin{figure}
\centering
\begin{subfigure}{.5\textwidth}
  \centering
  \includegraphics[width= \linewidth]{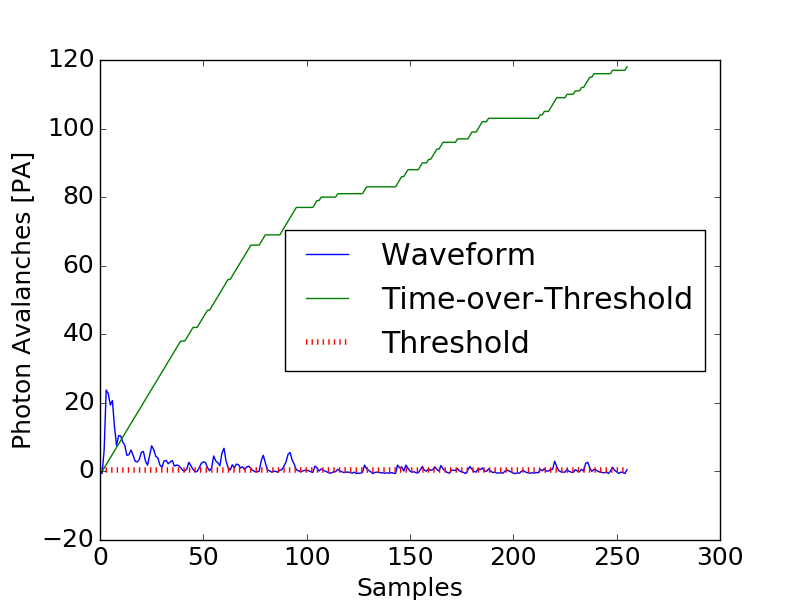}
  \caption{Neutron signal.}
  \label{fig:sub1}
\end{subfigure}%
\begin{subfigure}{.5\textwidth}
  \centering
  \includegraphics[width= \linewidth]{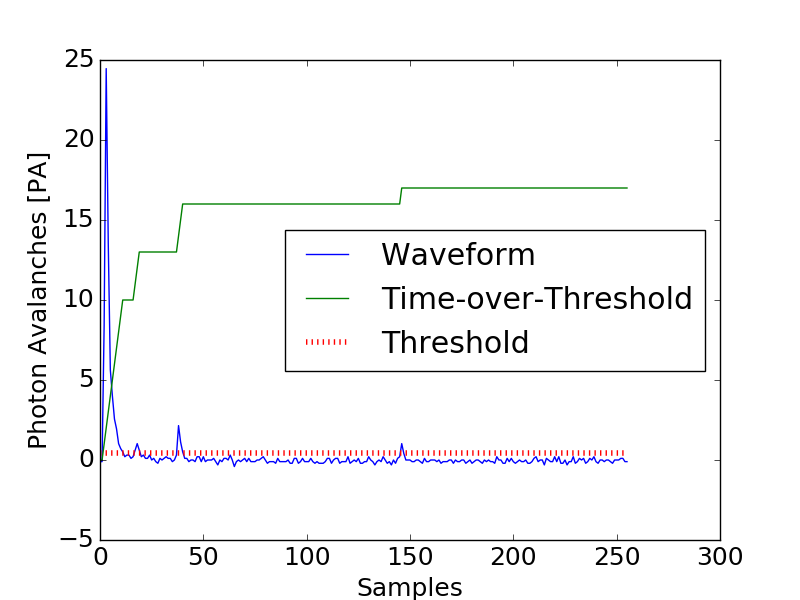}
  \caption{Photon/dark count (EM) signal.}
  \label{fig:sub2}
\end{subfigure}
\caption{Time-over-Threshold algorithm.}
\label{fig:totex}
\end{figure}
The Time-over-Threshold measures the number of samples (i.e. length of time) a signal is above a certain threshold $\theta$ (see figure~\ref{fig:totex}). This takes advantage of the fact that neutron signal last much longer than EM signals.

It is expressed as a a function of a continuous signal as
\begin{equation}
\label{eq:tot}
g_c\left(u\right)={\int \delta(t)\, dt} \land {\delta(t) = \begin{cases}
  1, & \text{if } u(t) > \theta, \\
  0, & \text{otherwise}
\end{cases}}
\end{equation}
and in discrete space as
\begin{equation}
\label{eq:totd}
g_i\left(X\right)=
{
\sum_{t=1}^{m}\delta[t]} \land {\delta[t] = \begin{cases}
  1, & \text{if } X[t] > \theta, \\
  0, & \text{otherwise}.
\end{cases}}
\end{equation}

\paragraph{Decay Time} 
\begin{figure}
\centering
\begin{subfigure}{.5\textwidth}
  \centering
  \includegraphics[width= \linewidth]{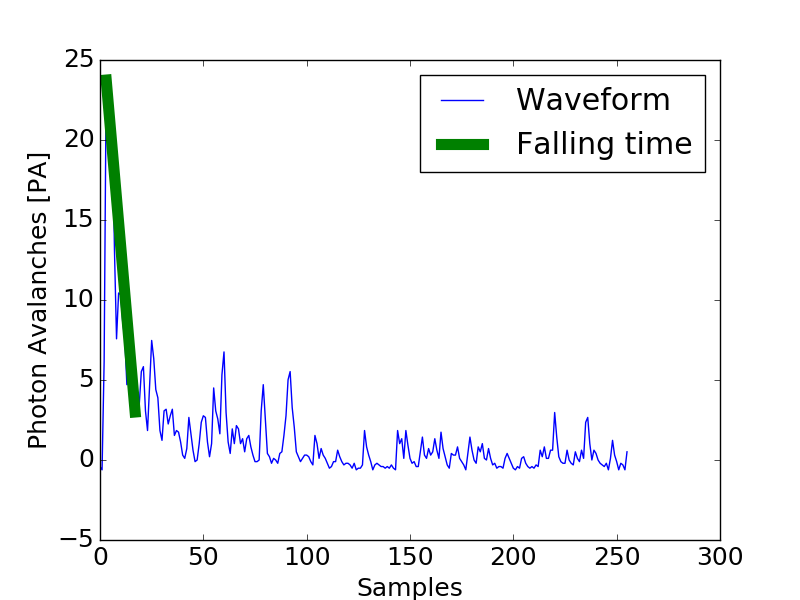}
  \caption{Neutron signal.}
  \label{fig:sub19}
\end{subfigure}%
\begin{subfigure}{.5\textwidth}
  \centering
  \includegraphics[width= \linewidth]{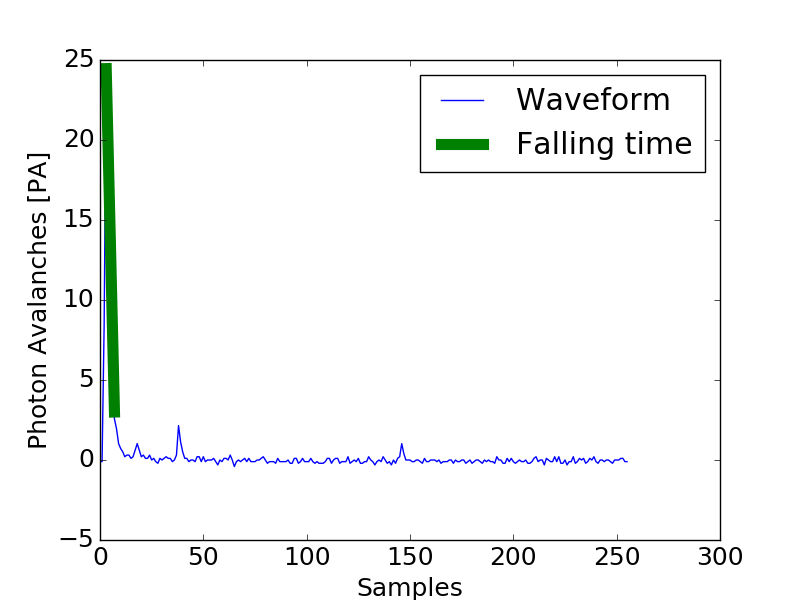}
  \caption{Photon/dark count (EM) signal.}
  \label{fig:sub29}
\end{subfigure}
\caption{Falling time algorithm. The value is given as the $x$-axis projection of the green line.}
\label{fig:decaytime}
\end{figure}
\textit{Falling time} means the time the signal undertakes until it falls below a certain percentage of its maximum value (see figure~\ref{fig:decaytime} for an example).
\paragraph{Integral-over-Threshold and Integral}
The integral-over-threshold is calculated in continuous space as
\begin{equation}
\label{eq:integral}
g_c\left(u\right)={\int u(t)\delta(t)\, dt},
\end{equation}
with $\delta(t)$ defined as in equation~\ref{eq:tot}.
This corresponds to the sum in discrete terms:
\begin{equation}
\label{eq:integral2}
g_i\left(X\right)=
{
\sum_{t=1}^{m} X[t]\delta[t]},
\end{equation}
with $\delta[t]$ defined as in equation~\ref{eq:totd}.

For the integral feature, this simplifies to
\begin{equation}
\label{eq:featintegral}
g_c\left(u\right)={\int u(t)\, dt}
\end{equation}
and
\begin{equation}
\label{eq:featintegrald}
g_i\left(X\right)=
{
\sum_{t=1}^{m} X[t]}.
\end{equation}

\paragraph{Integral-over-Amplitude}
The Integral-over-Amplitude (or Integral$/$maximum Amplitude) divides the value obtained by equations~\ref{eq:integral} and~\ref{eq:integral2} by division by maximum amplitude (equations~\ref{eq:maxa} and~\ref{eq:maxa2}), i.e.
\begin{equation}
\label{eq:integralamplitude}
g_c\left(u\right)=\frac{\int u(t)\, dt}{\max u(t)}.
\end{equation}
This corresponds to the sum and a division in discrete terms:
\begin{equation}
\label{eq:integralamplitude2}
g_i\left(X\right)=\frac
{
\sum_{t=1}^{m} X[t]}{\max X}.
\end{equation}

\paragraph{Statistical Moments}

Statistical moments with their common definitions are considered such as \textit{mean},
\begin{equation}
{\displaystyle g_i(X)={\frac {1}{m}}\sum _{t=1}^{m}X[t],}
\end{equation}

and \textit{standard deviation}:
\begin{equation}
g_i(X)=\sigma ={\sqrt {{\frac {1}{m}}\sum _{t=1}^{m}(X[t]-\mu )^{2}}},{\rm {\ \ where\ \ }}\mu ={\frac {1}{m}}\sum _{t=1}^{m}X[t].
\end{equation}

\subsubsection{Machine Learning Algorithms}
\label{sec:machinelearningalgorithms}
Machine learning algorithms classify data using the features extracted. While feature extraction algorithms correspond to function $g(\textbf{X})$ in equation~\ref{eq:composition}, the machine learning algorithm corresponds to $h(\textbf{F})$.
\paragraph{Threshold Function}
\label{sec:thresholdfunction}
The simplest algorithm uses only one feature and triggers on this particular value. The threshold function is a step function \cite{rojas} and in fact is the same as the function used for evaluation in equation~\ref{eq:trigger}. As evaluation and machine learning algorithms are the same in this case, the metaheuristics used for feature selection is the wrapper method rather than the filter method (see section~\ref{sec:metaheuristics}).
\paragraph{Perceptron}
\label{sec:perceptron}
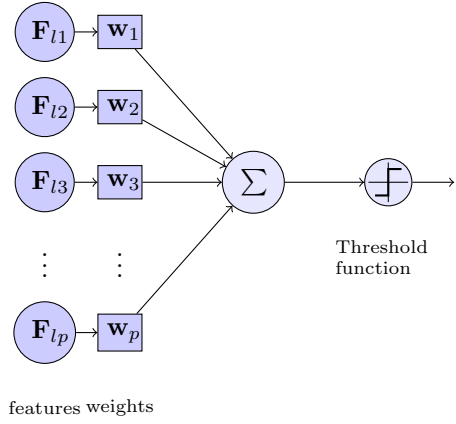
\begin{figure}
\centering
    \begin{tikzpicture}
        \node[functions] (center) {};
        \node[below of=center,font=\scriptsize,text width=4em] {Threshold function};
        \draw[thick] (0.5em,0.5em) -- (0,0.5em) -- (0,-0.5em) -- (-0.5em,-0.5em);
        \draw (0em,0.75em) -- (0em,-0.75em);
        \draw (0.75em,0em) -- (-0.75em,0em);
        \node[right of=center] (right) {};
            \path[draw,->] (center) -- (right);
        \node[functions,left=3em of center] (left) {$\sum$};
            \path[draw,->] (left) -- (center);
        \node[weights,left=3em of left] (2) {$\mathbf{w}_3$} -- (2) node[input,left of=2] (l2) {$\mathbf{F}_{l3}$};
            \path[draw,->] (l2) -- (2);
            \path[draw,->] (2) -- (left);
        \node[below of=2] (dots) {$\vdots$} -- (dots) node[left of=dots] (ldots) {$\vdots$};
        \node[weights,below of=dots] (p) {$\mathbf{w}_p$} -- (p) node[input,left of=p] (lp) {$\mathbf{F}_{lp}$};
            \path[draw,->] (lp) -- (p);
            \path[draw,->] (p) -- (left);
        \node[weights,above of=2] (1) {$\mathbf{w}_2$} -- (1) node[input,left of=1] (l1) {$\mathbf{F}_{l2}$};
            \path[draw,->] (l1) -- (1);
            \path[draw,->] (1) -- (left);
        \node[weights,above of=1] (0) {$\mathbf{w}_1$} -- (0) node[input,left of=0] (l0) {$\mathbf{F}_{l1}$};
            \path[draw,->] (l0) -- (0);
            \path[draw,->] (0) -- (left);
        \node[below of=lp,font=\scriptsize] {features};
        \node[below of=p,font=\scriptsize] {weights};
    \end{tikzpicture}
\caption{Block diagram of a perceptron.}
\label{fig:perceptron}
\end{figure}
The single-layer perceptron, also called single-layer neural network or artificial neuron and suggested by Rosenblatt in 1958 \cite{Rosenblatt58theperceptron}\cite{Rosenblatt62}, triggers with a threshold $\theta$ on the inner product of the input or feature vector $\mathbf{F}_{l.}$ with a weight vector $\mathbf{w}$. This type of Artificial Neural Network (ANN) is the same as a linear classifier.

The feature vector $\mathbf{F}_{l.}$ of a single time window $l$ of the feature matrix $\mathbf{F}$ is considered.
The space spanned by the feature vector $\mathbf{F}_{l.}$ can be seen as an $\bm{R}^p$ manifold on the underlying analogue signal. This allows to apply the definition of the inner product, \cite{9780521780193}
\begin{equation}
\label{eq:innerproduct}
\left\langle {x,y}\right\rangle =x^{\text{T}}y=\sum _{i=1}^{n}x_{i}y_{i}=x_{1}y_{1}+\cdots +x_{n}y_{n},
\end{equation}
in order to re-define the threshold function in equation~\ref{eq:trigger} in order to get the perceptron function,
\begin{equation}
\label{eq:perceptron}
h(\mathbf{F}_{l.},\theta)= 
\begin{cases}
    \text{1},& \text{if } \left\langle {\mathbf{F}_{l.},\mathbf{w}}\right\rangle\geq \theta\\
    \text{0},              & \text{otherwise},
\end{cases}
\end{equation}
or in other words: the single-layer perceptron triggers on the weighted sum of the individual features. A block diagram of the perceptron can be seen in figure~\ref{fig:perceptron}. The weight vector has to be trained in terms of classification accuracy. \cite{cohen}

\paragraph{Feed-Forward Neural Network}
\def\layersep{2.5cm}
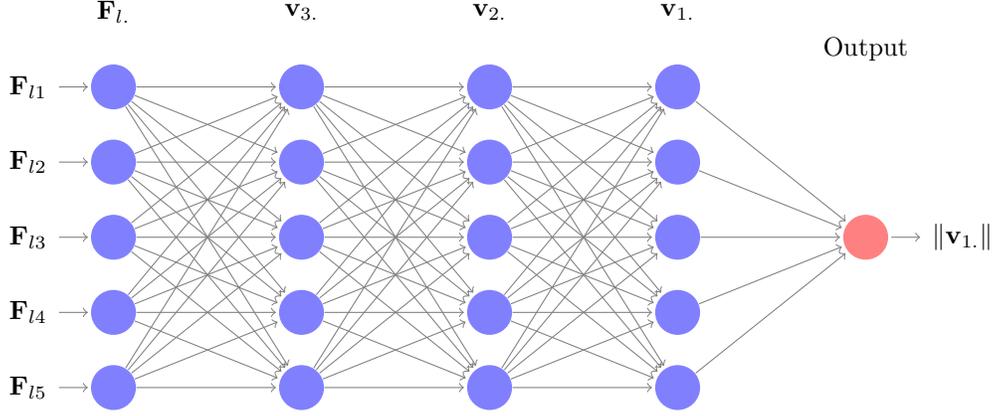
\begin{figure}
\centering
\begin{tikzpicture}[shorten >=1pt,->,draw=black!50, node distance=\layersep]
    \tikzstyle{every pin edge}=[<-,shorten <=1pt]
    \tikzstyle{neuron}=[circle,fill=black!25,minimum size=17pt,inner sep=0pt]
    \tikzstyle{input neuron}=[neuron, fill=blue!50];
    \tikzstyle{output neuron}=[neuron, fill=red!50];
    \tikzstyle{hidden neuron}=[neuron, fill=blue!50];
    \tikzstyle{hidden neuron2}=[neuron, fill=blue!50];
    \tikzstyle{hidden neuron3}=[neuron, fill=blue!50];
    \tikzstyle{annot} = [text width=4em, text centered]

    \foreach \name / \y in {1,...,5}
        \node[input neuron, pin=left:$\mathbf{F}_{l\y}$] (I-\name) at (0,-\y) {};

    \foreach \name / \y in {1,...,5}
        \path[yshift=0.0cm]
            node[hidden neuron] (A-\name) at (\layersep,-\y cm) {};
    \foreach \name / \y in {1,...,5}
        \path[yshift=0.0cm]
            node[hidden neuron2] (B-\name) at (2*\layersep,-\y cm) {};
    \foreach \name / \y in {1,...,5}
        \path[yshift=0.0cm]
            node[hidden neuron3] (H-\name) at (3*\layersep,-\y cm) {};

    \node[output neuron,pin={[pin edge={->}]right:$\left\| \mathbf{v}_{1.} \right\|$}, right of=H-3] (O) {};

    \foreach \source in {1,...,5}
        \foreach \dest in {1,...,5}
            \path (I-\source) edge (A-\dest);
    \foreach \source in {1,...,5}
        \foreach \dest in {1,...,5}
            \path (A-\source) edge (B-\dest);
    \foreach \source in {1,...,5}
        \foreach \dest in {1,...,5}
            \path (B-\source) edge (H-\dest);

    \foreach \source in {1,...,5}
        \path (H-\source) edge (O);

    \node[annot,above of=A-1, node distance=1cm] (hl) {$\mathbf{v}_{3.}$};
    \node[annot,above of=B-1, node distance=1cm] (hl) {$\mathbf{v}_{2.}$};
    \node[annot,above of=H-1, node distance=1cm] (hl) {$\mathbf{v}_{1.}$};
    \node[annot,above of=I-1, node distance=1cm] (hl) {$\mathbf{F}_{l.}$};
    \node[annot,above of=O] {Output};
\end{tikzpicture}
\caption{Block diagram of a feed-forward neural network.}
\label{fig:fann}
\end{figure}

The feed-forward neural network is an artificial neural network consisting of multiple layers\cite{bebis1994feed}. Each of the nodes $v_{ai}$ of one layer $a$ connects to all the nodes of the neighbouring layers $a-1,a+1$. To each of the connections, a weight $w_{aij}$ is assigned. Applying $\bm{L}_1$ metrics to $\mathbf{v}_{1.}$ (the norm of a vector is defined as the sum of its elements), the output equates to the norm of the last layer $\mathbf{v}_{1.}$. A threshold function, as in equation~\ref{eq:trigger}, then can trigger on the output value.

The feed-forward neural network with number of layers $\chi$ can be formalised as
\begin{equation}
h(\mathbf{F_l},\theta)=
\begin{cases}
    \text{1},& \text{if } \left\| \mathbf{v}_{1.} \right\| \geq \theta\\
    \text{0},              & \text{otherwise},
\end{cases}
\end{equation}
with the recursive definition, again taking advantage of inner products (equation~\ref{eq:innerproduct}),
\begin{equation}
\label{eq:ffnn}
\mathbf{v}_{ai}= 
\begin{cases}
    \left\langle \mathbf{v}_{a+1}, \mathbf{w}_{ai.}\right\rangle,& \text{if } a+1<\chi\\
    \left\langle \mathbf{F}_{j.}, \mathbf{w}_{ai.}\right\rangle,              & \text{otherwise}.
\end{cases}
\end{equation}
Dimensionality of the weight tensor $\mathbf{w}$ increases exponentially with each added input variable or layer, leading to high computational demands for multi-layer, multi-input feed-forward neural networks. A block diagram of the feed-forward neural network is shown in figure~\ref{fig:fann}.

\subsection{Results}
\label{sec:results}

\subsubsection{Test Setup}
\label{sec:testsetup}
\begin{figure}
\centering
  \includegraphics[width= 0.8\linewidth]{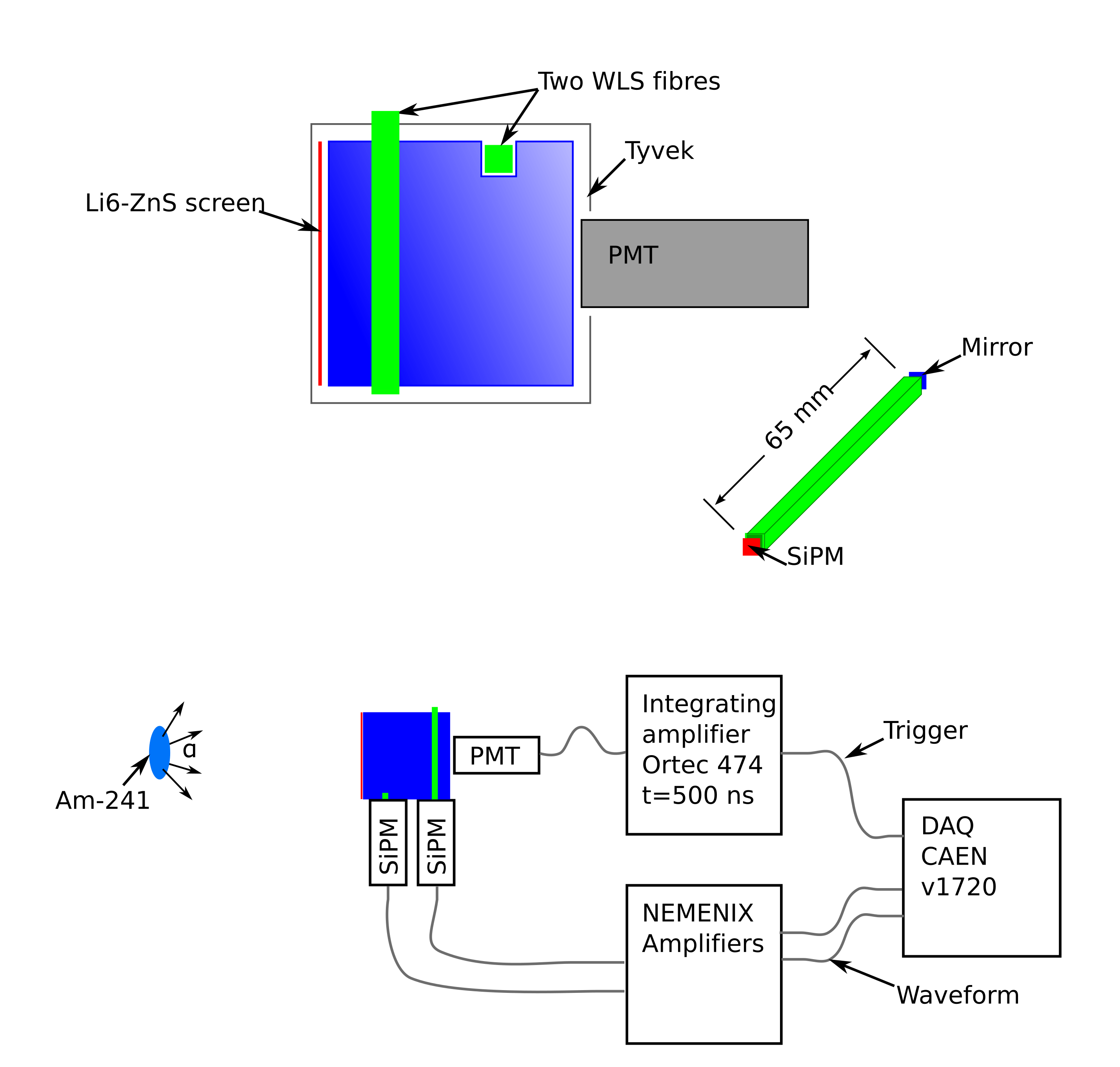}
\caption{Test setup for taking data of an $\mathrm{AmBe}$ source for neutron trigger evaluation.}
\label{fig:testsetup}
\end{figure}
Test data have been used for feature evaluation, as data from the SM1 did not have sufficient quality.  The test setup uses an $\prescript{241}{}{\mathrm{AmBe}}$ $\alpha$-particle source, located $\sim 3\si{\centi\meter}$ from a PVT cube. The test detector catches signals with two SiPMs and a PMT.  The test setup is sketched in figure~\ref{fig:testsetup}. $\alpha$-particle signals are very similar to neutron signals as both are caught by the \LiZnS-layer \cite{Stoykov:2014uca}, thus can be used to emulate IBD neutrons. Data read-out was triggered using a PMT signal, as the PMT delivers much more accurate signals than the SiPMs. Based on the PMT signal, the answer vector $\mathbf{E}$ was created stating whether a signal is a neutron or not. The algorithm for creating the reference vector uses a maximum amplitude threshold as well as Time-over-Threshold for PMT neutron identification. $120000$ events were acquired for the test data set, of which $2400$ are neutron signals and the rest are electro-magnetic/dark-count signals.

Using the test setup gives rise to two uncertainties: One is that the degree is unknown on which test conditions reflect conditions faced during the actual reactor-on phase. The other is uncertainty about how accurately the PMT signal classifies signals into neutrons. However it is assumed that algorithms being evaluated as having good performance in test data will also perform well at the reactor site, and that mis-classification by using the PMT signal is negligible.

\subsubsection{Feature Extraction Algorithms}
\label{sec:fealgorithms}
\begin{figure}
\centering
\begin{subfigure}{.55\textwidth}
  \centering
  \includegraphics[width= \linewidth]{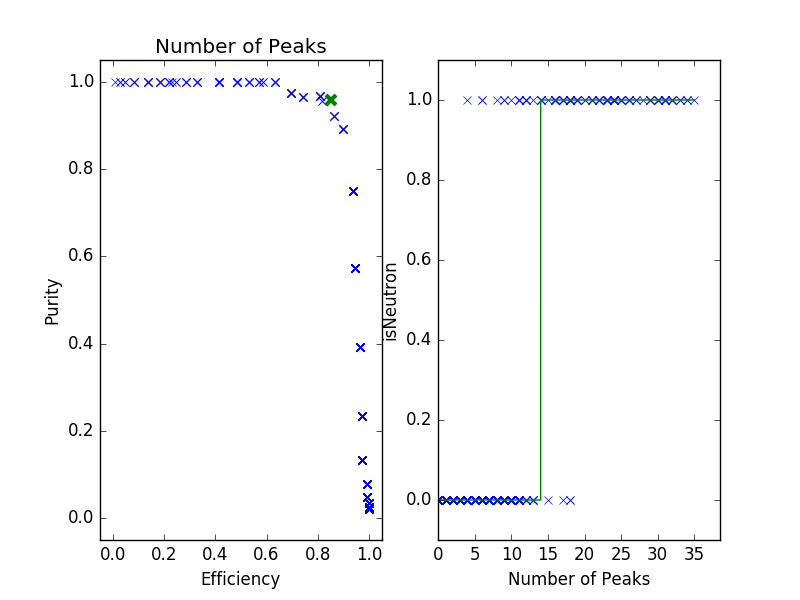}
  \caption{Number-of-Peaks.}
  \label{fig:sub18}
\end{subfigure}%
\begin{subfigure}{.55\textwidth}
  \centering
  \includegraphics[width= \linewidth]{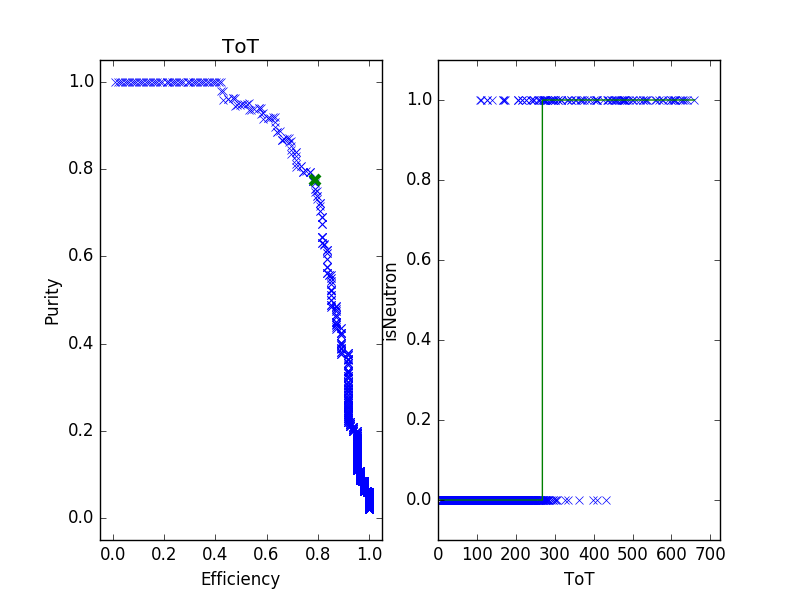}
  \caption{Time-over-Threshold (non-optimum $\theta$).}
  \label{fig:sub28}
\end{subfigure}%
\vskip\baselineskip
\begin{subfigure}{.55\textwidth}
  \centering
  \includegraphics[width= \linewidth]{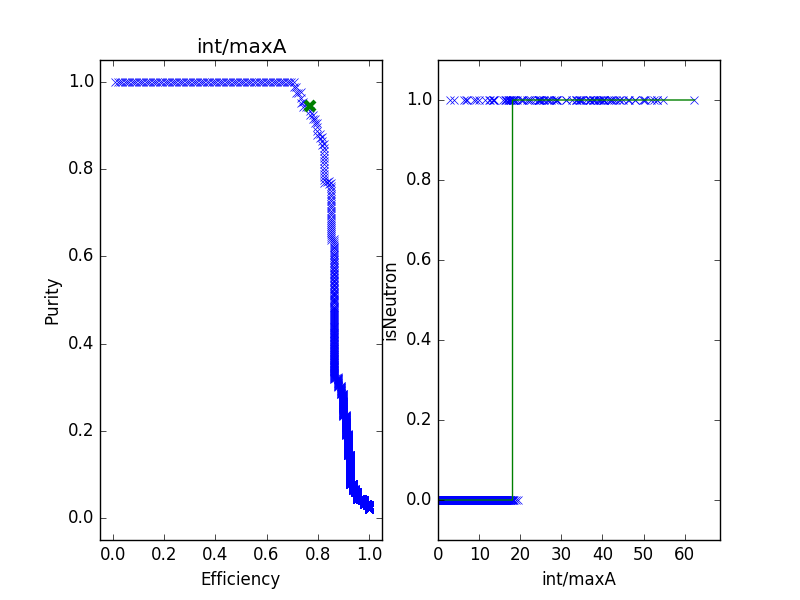}
  \caption{Integral-over-Amplitude.}
  \label{fig:sub38}
\end{subfigure}%
\begin{subfigure}{.55\textwidth}
  \centering
  \includegraphics[width= \linewidth]{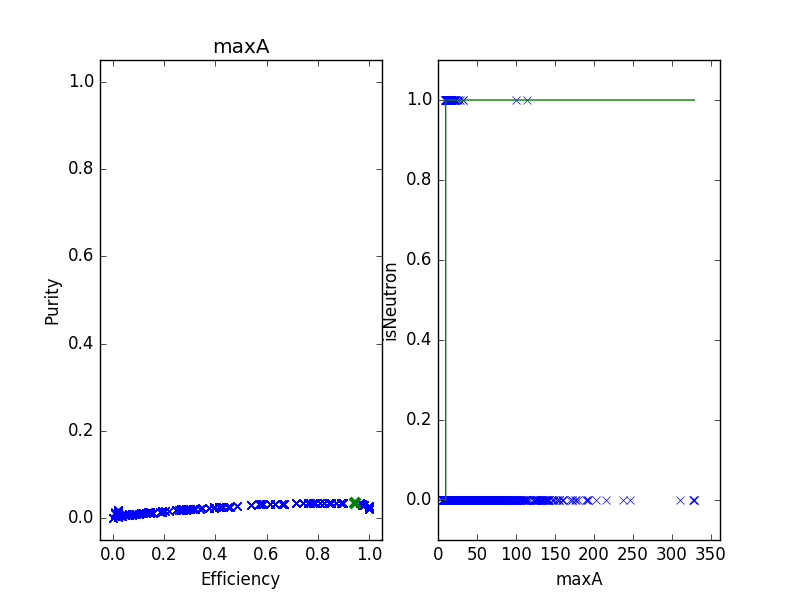}
  \caption{Maximum Amplitude.}
  \label{fig:sub48}
\end{subfigure}
\caption{Evaluation of individual features for the Number-of-Peaks (a), Time-over-Threshold (b), Integral-over-Amplitude (c) and maximum amplitude (d) algorithms on the efficiency-purity plane (each left) and for categorisation (see figure~\ref{fig:solexample}) (each right). The green cross and green horizontal line indicate the optimum trigger threshold. }
\label{fig:indfeat}
\end{figure}

Features are assessed using the evaluation outlined in section~\ref{sec:evaluation}. The threshold $\theta_i$ is swept from the minimum to the maximum value within the feature set in order to extract the corresponding curve. The procedure for selected features can be seen in figure~\ref{fig:indfeat}: Each trigger threshold will produce a value on the purity-efficiency plane. Higher thresholds usually correspond to higher purity and lower efficiency, while lower thresholds correspond to higher efficiency and lower purity. Those features that perform well both in terms of efficiency and purity can be considered suitable for triggering.
\paragraph{Tuning}
\begin{figure}
\centering
  \includegraphics[width= 0.85\linewidth]{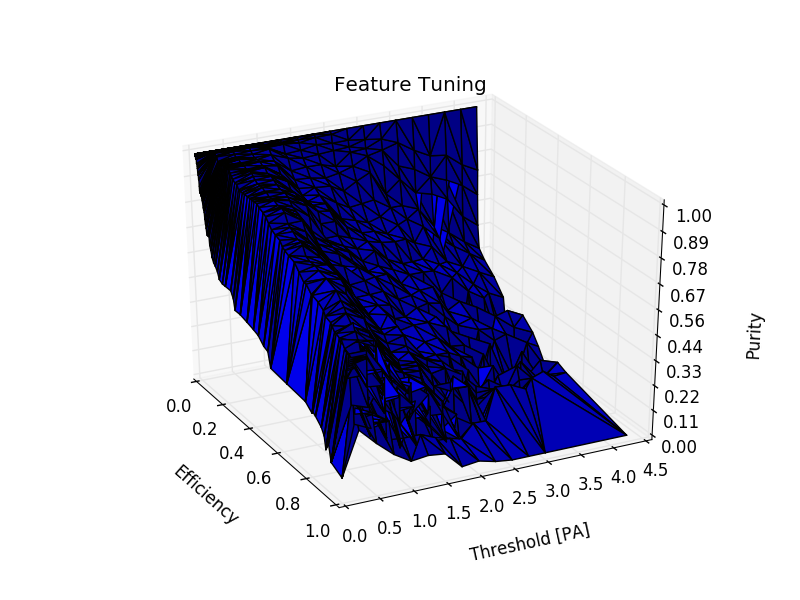}
\caption{Surface plot for Time-over-Threshold feature tuning. Threshold is swept from $0$ to $4.5\mathrm{PA}$ with a mountain visible in the range of $0.5\mathrm{PA}$, indicating the optimum value.}
\label{fig:tottuning}
\end{figure}
\begin{figure}
\centering
  \includegraphics[width= 0.85\linewidth]{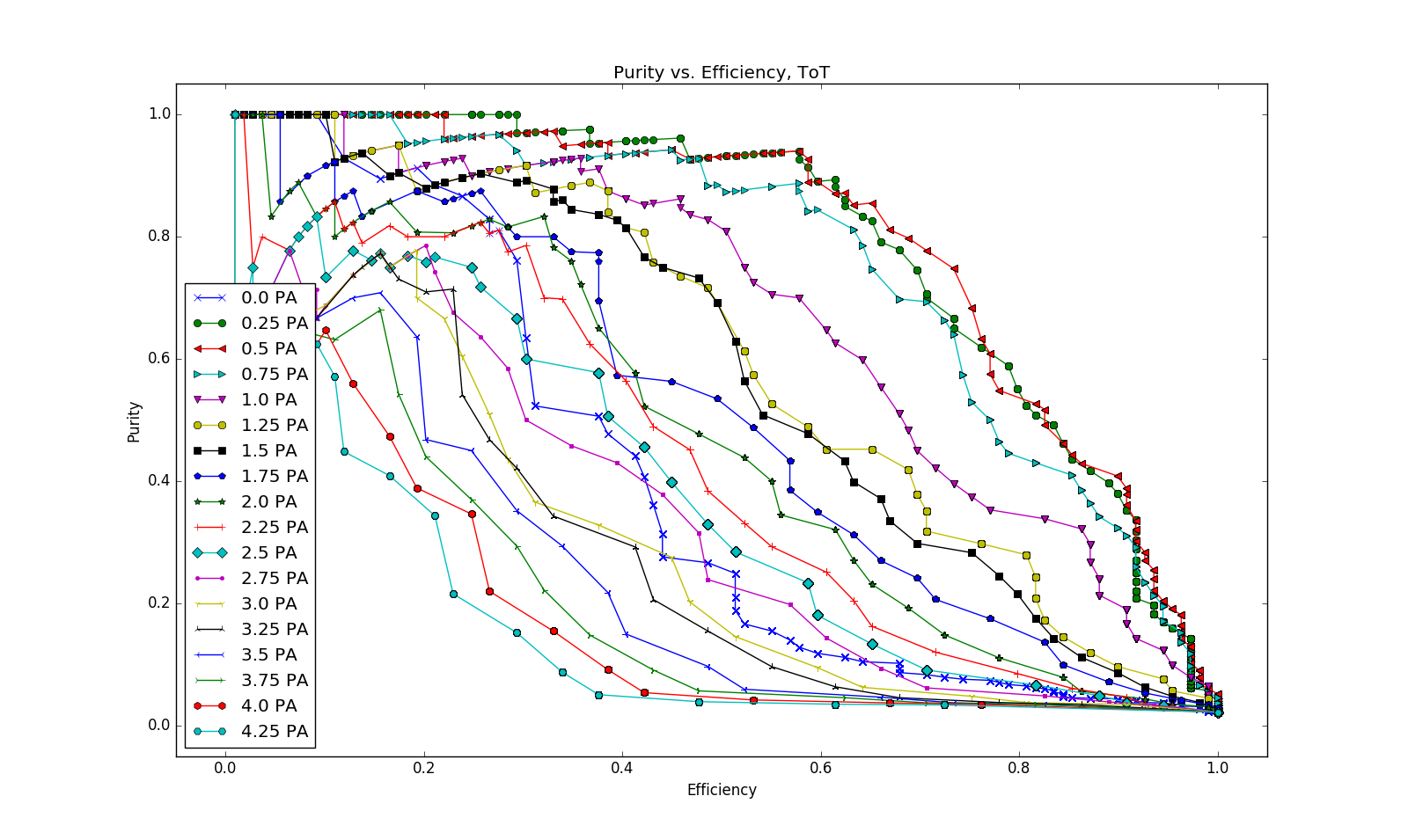}
\caption{Line plot for Time-over-Threshold feature tuning.}
\label{fig:tottuning_lines}
\end{figure}
Many features have free parameters, such as the threshold $\theta$ in the Number-of-Peaks (equation~\ref{eq:npeaks}) and Time-over-Threshold (equation~\ref{eq:tot}) algorithms. These parameters have to be tuned for the optimum value. This can be achieved by sweeping the free parameter \cite{Eiben11parametertuning}. 

An illustration of feature tuning on the example of the Time-over-Threshold algorithm is shown in figures~\ref{fig:tottuning} and~\ref{fig:tottuning_lines}, with the threshold value being swept from $0$ to $4.5$. On the surface plot a peak is visible in the range of $0.5\mathrm{PA}$. This indicates that optimum performance is achieved with the threshold in this range. 

The optimum thresholds are listed in table~\ref{tab:optimumthresholds}.
\begin{table}[]
\centering
\begin{tabular}{ll}
\hline\hline
\textbf{Algorithm}  & \textbf{Optimum threshold $\theta [\mathrm{PA}]$}  \\ \hline
Number-of-Peaks & $0.35$ \\
Time-over-Threshold & $0.45$ \\
\hline\hline
\end{tabular}
\caption{Optimum thresholds $\theta$ for the Number-of-Peaks and Time-over-Threshold algorithms acquired by parameter sweeps.}
\label{tab:optimumthresholds}
\end{table}
\paragraph{Comparison}
\begin{figure}
\centering
\begin{subfigure}{.5\textwidth}
  \centering
  \includegraphics[width= \linewidth]{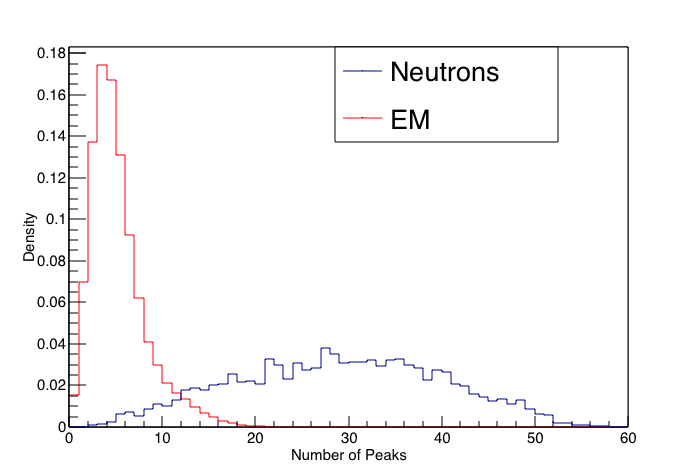}
  \caption{Number-of-Peaks.}
  \label{fig:sub13}
\end{subfigure}%
\begin{subfigure}{.5\textwidth}
  \centering
  \includegraphics[width= \linewidth]{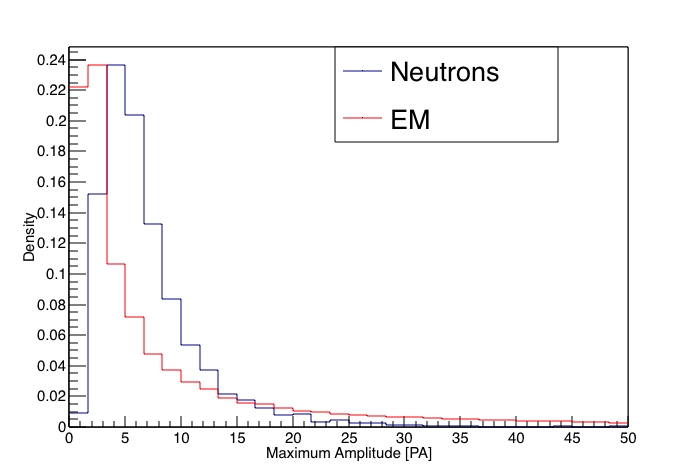}
  \caption{Maximum Amplitude.}
  \label{fig:sub23}
\end{subfigure}
\caption{Histograms for the number-of-peak (a) and for the maximum amplitude (b) values on the test data. Neutron signals and photon/dark count signals are clearly distinguishable, in contrast, the maximum amplitude feature values overlap for the most part.}
\label{fig:npeakhisto}
\end{figure}
\begin{figure}
\centering
  \includegraphics[width= 1.15\linewidth]{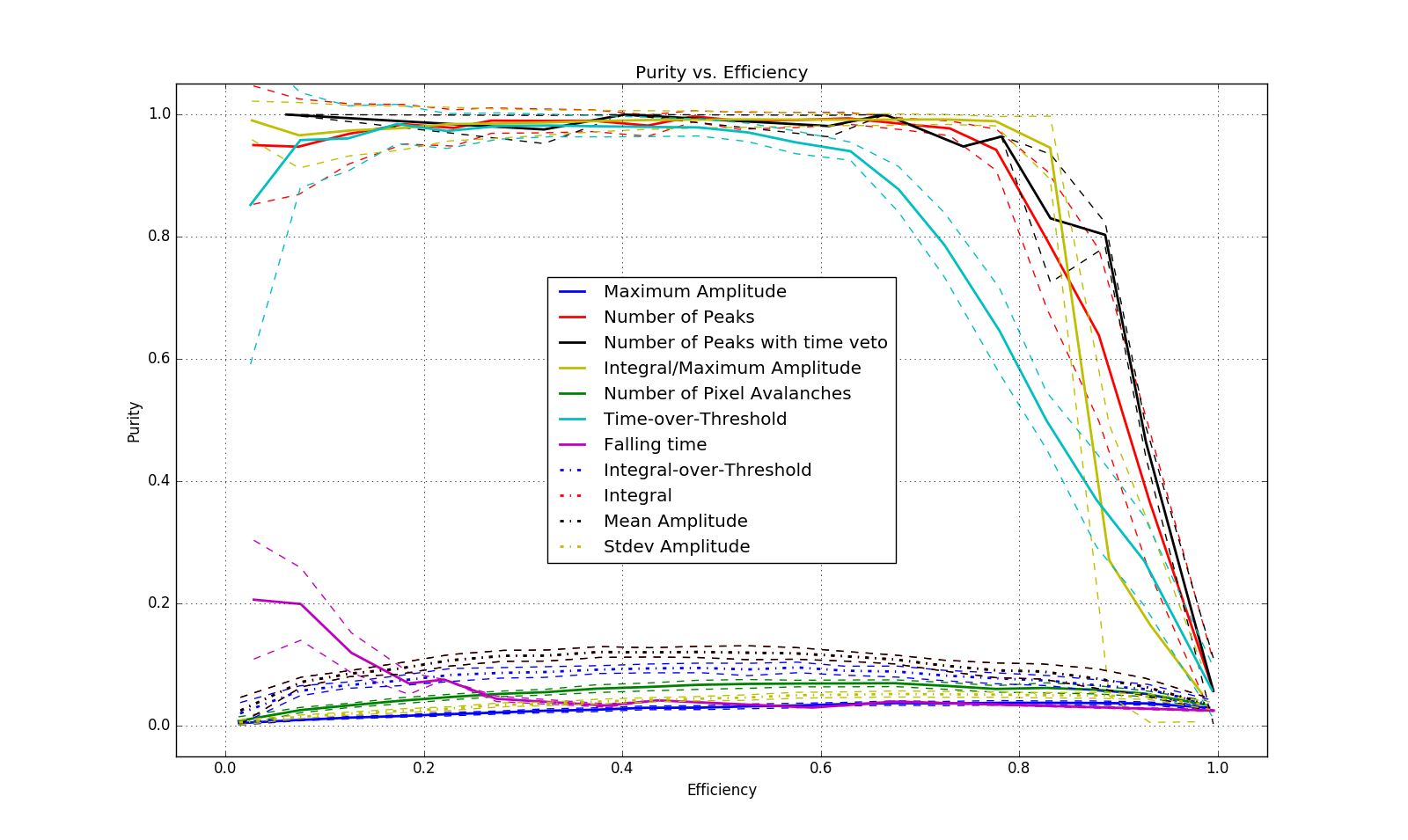}
\caption{Results for feature evaluation on the purity-efficiency plane.}
\label{fig:comp1}
\end{figure}
\begin{figure}
\centering
  \includegraphics[width= 1.15\linewidth]{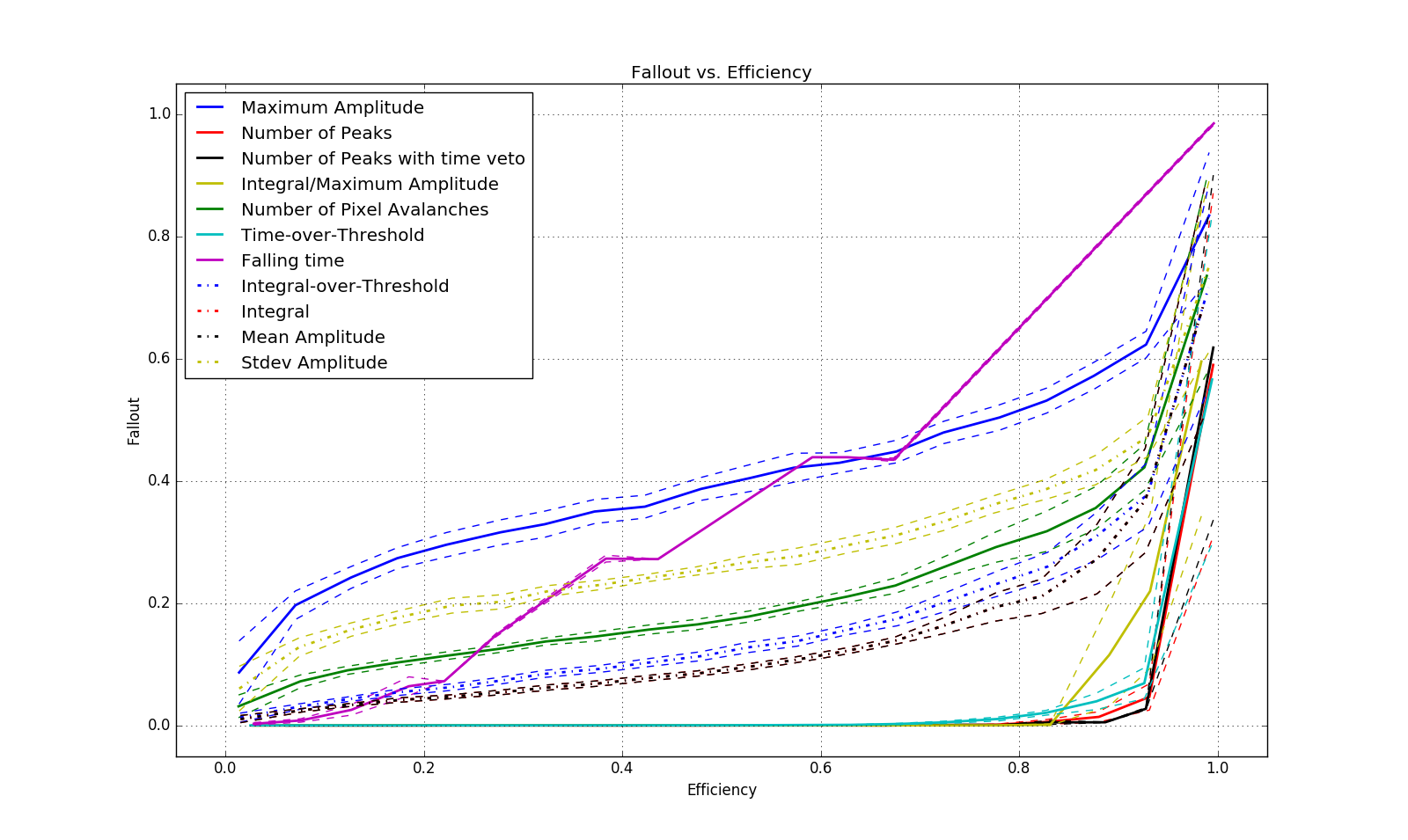}
\caption{Results for feature evaluation on the purity-fake rate (false positive rate, fallout) plane. The ROC curve summarising fewer algorithms is shown in figure~\ref{fig:comp3}.}
\label{fig:comp2}
\end{figure}

 The overall comparison of features is used to reduce the number of potential features by short-listing those that perform well, i.e. that yield high efficiency, high purity and low fake rate simultaneously. 
Features yielding these characteristics allow a good distinction between neutrons and non-neutrons, in contrast to those having bad performance (a visualisation can be seen in figure~\ref{fig:npeakhisto}). 
High efficiency (see equation~\ref{eq:efficiency}) correlates with how many IBD physics events are caught. High purity (see equation~\ref{eq:purity}) correlates with contamination of IBD events with backgrounds. And a low fake rate means that few non-IBD events are acquired (see equation~\ref{eq:fakerate}). \newpage
The results on the purity-efficiency plane are shown in figure~\ref{fig:comp1}, and those on the fake rate-efficiency plane are shown in figure~\ref{fig:comp2}, with the corresponding $1\sigma$ level (the ROC curve is shown in the conclusions section in figure~\ref{fig:comp3}).
Results have been obtained on the same channel, but on  $4$ different data subsets.

The analysis on the final results allow to reduce the features to be considered to those performing well:
\begin{itemize}
\item Number-of-Peaks (with and without time veto),
\item Time-over-Threshold, and
\item Integral-over-Amplitude.
\end{itemize}
At $10\%$ fake rate, efficiency is above $90\%$ efficiency for these features, and efficiency is held above $65\%$ -- and even above $\sim 90\%$ without the Time-over-Threshold feature --  for purity above $90\%$.

\subsubsection{Machine Learning Algorithms}
\label{sec:mlalgorithms}
\paragraph{Correlations}
\begin{figure}
\centering
\begin{subfigure}{.55\textwidth}
  \centering
  \includegraphics[width= \linewidth]{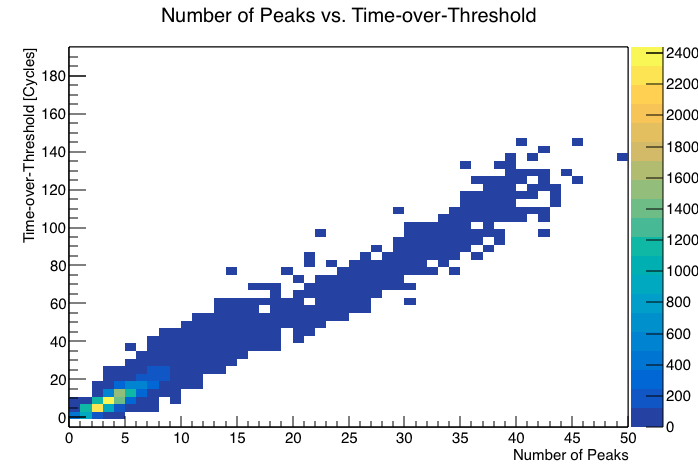}
  \label{fig:sub11}
\end{subfigure}%
\begin{subfigure}{.55\textwidth}
  \centering
  \includegraphics[width= \linewidth]{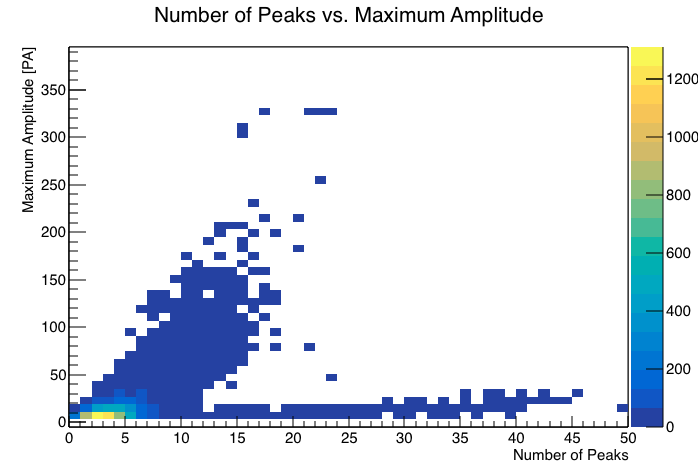}
  \label{fig:sub21}
\end{subfigure}%
\vskip\baselineskip
\begin{subfigure}{.55\textwidth}
  \centering
  \includegraphics[width= \linewidth]{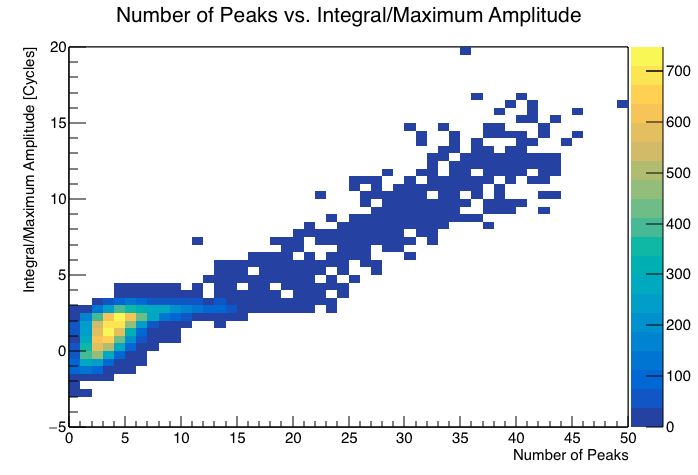}
  \label{fig:sub31}
\end{subfigure}%
\begin{subfigure}{.55\textwidth}
  \centering
  \includegraphics[width= \linewidth]{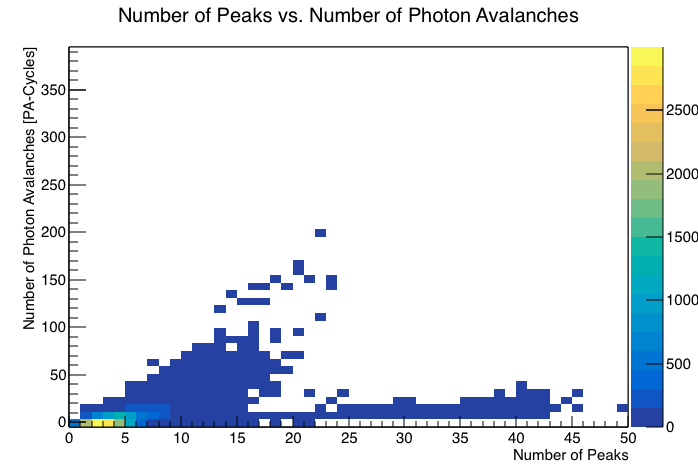}
  \label{fig:sub41}
\end{subfigure}
\caption{Two-dimensional histograms indicating correlation for both neutrons and non-neutrons between different features.}
\label{fig:pearson}
\end{figure}
\begin{table}[htbp]
\centering
  \includegraphics[width= 0.6\linewidth]{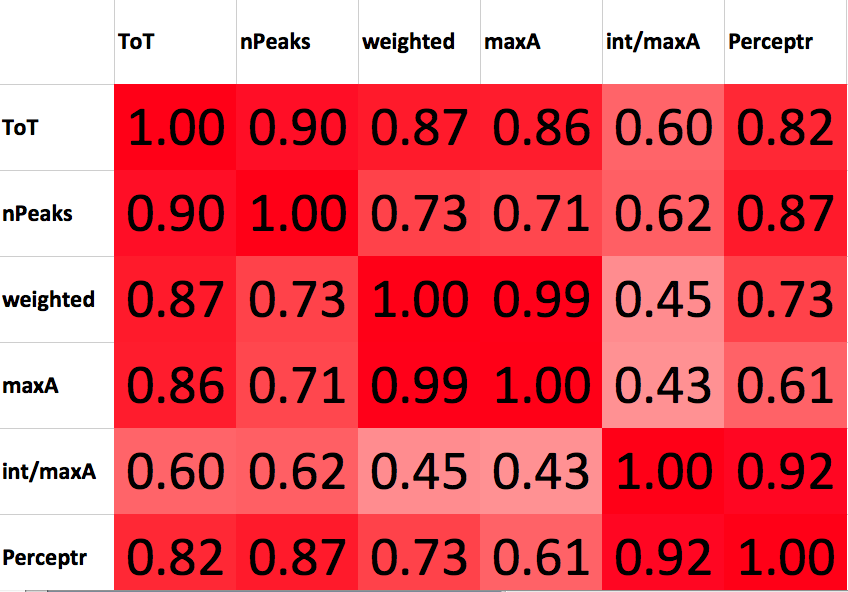}
\caption{Correlation matrix for selected features and the perceptron on both neutrons and non-neutrons. Dark red indicates highest, white lowest correlation ($p=1$, $p=0$). \textit{ToT:} Time-over-Threshold. \textit{nPeaks:} Number-of-Peaks. \textit{weighted:} Weighted Number-of-Peaks $=$ Number of Photon Avalanches. \textit{maxA}: Maximum Amplitude. \textit{int/maxA:} Integral-over-Amplitude. \textit{Perceptr:} Perceptron.}
\label{tab:pearson}
\end{table}
The more correlated features are and the worse they perform, the less information is gained by using a combination of different features. The decision whether or not one or several features are used is based on correlation between features and their individual performance. For correlation analysis, Pearson's $r$, defined as \cite{pearson1895note}
\begin{equation}
{\displaystyle r={\frac {\sum _{i=1}^{n}(x_{i}-{\bar {x}})(y_{i}-{\bar {y}})}{{\sqrt {\sum _{i=1}^{n}(x_{i}-{\bar {x}})^{2}}}{\sqrt {\sum _{i=1}^{n}(y_{i}-{\bar {y}})^{2}}}}}\\
={\frac {\operatorname {cov} (X,Y)}{\sigma _{X}\sigma _{Y}}}}
\end{equation}
 for data sets $X=\{x_1,x_2,\ldots,x_n\},Y=\{y_1,y_2,\ldots,y_n\}$ is commonly used to indicate correlation. Results of the correlation analysis, mostly on features with good performance, on the data set including both neutrons and non-neutrons can be seen in table~\ref{tab:pearson}, with some of the corresponding two-dimensional histograms shown in figure~\ref{fig:pearson}. The correlation analysis has also been undertaken on data sets consisting only of neutrons and only of non-neutrons. The features examined are correlated to a very high degree, except the Integral-over-Amplitude, indicating that redundant information would be obtained when using more than one feature. From the results it can be concluded that no significant increase of performance can be expected when using more than one feature that is not the Integral-over-Amplitude feature.
In section~\ref{sec:implementation} it will be shown why the Integral-over-Amplitude feature cannot be used on the firmware-level trigger.

\paragraph{Perceptron}
\begin{figure}
\centering
\begin{subfigure}{.5\textwidth}
  \centering
  \includegraphics[width= \linewidth]{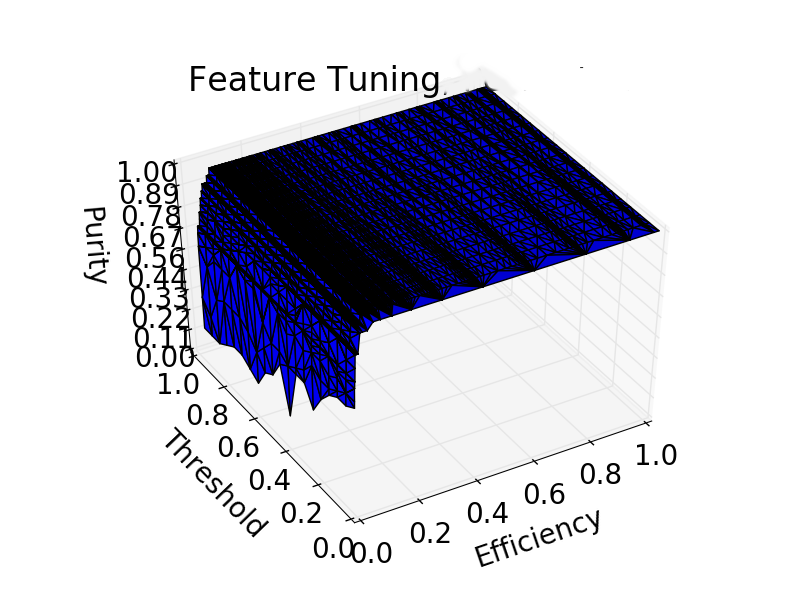}
  \caption{Perceptron with Number-of-Peaks and Time-over-Threshold.}
  \label{fig:ssub1a}
\end{subfigure}%
\begin{subfigure}{.5\textwidth}
  \centering
  \includegraphics[width= \linewidth]{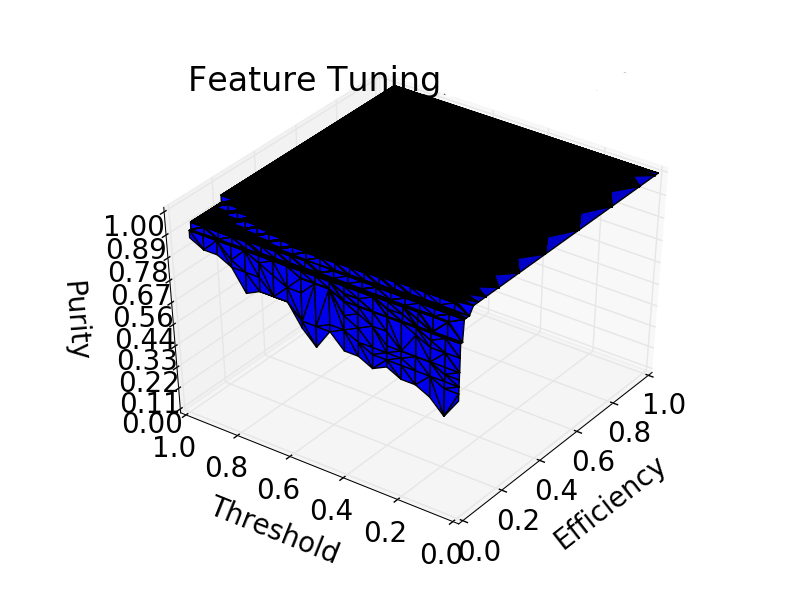}
  \caption{Perceptron with Number-of-Peaks and Integral-over-Amplitude.}
  \label{fig:ssub2a}
\end{subfigure}
\\
\begin{subfigure}{\textwidth}
  \centering
  \includegraphics[width= \linewidth]{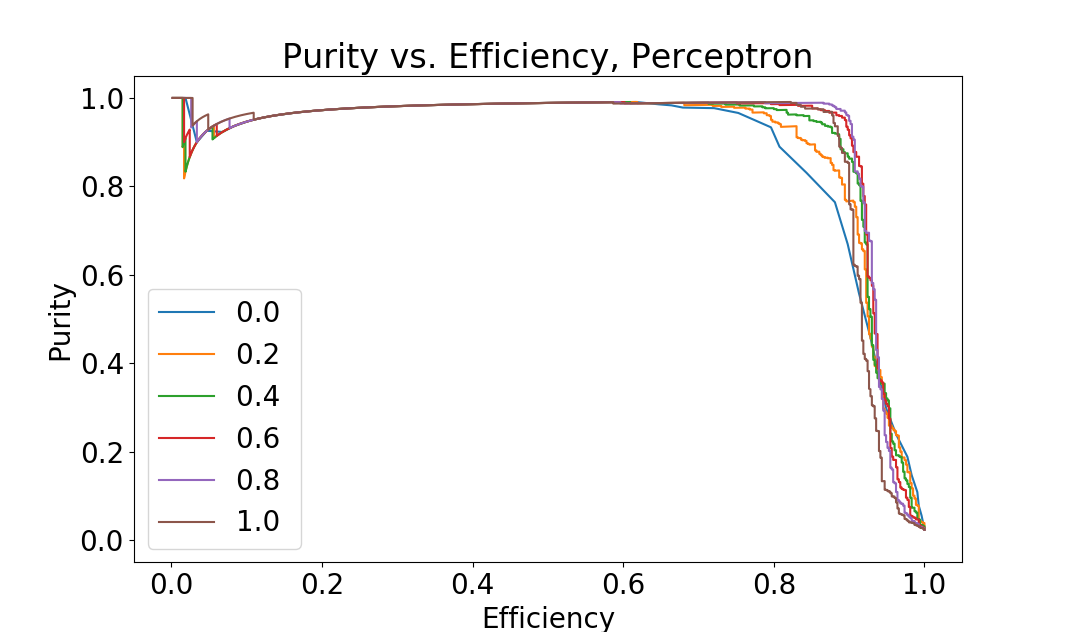}
  \caption{One-dimensional plot corresponding to the perceptron performance shown in \ref{fig:ssub2a}. Weight $w_1$ is sweeped from 0 (only Number-of-Peaks) to 1 (only Integral-over-Amplitude). The difference in purity between $w_1=1$ to the optimum at $w_1=0.8$ is less than $2\%$.}
  \label{fig:ssub3a}
\end{subfigure}
\caption{Perceptron evaluation.}
\label{fig:perceptronparameter}
\end{figure}
 Correlation analysis suggests that using a perceptron does not yield much better performance as information of different features are redundant. However, the perceptron has been evaluated using a normalised two input-node setup (one with Number-of-Peaks and Time-over-Threshold as inputs, one with Number-of-Peaks and Integral-over-Amplitude), in order to have only one parameter to sweep, $w_1$. With the weight vector defined as $\mathbf{w}=(w_1,1-w_1)$, the parameter sweep shows clearly that changes in performance by using the perceptron are tiny (see figure~\ref{fig:perceptronparameter}).
\paragraph{Feed-Forward Neural Network} 
\begin{figure}
\centering
\begin{subfigure}{.5\textwidth}
  \centering
  \includegraphics[width= \linewidth]{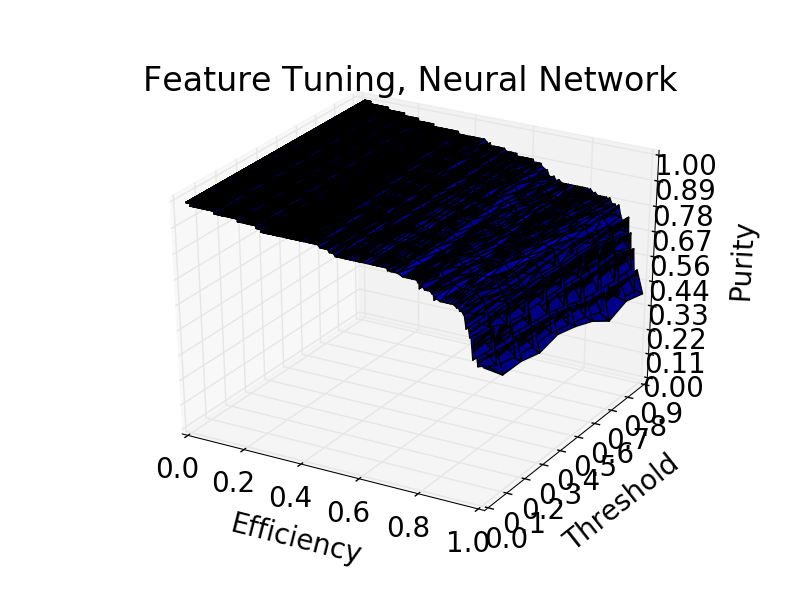}
\caption{Sweep of $w_2$.  $w_1=0.25$ }
  \label{fig:siub11}
\end{subfigure}%
\begin{subfigure}{.5\textwidth}
  \centering
  \includegraphics[width= \linewidth]{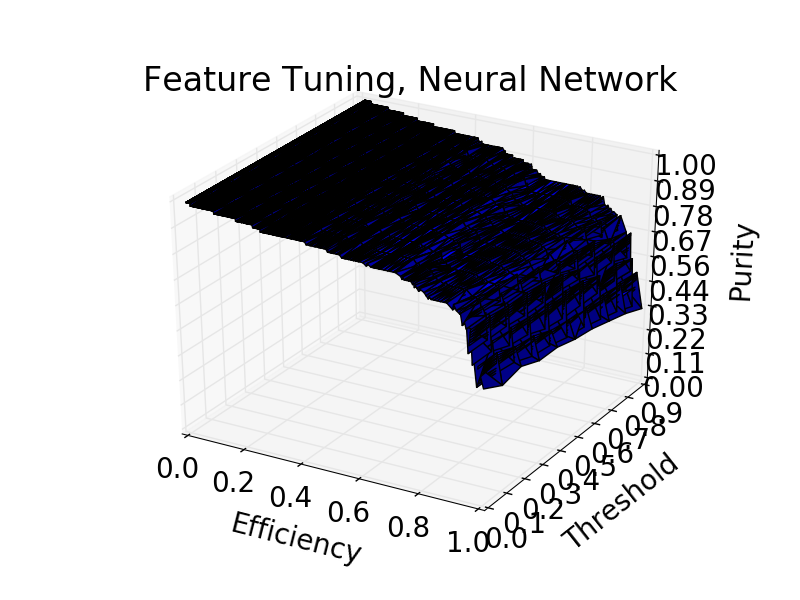}
\caption{Sweep of $w_2$. $w_1=0.5$}
  \label{fig:siub21}
\end{subfigure}%
\vskip\baselineskip
\begin{subfigure}{.5\textwidth}
  \centering
  \includegraphics[width= \linewidth]{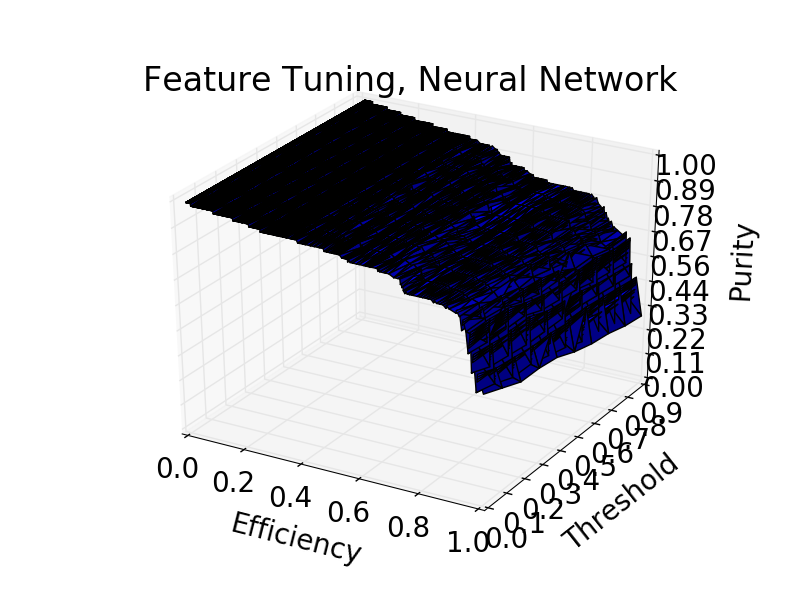}
\caption{Sweep of $w_2$. $w_1=0.75$}
  \label{fig:siub31}
\end{subfigure}%
\begin{subfigure}{.5\textwidth}
  \centering
  \includegraphics[width= \linewidth]{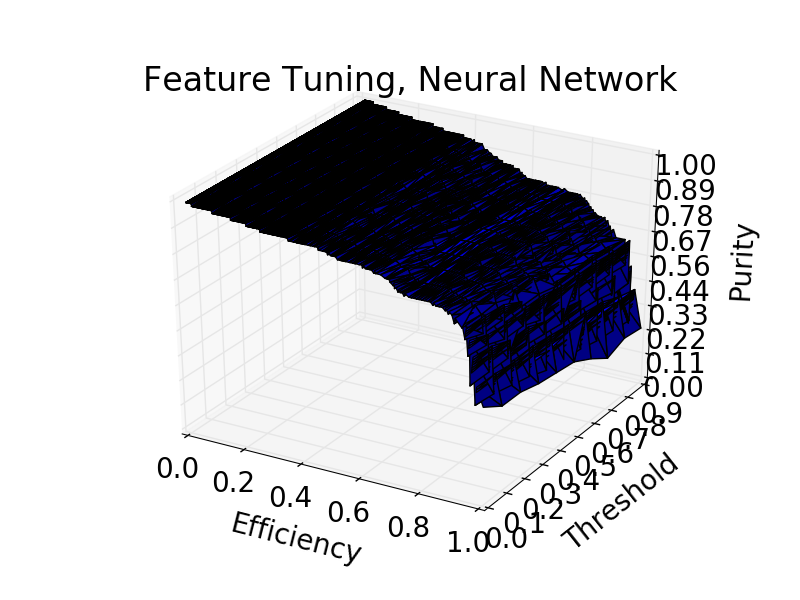}
\caption{Sweep of $w_2$. $w_1=1$}
  \label{fig:siub41}
\end{subfigure}
\\
\begin{subfigure}{0.7\textwidth}
  \centering
  \includegraphics[width= \linewidth]{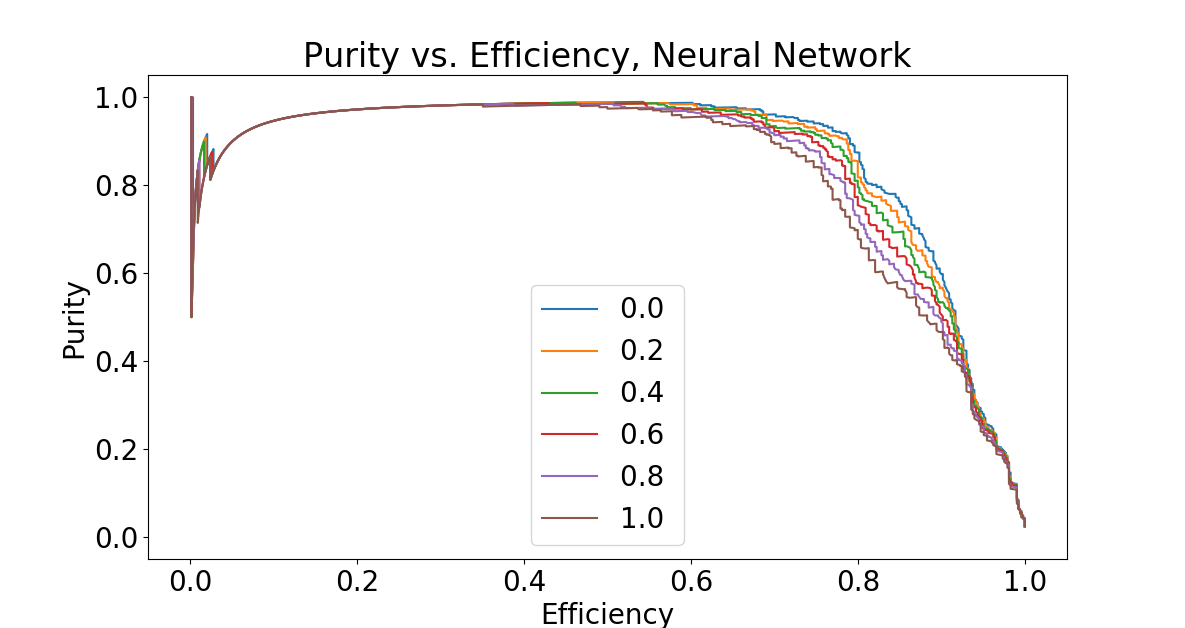}
  \caption{$w_2=0.5$, sweep of parameter $w_1$ from 0 to 1. The optimum value is $w_1=0$, giving full weight to Number-of-Peaks.}
  \label{fig:ssub1b}
\end{subfigure}%
\caption{Two-dimensional parameter sweep for a two-layer two-input (Number-of-Peaks and Time-over-Threshold) feed-forward neural network. No significant improvement in performance is achieved.}
\label{fig:ffnn}
\end{figure}

The same conclusions as for the perceptron hold for the feed-forward neural network: Since features are highly redundant, only a tiny amount of information is gained. A two-layer two-input feed-forward neural network with Number-of-Peaks and Time-over-Threshold as input has been benchmarked, with the weight matrix (as in equation~\ref{eq:ffnn})
\begin{equation}
\mathbf{w}=\left(\begin{matrix} w_1 & w_2 \\ 1-w_1 & 1-w2 \end{matrix}\right)
\end{equation}
with two free parameters $w_1,w_2$, resulting in the confirmation of the assumption that no significant increase in classification performance is achieved (see figures~\ref{fig:ffnn}). For the ANN, no advanced training method has been used, as the number of free parameters were chosen low in order to have a model that can be evaluated by parameter sweeps.
\paragraph{Triggering} The conclusion from correlation analysis is that only one feature will be used, hence a simple binary trigger (see equation~\ref{eq:trigger}) is implemented as the classification algorithm.

\subsection{Implementation}
\label{sec:implementation}
\begin{figure}
\centering
  \includegraphics[width= 1.05\linewidth]{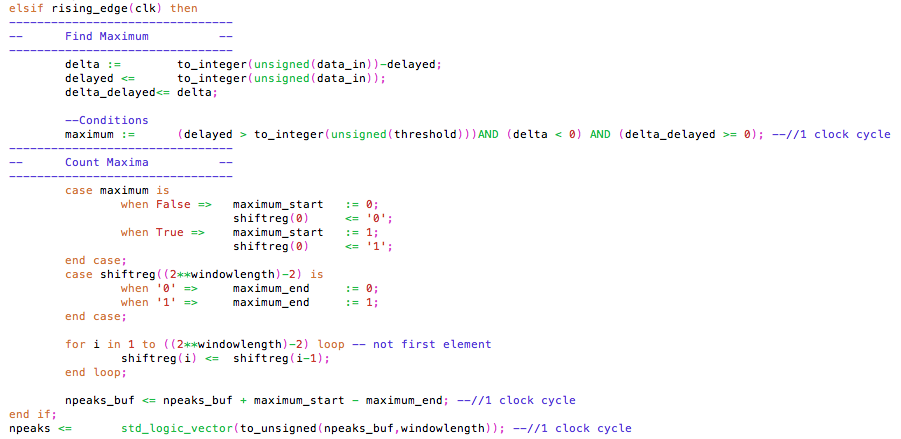}
\caption{Code snipped from the VHDL Number-of-Peaks feature extraction source code.}
\label{fig:workflow1}
\end{figure}
\begin{figure}
\centering
\begin{subfigure}{.5\textwidth}
  \centering
  \includegraphics[width= \linewidth]{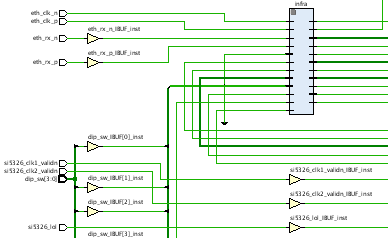}
  \caption{Part of the RTL netlist generated from the VHDL code.}
  \label{fig:ssub1}
\end{subfigure}%
\begin{subfigure}{.5\textwidth}
  \centering
  \includegraphics[width= 0.77\linewidth]{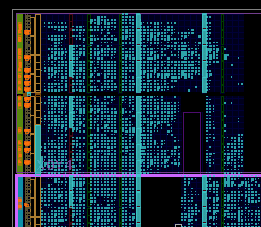}
  \caption{Snippet of the implementation of the RTL onto an Artix-7 FPGA.}
  \label{fig:ssub2}
\end{subfigure}
\caption{From the code, an RTL netlist is generated (a) that is then implemented into FPGA (b).}
\label{fig:workflow2}
\end{figure}

The selected trigger algorithms are implemented via synthesis and routing from VHDL code, a hardware description language synthesisable into digital circuits \cite{skahill1996vhdl}, into the FPGA. A netlist (RTL) is generated from the HDL code input, and implementation and bitstream containing the FPGA circuit configuration is created using the netlist \cite{pichler} \cite{9780534384623}. For the toolchain, \textsc{Xilinx} Vivado has been used for design and synthesis and \textsc{Mentor Graphics} Questasim for simulation and verification.

As 64 channel triggers plus the other firmware elements have to be synthesised on a single FPGA, hardware resources are strictly limited (see table~\ref{tab:artix}). Therefore each of the algorithms has to be checked whether their resource usage fits the device's resource availability. VHDL implementation of algorithms and their synthesis has been undertaken for the Time-over-Threshold, Number-of-Peaks (with and without time veto) and Integral-over-Amplitude features, as well as for the Integral feature. An example of an implementation workflow is shown in figures~\ref{fig:workflow1} and \ref{fig:workflow2}.

\subsubsection{Resource Usage}
\label{sec:resource}
\begin{figure}
\centering
  \includegraphics[width= 0.66\linewidth]{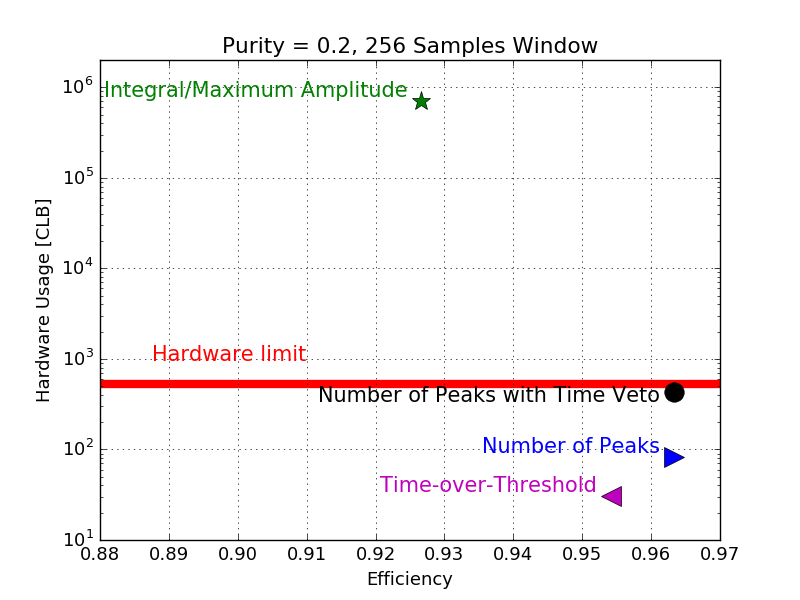}
\caption{Hardware usage of different feature extraction algorithms versus efficiency at a $20\%$ purity level and on a window size of $256$ samples. The red line indicates hardware resource availability per channel.}
\label{fig:usagevsefficiency}
\end{figure}
\begin{figure}
\centering
  \includegraphics[width= 0.91\linewidth]{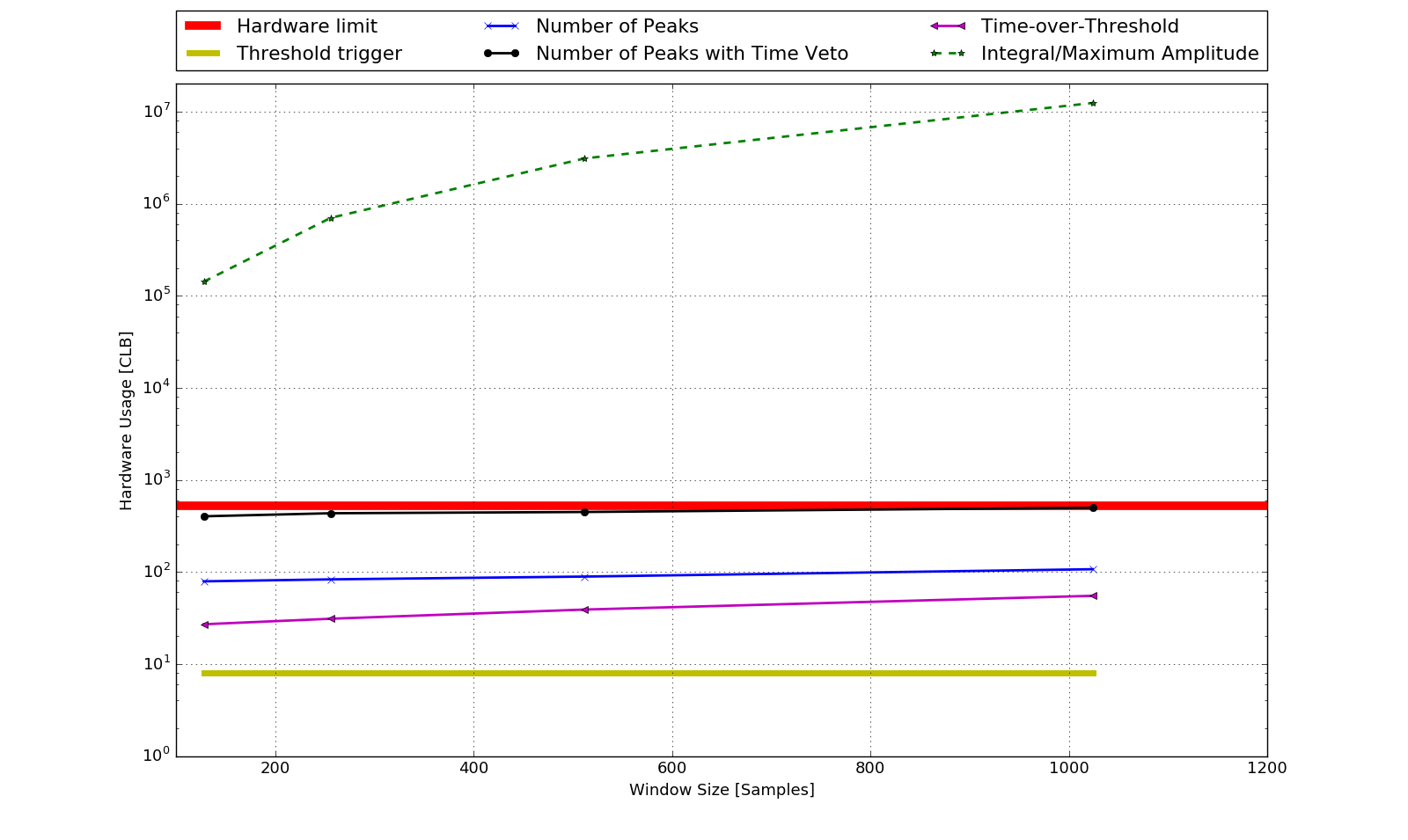}
\caption{Hardware usage of different feature extraction algorithms versus time window size. The expected window size is $256$ samples. The red line indicates hardware availability per channel.}
\label{fig:usagevswindowsize}
\end{figure}
As can be seen in equations~\ref{eq:totd}, \ref{eq:npeaksd} and  \ref{eq:npeaksd2} and \ref{eq:featintegrald}, the Integral, Time-over-Threshold, Number-of-Peaks features rely on comparators, sums and differences only, as integrals will turn to sums in discretisation, and differentials will turn to differences. Sums and differences can be easily implemented as adder-subtractors into digital logic \cite{nagaraj2012fpga}.
However, the Integral-over-Amplitude algorithm contains a division, as can be seen in equation~\ref{eq:integralamplitude2}, an operation that is resource-intense on the FPGA. 
The usage of hardware resources is compared in figure~\ref{fig:usagevsefficiency} as a function of efficiency at a $20\%$ purity level and in figure~\ref{fig:usagevswindowsize} as a function of window size. 
It can be seen that the Integral-over-Amplitude algorithm exceeds the hardware limit by far. 
The time veto-version of the Number-of-Peaks algorithms approaches the limit of what is implementable. 
Despite it is the best-performing feature, it has been dropped because of this issue, namely in consideration to the logic used for other functions of the FPGA.

The Number-of-Peaks algorithm has been tested both in the variation of equation~\ref{eq:npeaksd} (using the zero crossing of the differential signal) and \ref{eq:npeaksd2} (using amplitude comparison only), and the exact same logic usage has been determined for both algorithms. As both lead to the same results, performance stays the same for both variants. 

Therefore, implementation into the firmware has been made for the Number-of-Peaks algorithm according to equation~\ref{eq:npeaksd} without time veto, and the Time-over-Threshold feature extraction method.

\subsubsection{Power Consumption}
\begin{table}[]
\centering
\begin{tabular}{llll}
\hline\hline
 Algorithm&Number of Channels& \textbf{Current $[\si{\ampere}]$}& \textbf{Power consumption $[\si{\watt}]$} \\ \hline
Number-of-Peaks			&	$2$ 	& $1.31$&	\\
Number-of-Peaks			&	$4$ 	& $1.32$&	\\
Number-of-Peaks			&	$6$ 	& $1.33$&	\\
Number-of-Peaks			&	$8$ 	& $1.33$&	\\
Number-of-Peaks			&	$10$ 	& $1.34$&	\\
Number-of-Peaks			&	$12$ 	& $1.35$&	\\ \hline
Time-over-Threshold		&	$2$ 	& $1.48$&	\\
Time-over-Threshold		&	$4$ 	& $1.49$&	\\
Time-over-Threshold		&	$6$ 	& $1.49$&	\\
Time-over-Threshold		&	$8$ 	& $1.50$&	\\
Time-over-Threshold		&	$10$ 	& $1.51$&	\\
Time-over-Threshold		&	$12$ 	& $1.51$&	\\ \hline
Number-of-Peaks	&	$1$ (extrapolated) 	& $0.007$&	\\
Time-over-Threshold&	$1$ (extrapolated) 	& $0.005$&	\\ \hline
\textbf{Number-of-Peaks}	&	$64$ (extrapolated) 	& $1.72$&$8.6$	\\
\textbf{Time-over-Threshold}&	$64$ (extrapolated) 	& $1.79$&$9.0$	\\
\hline\hline
\end{tabular}
\caption{Current and power consumption for the trigger algorithms for different number of channels.}
\label{tab:power}
\end{table}
Power consumption is limited to $12\si{\watt}$ by the DC-DC converter on the \textsc{Trenz} module \cite{dcdc}. Therefore it has to be verified that FPGA power consumption meets the requirements. 
The full firmware including other firmware components has been synthesised in multiple variants, from 2 to 12 channels (the maximum number of channels that were available at the time of measurement), and the consumed power has been extrapolated to a single and to 64 channels. 
The results are shown in table~\ref{tab:power}, according to which power consumption should meet the requirements set by the DC-DC converter. However some uncertainty exists depending on the increase of power consumption when in operation.

\subsection{Plane-level Trigger}
\label{sec:prospects}
\begin{table}
\centering
\begin{tabular}{l l l}
\hline\hline
 & \textbf{Efficiency [\%]} & \textbf{Purity [\%]}\\ \hline
\textbf{Channel 0} & 96.2 & 20.4 \\
\textbf{Channel 0} $\lor$ \textbf{Channel 1} & 97.6 & 12.3 \\
\textbf{Channel 0} $\land$ \textbf{Channel 1} & 93.8 & 55.5 \\
\hline\hline
\end{tabular}
\caption{Efficiency and purity values on Number-of-Peaks feature for the same threshold value and data set. In this example, by using logical conjunction, a small drop in efficiency is contrasted by a high increase in purity.}
\label{tab:prospects}
\end{table}
The implemented trigger is a per-channel trigger. It decides on a rolling basis whether in the past $m$ samples a neutron event has possibly occured. If so, the block is tagged as trigger-positive and this decision forwarded to the plane-level trigger. 

At the moment, a trigger on each channel causes a read-out, or in other words, the 64 channel triggers are logically disjunct ($\lor=$ \textsc{or},\cite{andrews2002introduction}):
\begin{equation}
\hat{\mathbf{y}}_\text{plane}=\bigvee\limits_{i=1}^{64} \hat{\mathbf{Y}}_i.
\end{equation}
Several other logical connections can be implemented, such as the logical conjunction ($\land=$ \textsc{and},\cite{andrews2002introduction}) of fibres in the same cube (two on the $x$ and two on the $y$ axis):
\begin{equation}
\hat{\mathbf{y}}_\text{plane}=\left(\bigvee\limits_{i=1}^{16}\bigwedge\limits_{j=2i-1}^{2i} \hat{\mathbf{Y}}_j\right) \land\left(\bigvee\limits_{i=33}^{48}\bigwedge\limits_{j=2i-1}^{2i} \hat{\mathbf{Y}}_j  \right).
\end{equation}
As only two channels on the same cube have been available on the test data, changes in performance has been evaluated using just these two highly correlated data sets. Results presented in table~\ref{tab:prospects} show that, in the present example, when using logical conjunction instead of logical disjunction, a small drop in efficiency contrasts a high increase in purity, making this logical linking  advisable for implementation on the plane level.

\clearpage
\section{Conclusion}
\label{sec:conclusion} 
\begin{figure}[H]
\centering
  \includegraphics[width= 0.86\linewidth]{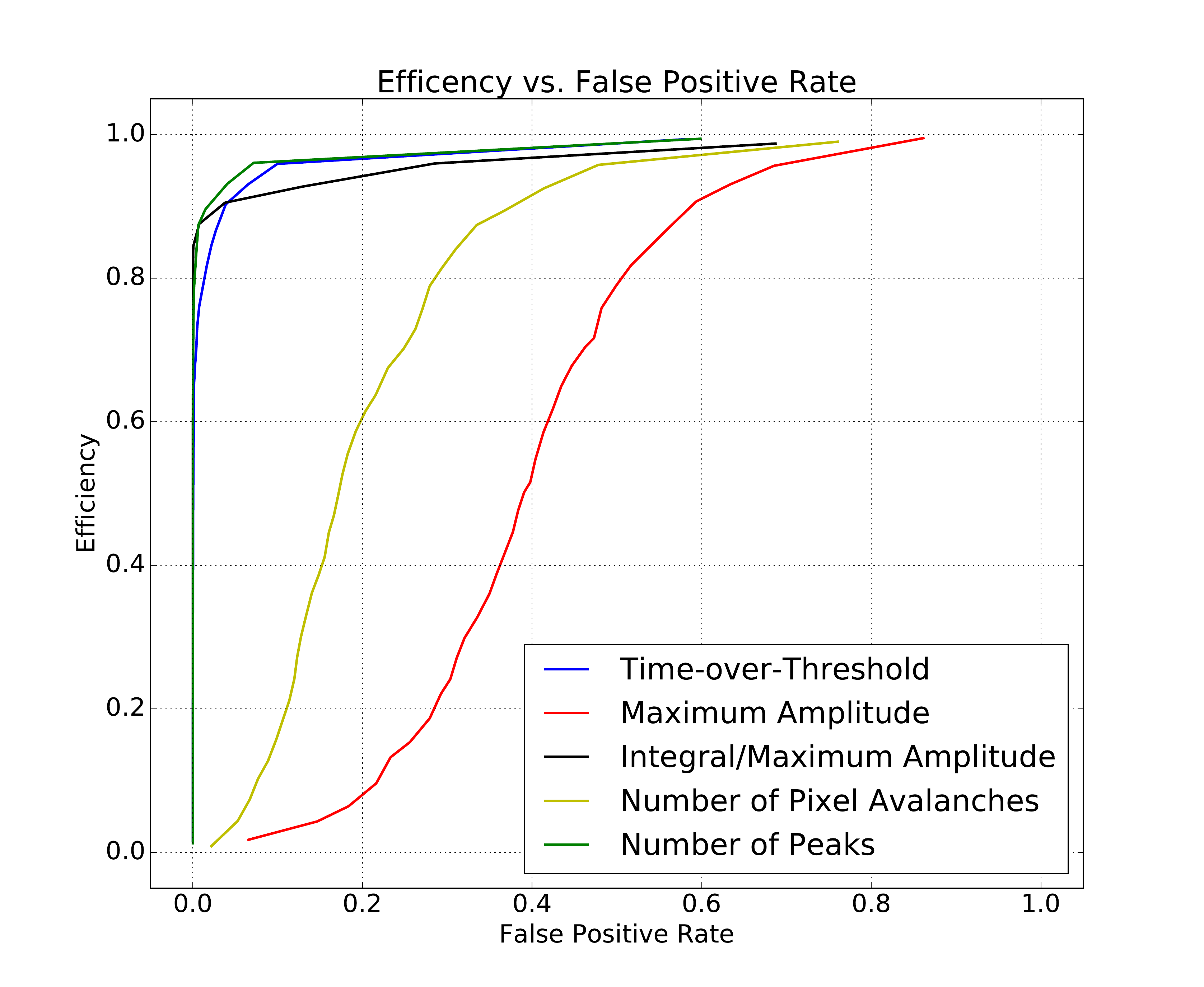}
\caption{Summarising ROC curve of different features in terms of efficiency (TPR) and fake rate (fallout, FPR), with the Number-of-Peaks algorithm performing best on the test data set.}
\label{fig:comp3}
\end{figure}

In order to define the channel firmware trigger, various feature extraction and machine learning algorithms have been evaluated. This includes a variety of feature extraction approaches, and a threshold trigger and two sorts of neural networks on the machine learning side.
According to the findings, the most suitable trigger has been implemented, that are now the Number-of-Peaks and the Time-over-Threshold feature extraction algorithms (see summarising figure~\ref{fig:comp3}) and a threshold trigger on the feature extraction output value. These algorithms have been implemented and synthesised into firmware, and embedded into the \textit{IPbus} framework. The commissioning of the trigger is still to come, and the surrounding software and firmware have to be made ready to write to the trigger parameter registers.

The trigger performance evaluation was carried out relative to the test data, and conditions at the reactor are not known at the moment. Therefore the decision on the trigger algorithm relies on the assumption that the best performing algorithms on the test environment also perform well on the reactor site. However this might be wrong if background signals are qualitatively different from the test signals. This can be considered unlikely though, as the same detection scheme underlies both setups. The actual background rate determines at which efficiency level the trigger can operate. The tuning of the parameters for the trigger is performed at \textbf{calibration}. Calibration will aim not only for highest efficiency, but also for uniformity along the different channels. As trigger efficiency directly propagates to detector efficiency, optimal trigger calibration is essential for the quality of the experiment.

The hunt for sterile neutrinos in the very-short baseline range has started, and SoLid is about to start its first large-scale physics run. Whether or not the reactor anomaly -- the deficit of neutrinos in the short-baseline range of neutrinos -- can be reproduced very close to the reactor core is crucial in deciding the fate of the three-neutrino oscillation model. 
In the case the reactor anomaly cannot be reconfirmed, the results would fit nicely not only with the three-flavour neutrino oscillation model, but also with recent results from accelerator- and atmospheric-based neutrino experiments. In case SoLid, in line with other very short-baseline experiments, does find a deficit, a theorist has to seriously consider an update of the neutrino model currently most favoured, from having three to having four or even more neutrino flavour states by adding sterile neutrinos.



\clearpage

\clearpage


\bibliographystyle{unsrt}
\bibliography{Thesis}

\begin{thebibliography}{100}

\bibitem{Agashe:2014kda}
K.A. Olive et~al.
\newblock {Review of Particle Physics}.
\newblock {\em Chin. Phys.}, C38:090001, 2014.

\bibitem{Langacker:2009my}
P.~Langacker.
\newblock {Introduction to the Standard Model and Electroweak Physics}.
\newblock In {\em {Proceedings of Theoretical Advanced Study Institute in
  Elementary Particle Physics on The dawn of the LHC era (TASI 2008), Boulder,
  USA, 2-27 Jun}}, 2008.

\bibitem{Herrero:1998eq}
M.~Herrero.
\newblock {The Standard model}.
\newblock {\em NATO Sci. Ser. C}, 534:1--59, 1999.

\bibitem{Fukuda:1998fd}
Y.~Fukuda et~al.
\newblock {Measurements of the solar neutrino flux from Super-Kamiokande's
  first 300 days}.
\newblock {\em Phys. Rev. Lett.}, 81:1158--1162, 1998.
\newblock [Erratum: Phys. Rev. Lett.81,4279(1998)].

\bibitem{Zhang:2015wua}
J.~Zhang and S.~Zhou.
\newblock {Relic Right-handed Dirac Neutrinos and Implications for Detection of
  Cosmic Neutrino Background}.
\newblock {\em Nucl. Phys.}, B903:211--225, 2016.

\bibitem{Lipari:2001is}
P.~Lipari.
\newblock {Introduction to neutrino physics}.
\newblock In {\em {2001 CERN-CLAF School of High-Energy Physics, Itacuruca,
  Brazil, 6-19 May}}, pages 115--199, 2001.

\bibitem{Dobrynina:2016rwy}
A.~Dobrynina, A.~Kartavtsev, and G.~Raffelt.
\newblock {Helicity oscillations of Dirac and Majorana neutrinos}.
\newblock {\em Phys. Rev.}, D93(12):125030, 2016.

\bibitem{Goldhaber:1958nb}
M.~Goldhaber, L.~Grodzins, and A.W. Sunyar.
\newblock {Helicity of Neutrinos}.
\newblock {\em Phys. Rev.}, 109:1015--1017, 1958.

\bibitem{King:2008vg}
S.F. King.
\newblock {Neutrino Mass Models: A Road map}.
\newblock {\em J. Phys. Conf. Ser.}, 136:022038, 2008.

\bibitem{Pontecorvo:1967fh}
B.~Pontecorvo.
\newblock {Neutrino Experiments and the Problem of Conservation of Leptonic
  Charge}.
\newblock {\em Sov. Phys. JETP}, 26:984--988, 1968.
\newblock [Zh. Eksp. Teor. Fiz.53,1717(1967)].

\bibitem{Ahmad:2002jz}
Q.R. Ahmad et~al.
\newblock {Direct evidence for neutrino flavor transformation from neutral
  current interactions in the Sudbury Neutrino Observatory}.
\newblock {\em Phys. Rev. Lett.}, 89:011301, 2002.

\bibitem{Fukuda:1998mi}
Y.~Fukuda et~al.
\newblock {Evidence for oscillation of atmospheric neutrinos}.
\newblock {\em Phys. Rev. Lett.}, 81:1562--1567, 1998.

\bibitem{Maki:1962mu}
Z.~Maki, M.~Nakagawa, and S.~Sakata.
\newblock {Remarks on the unified model of elementary particles}.
\newblock {\em Prog. Theor. Phys.}, 28:870--880, 1962.

\bibitem{Kobayashi:1973fv}
M.~Kobayashi and T.~Maskawa.
\newblock {CP Violation in the Renormalizable Theory of Weak Interaction}.
\newblock {\em Prog. Theor. Phys.}, 49:652--657, 1973.

\bibitem{EIDELMAN20041}
S.~Eidelman, K.G. Hayes, K.A. Olive, et~al.
\newblock Review of particle physics.
\newblock {\em Physics Letters B}, 592(1):1 -- 5, 2004.

\bibitem{Czakon:1999ed}
M.~Czakon, M.~Zralek, and J.~Gluza.
\newblock {Are neutrinos Dirac or Majorana particles?}
\newblock {\em Acta Phys. Polon.}, B30:3121--3138, 1999.

\bibitem{xing}
Z.Z. Xing.
\newblock Neutrino masses \& flavor mixing.
\newblock In {\em 48. Internationale Universit\"{a}tswochen für Theoretische
  Physik, Schladming, Austria, 27 Feb-6 Mar}, 2010.

\bibitem{Olive:2016xmw}
C.~Patrignani et~al.
\newblock {Review of Particle Physics}.
\newblock {\em Chin. Phys.}, C40(10):100001, 2016.

\bibitem{Capozzi2016218}
F.~Capozzi, E.~Lisi, A.~Marrone, D.~Montanino, and A.~Palazzo.
\newblock Neutrino masses and mixings: Status of known and unknown 3$\nu$
  parameters.
\newblock {\em Nuclear Physics B}, 908:218 -- 234, 2016.
\newblock Neutrino Oscillations: Celebrating the Nobel Prize in Physics 2015.

\bibitem{giuntibook}
C.~Giunti and C.W. Kim.
\newblock {\em {Fundamentals of Neutrino Physics and Astrophysics}}.
\newblock {Oxford University Press}, 2007.

\bibitem{kopp}
J.~Kopp.
\newblock {\em Phenomenology of Three-Flavour Neutrino Oscillations}.
\newblock PhD thesis, {Technische Universit\"at M\"unchen}, 2006.

\bibitem{Abdurashitov:2009tn}
J.N. Abdurashitov et~al.
\newblock {Measurement of the solar neutrino capture rate with gallium metal.
  III: Results for the 2002--2007 data-taking period}.
\newblock {\em Phys. Rev.}, C80:015807, 2009.

\bibitem{Altmann:2005ix}
M.~Altmann et~al.
\newblock {Complete results for five years of GNO solar neutrino observations}.
\newblock {\em Phys. Lett.}, B616:174--190, 2005.

\bibitem{Eguchi:2002dm}
K.~Eguchi et~al.
\newblock {First results from KamLAND: Evidence for reactor anti-neutrino
  disappearance}.
\newblock {\em Phys. Rev. Lett.}, 90:021802, 2003.

\bibitem{Ahn:2006zza}
M.H. Ahn et~al.
\newblock {Measurement of Neutrino Oscillation by the K2K Experiment}.
\newblock {\em Phys. Rev.}, D74:072003, 2006.

\bibitem{Adamson:2013whj}
P.~Adamson et~al.
\newblock {Measurement of Neutrino and Antineutrino Oscillations Using Beam and
  Atmospheric Data in MINOS}.
\newblock {\em Phys. Rev. Lett.}, 110(25):251801, 2013.

\bibitem{Abe:2012gx}
K.~Abe et~al.
\newblock {First Muon-Neutrino Disappearance Study with an Off-Axis Beam}.
\newblock {\em Phys. Rev.}, D85:031103, 2012.

\bibitem{Abe:2012jj}
K.~Abe et~al.
\newblock {Evidence for the Appearance of Atmospheric Tau Neutrinos in
  Super-Kamiokande}.
\newblock {\em Phys. Rev. Lett.}, 110(18):181802, 2013.

\bibitem{Agafonova:2015jxn}
N.~Agafonova et~al.
\newblock {Discovery of $\tau$ Neutrino Appearance in the CNGS Neutrino Beam
  with the OPERA Experiment}.
\newblock {\em Phys. Rev. Lett.}, 115(12):121802, 2015.

\bibitem{Abreu:2017bpe}
Y.~Abreu, Y.~Amhis, L.~Arnold, et~al.
\newblock {A novel segmented-scintillator antineutrino detector}.
\newblock {\em JINST}, 12(04):P04024, 2017.

\bibitem{Arnold:2017lph}
L.~Arnold, W.~Beaumont, D.~Cussans, D.~Newbold, N.~Ryder, and A.~Weber.
\newblock {The SoLid anti-neutrino detector's readout system}.
\newblock {\em JINST}, 12(02):C02012, 2017.

\bibitem{Boireau:2015dda}
G.~Boireau et~al.
\newblock {Online Monitoring of the Osiris Reactor with the Nucifer Neutrino
  Detector}.
\newblock {\em Phys. Rev.}, D93(11):112006, 2016.

\bibitem{Derbin:2012kf}
A.V. Derbin, A.S. Kayunov, and V.N. Muratova.
\newblock {Search for Neutrino Oscillations at a Research Reactor}.
\newblock 2012.

\bibitem{Lhuillier}
D.~Lhuillier.
\newblock {Future Short-Baseline Sterile Neutrino Searches with Reactors}.
\newblock In {\em {XXVI International Conference on Neutrino Physics and
  Astrophysics (Neutrino 2014), Boston, USA, 2-7 Jul}}, 2014.

\bibitem{Haser:2016xlb}
J.~Haser.
\newblock {Search for eV sterile neutrinos at a nuclear reactor — the Stereo
  project}.
\newblock {\em J. Phys. Conf. Ser.}, 718(6):062023, 2016.

\bibitem{Serebrov:2013}
A.~Serebrov et~al.
\newblock On possibility of realization {NEUTRINO-4} experiment on search for
  oscillations of the reactor antineutrino into a sterile state.
\newblock 2013.

\bibitem{Serebrov:2016}
A.~Serebrov et~al.
\newblock {Experiment for search for sterile neutrino at SM-3 reactor}.
\newblock {\em Phys. Part. Nucl.}, 47(6):1014--1023, 2016.

\bibitem{Yeo:2014spa}
I.S. Yeo et~al.
\newblock {Development of a gadolinium-loaded liquid scintillator for the
  Hanaro short baseline prototype detector}.
\newblock {\em J. Korean Phys. Soc.}, 64(3):377--381, 2014.

\bibitem{Alekseev:2016llm}
I.~Alekseev et~al.
\newblock {DANSS: Detector of the reactor AntiNeutrino based on Solid
  Scintillator}.
\newblock {\em JINST}, 11(11):P11011, 2016.

\bibitem{Ashenfelter:2013oaa}
J.~Ashenfelter et~al.
\newblock {PROSPECT - A Precision Reactor Oscillation and Spectrum Experiment
  at Short Baselines}.
\newblock In {\em {Proceedings, Community Summer Study 2013: Snowmass on the
  Mississippi (CSS2013), Minneapolis, MN, USA, 29 Jul-6 Aug}}, 2013.

\bibitem{Lane:2015alq}
C.~Lane et~al.
\newblock {A new type of Neutrino Detector for Sterile Neutrino Search at
  Nuclear Reactors and Nuclear Nonproliferation Applications}.
\newblock 2015.

\bibitem{Abe:2011fz}
Y.~Abe et~al.
\newblock {Indication for the disappearance of reactor electron antineutrinos
  in the Double Chooz experiment}.
\newblock {\em Phys. Rev. Lett.}, 108:131801, 2012.

\bibitem{Wang:2014nta}
Z.~Wang.
\newblock {Highlights from the Daya Bay Neutrino Experiment}.
\newblock {\em EPJ Web Conf.}, 71:00138, 2014.

\bibitem{Ahn:2012nd}
J.K. Ahn et~al.
\newblock {Observation of Reactor Electron Antineutrino Disappearance in the
  RENO Experiment}.
\newblock {\em Phys. Rev. Lett.}, 108:191802, 2012.

\bibitem{Lasserre:2005qw}
T.~Lasserre and H.W. Sobel.
\newblock {Reactor neutrinos}.
\newblock {\em Comptes Rendus Physique}, 6:749--757, 2005.

\bibitem{Gratta:1999ny}
G.~Gratta.
\newblock {Neutrino oscillation experiments at nuclear reactors}.
\newblock {\em Nucl. Phys. Proc. Suppl.}, 85:72--77, 2000.

\bibitem{Shirai:2005zz}
J.~Shirai.
\newblock {Reactor neutrinos and KamLAND}.
\newblock In {\em {Particle physics at the year of 250th anniversary of Moscow
  University. Proceedings, 12th Lomonosov Conference on elementary particle
  physics, Moscow, Russia, 25-31 Aug}}, pages 29--36, 2005.

\bibitem{DiLella:1999ar}
L.~Di~Lella.
\newblock {Accelerator and reactor neutrino experiments}.
\newblock {\em Int. J. Mod. Phys.}, A15S1:257--282, 2000.
\newblock [,257(1999)].

\bibitem{Nieto:2003wd}
M.~Nieto, A.~Hayes, C.~Teeter, W.~Wilson, and W.~Stanbro.
\newblock {Detection of anti-neutrinos for nonproliferation}.
\newblock 2003.

\bibitem{Hayes:2016qnu}
A.C. Hayes and P.~Vogel.
\newblock {Reactor Neutrino Spectra}.
\newblock {\em Ann. Rev. Nucl. Part. Sci.}, 66:219--244, 2016.

\bibitem{Cao:2011gb}
J.~Cao.
\newblock {Determining Reactor Neutrino Flux}.
\newblock {\em Nucl. Phys. Proc. Suppl.}, 229-232:205--209, 2012.

\bibitem{Mueller:2011nm}
T.A. Mueller, D.~Lhuillier, M.~Fallot, A.~Letourneau, S.~Cormon, M.~Fechner,
  L.~Giot, T.~Lasserre, J.~Martino, G.~Mention, A.~Porta, and F.~Yermia.
\newblock {Improved Predictions of Reactor Antineutrino Spectra}.
\newblock {\em Phys. Rev.}, C83:054615, 2011.

\bibitem{Mention:2011}
G.~Mention, M.~Fechner, T.~Lasserre, T.A. Mueller, D.~Lhuillier, M.~Cribier,
  and A.~Letourneau.
\newblock {The Reactor Antineutrino Anomaly}.
\newblock {\em Phys. Rev.}, D83:073006, 2011.

\bibitem{Gariazzo:2016lsd}
S.~Gariazzo.
\newblock {Light sterile neutrinos and pseudoscalar interactions in cosmology}.
\newblock In {\em {Neutrino Oscillation Workshop (NOW) 2016, Otranto, Italy,
  5-10 Sep}}, 2016.

\bibitem{Abe:2014bwa}
Y.~Abe et~al.
\newblock {Improved measurements of the neutrino mixing angle $\theta_{13}$
  with the Double Chooz detector}.
\newblock {\em JHEP}, 10:086, 2014.
\newblock [Erratum: JHEP02,074(2015)].

\bibitem{Kim:2014rfa}
S.B. Kim.
\newblock {New results from RENO and prospects with RENO-50}.
\newblock {\em Nucl. Part. Phys. Proc.}, 265-266:93--98, 2015.

\bibitem{PhysRevLett.116.061801}
F.P. An.
\newblock {Measurement of the Reactor Antineutrino Flux and Spectrum at Daya
  Bay}.
\newblock {\em Phys. Rev. Lett.}, 116:061801, Feb 2016.

\bibitem{Kaether:2010ag}
F.~Kaether, W.~Hampel, G.~Heusser, J.~Kiko, and T.~Kirsten.
\newblock {Reanalysis of the GALLEX solar neutrino flux and source
  experiments}.
\newblock {\em Phys. Lett.}, B685:47--54, 2010.

\bibitem{Giunti:2010zu}
C.~Giunti and M.~Laveder.
\newblock {Statistical Significance of the Gallium Anomaly}.
\newblock {\em Phys. Rev.}, C83:065504, 2011.

\bibitem{Kopp:2013vaa}
J.~Kopp, P.A.N. Machado, M.~Maltoni, and Th. Schwetz.
\newblock {Sterile Neutrino Oscillations: The Global Picture}.
\newblock {\em JHEP}, 05:050, 2013.

\bibitem{PhysRevLett.102.101802}
A.A. Aguilar et~al.
\newblock {Unexplained Excess of Electronlike Events from a 1-GeV Neutrino
  Beam}.
\newblock {\em Phys. Rev. Lett.}, 102:101802, Mar 2009.

\bibitem{PhysRevLett.110.161801}
A.A. Aguilar et~al.
\newblock {Improved Search for
  ${\overline{\ensuremath{\nu}}}_{\ensuremath{\mu}}\ensuremath{\rightarrow}{\overline{\ensuremath{\nu}}}_{e}$
  Oscillations in the MiniBooNE Experiment}.
\newblock {\em Phys. Rev. Lett.}, 110:161801, Apr 2013.

\bibitem{PhysRevD.65.112001}
B.~Armbruster et~al.
\newblock Upper limits for neutrino oscillations
  ${\overline{\ensuremath{\nu}}}_{\ensuremath{\mu}}\ensuremath{\rightarrow}{\overline{\ensuremath{\nu}}}_{e}$
  from muon decay at rest.
\newblock {\em Phys. Rev. D}, 65:112001, Jun 2002.

\bibitem{Dydak:1983zq}
F.~Dydak et~al.
\newblock {A Search for Muon-neutrino Oscillations in the $\Delta m^2$ Range
  $0.3 \mathrm{eV}^2$ to $90 \mathrm{eV}^2$}.
\newblock {\em Phys. Lett.}, B134:281, 1984.

\bibitem{Adamson:2011ku}
P.~Adamson et~al.
\newblock {Active to sterile neutrino mixing limits from neutral-current
  interactions in MINOS}.
\newblock {\em Phys. Rev. Lett.}, 107:011802, 2011.

\bibitem{Agafonova2013}
N.~Agafonova et~al.
\newblock {Search for $\nu_\mu\rightarrow \nu_e$ oscillations with the OPERA
  experiment in the CNGS beam}.
\newblock {\em Journal of High Energy Physics}, 2013(7):4, 2013.

\bibitem{Antonello:2012pq}
M.~Antonello et~al.
\newblock {Experimental search for the “LSND anomaly” with the ICARUS
  detector in the CNGS neutrino beam}.
\newblock {\em Eur. Phys. J.}, C73(3):2345, 2013.

\bibitem{PhysRevD.86.010001}
J.~Beringer et~al.
\newblock Review of particle physics.
\newblock {\em Phys. Rev. D}, 86:010001, Jul 2012.

\bibitem{Ke:2015xka}
H.W. Ke, J.H. Zhou, Sh. Chen, T.~Liu, and X.Q. Li.
\newblock {The hidden symmetries in the PMNS matrix and the light sterile
  neutrino(s)}.
\newblock {\em Mod. Phys. Lett.}, A30(27):1550136, 2015.

\bibitem{Kang:2013gpa}
S.~K. Kang, Y.~D. Kim, Y.~Ko, and K.~Siyeon.
\newblock {Sterile neutrino analysis of reactor-neutrino oscillation}.
\newblock 2013.

\bibitem{Giunti:2011gz}
C.~Giunti and M.~Laveder.
\newblock {3+1 and 3+2 Sterile Neutrino Fits}.
\newblock {\em Phys. Rev.}, D84:073008, 2011.

\bibitem{Collin:2016rao}
G.~H. Collin, C.~A. Argüelles, J.~M. Conrad, and M.~H. Shaevitz.
\newblock {Sterile Neutrino Fits to Short Baseline Data}.
\newblock {\em Nucl. Phys.}, B908:354--365, 2016.

\bibitem{wouter}
W.~Van~de Pontseele.
\newblock Characterisation and modelling of correlated noise in silicon
  photomultipliers for the {SoLid} experiment.
\newblock Master's thesis, Universiteit Gent, 2016.

\bibitem{Gando:2013zla}
A.~Gando et~al.
\newblock {White paper: CeLAND - Investigation of the reactor antineutrino
  anomaly with an intense
  $\prescript{144}{}{\mathrm{Ce}}-\prescript{144}{}{\mathrm{Pr}}$ antineutrino
  source in KamLAND}.
\newblock 2013.

\bibitem{An:2012eh}
F.P. An et~al.
\newblock {Observation of electron-antineutrino disappearance at Daya Bay}.
\newblock {\em Phys. Rev. Lett.}, 108:171803, 2012.

\bibitem{Seo:2016uom}
J.H. Choi et~al.
\newblock Observation of energy and baseline dependent reactor antineutrino
  disappearance in the {RENO} experiment.
\newblock {\em Phys. Rev. Lett.}, 116:211801, May 2016.

\bibitem{Christensen:2014pva}
E.~Christensen, P.~Huber, P.~Jaffke, and T.E. Shea.
\newblock {Antineutrino Monitoring for Heavy Water Reactors}.
\newblock {\em Phys. Rev. Lett.}, 113(4):042503, 2014.

\bibitem{Cribier:2007zh}
M.~Cribier.
\newblock {Neutrinos and non-proliferation in Europe}.
\newblock {\em Earth Moon Planets}, 99:331--341, 2006.
\newblock [,331(2007)].

\bibitem{joppen}
F.~Joppen, E.~Koonen, and S.~Van~Dijk.
\newblock Periodic safety review of the {BR2} reactor.
\newblock In {\em International Conference on Research Reactors: Safe
  Management and Effective Utilization, Rabat, Morocco, 14-18 Nov}, 2011.

\bibitem{vsi}
Y.~Abreu.
\newblock {SoLid: An innovative antineutrino detector for searching
  oscillations at the SCK\textbullet CEN BR2 reactor (talk)}.
\newblock In {\em The 14th Vienna Conference on Instrumentation, Austria, 16
  Feb}, 2016.

\bibitem{Michiels:2016qui}
I.~Michiels.
\newblock {SoLid: Search for Oscillation with a $\prescript{6}{}{\mathrm{Li}}$
  Detector at the BR2 research reactor}.
\newblock In {\em {NuPhys2015: Prospects in Neutrino Physics (NuPhys) London,
  UK, 16-18 Dec}}, 2015.

\bibitem{kalcheva}
S.~Kalcheva, M.~Fallot, and L.~Giot.
\newblock Recent efforts for the {BR2} reactor simulation and anti-neutrino
  spectrum.
\newblock In {\em SoLid meeting, London, UK, 25 May}, 2016.

\bibitem{Boehm:2000va}
F.~Boehm.
\newblock Studies of neutrino oscillations at reactors.
\newblock In {\em Current Aspects of Neutrino Physics}, 2000.

\bibitem{Bemporad:2001qy}
C.~Bemporad, G.~Gratta, and P.~Vogel.
\newblock {Reactor based neutrino oscillation experiments}.
\newblock {\em Rev. Mod. Phys.}, 74:297, 2002.

\bibitem{Vidyakin:1987ue}
G.S. Vidyakin, V.N. Vyrodov, I.I. Gurevich, Y.V. Kozlov, V.P. Martemyanov, S.V.
  Sukhotin, V.G. Tarasenkov, and S.Kh. Khakimov.
\newblock {Detection of Anti-neutrinos in the Flux From Two Reactors}.
\newblock {\em Sov. Phys. JETP}, 66:243--247, 1987.
\newblock [Zh. Eksp. Teor. Fiz.93,424(1987)].

\bibitem{cubesy}
F.~Yermia, G.~Barber, J.~Nash, A.~Rose, A.~Baird, N.~Ryder, P.R. Scovell,
  A.~Vacheret, A.~Weber, S.~Bouvier, J.M. Buhour, A.S. Cucuanes, M.~Fallot,
  L.~Giot, G~Guilloux, B.~Guillon, B.~Coupé, S.~Kalcheva, and E.~Koonen.
\newblock {Search for Oscillation with Lithium-6 Detector at SCK\textbullet CEN
  BR2 research reactor}.
\newblock In {\em Seminar Conseil Scientifique IN2P3, Paris, France}, 2013.

\bibitem{Moortgat:2015bwg}
C.~Moortgat.
\newblock {Technology of the SoLid detector and construction of the first
  submodule}.
\newblock {\em PoS}, EPS-HEP2015:080, 2015.

\bibitem{michielsmaster}
I.~Michiels.
\newblock {Investigation of the Muon Background for the NEMENIX prototype of
  the SoLid experiment}.
\newblock Master's thesis, Universiteit Gent, 2015.

\bibitem{Ryder:2015sma}
N.~Ryder.
\newblock {First results of the deployment of a SoLid detector module at the
  SCK-CEN BR2 reactor}.
\newblock {\em PoS}, EPS-HEP2015:071, 2015.

\bibitem{Sweany:2014ena}
M.~Sweany, J.~Brennan, B.~Cabrera-Palmer, S.~Kiff, D.~Reyna, and
  D.~Throckmorton.
\newblock {Above-ground antineutrino detection for nuclear reactor monitoring}.
\newblock {\em Nucl. Instrum. Meth.}, A769:37--43, 2015.

\bibitem{Schluter:2011rv}
T.~Schluter, W.~Dunnweber, K.~Dhibar, M.~Faessler, R.~Geyer, J.F. Rajotte,
  Z.~Roushan, and H.~Wohrmann.
\newblock {Large-Area Sandwich Veto Detector with WLS Fibre Readout for Hadron
  Spectroscopy at COMPASS}.
\newblock {\em Nucl. Instrum. Meth.}, A654:219--224, 2011.

\bibitem{verstraten}
M.~Verstraeten.
\newblock Optical transport in the {SoLid} detector.
\newblock Master's thesis, Universiteit Antwerpen, 2016.

\bibitem{Oliveira20102098}
P.H.F. Oliveira, S.T. Amancio-Filho, J.F. dos Santos, and E.~Hage~Jr.
\newblock Preliminary study on the feasibility of friction spot welding in
  {PMMA}.
\newblock {\em Materials Letters}, 64(19):2098 -- 2101, 2010.

\bibitem{saintgobain}
\textsc{Saint-Gobain}.
\newblock {\em {Brochure Scintillating Optical Fibers}}, 2014.

\bibitem{hamamatsudata}
\textsc{Hamamatsu}.
\newblock {\em {Datasheet MPPC\textregistered (multi-pixel photon counter)
  S12572-025, -050, -100C/P}}, 2015.

\bibitem{Buzhan:2001xq}
P.~Buzhan, B.~Dolgoshein, A.~Ilyin, V.~Kantserov, V.~Kaplin, A.~Karakash,
  A.~Pleshko, E.~Popova, S.~Smirnov, and Y.~Volkov.
\newblock {An advanced study of silicon photomultiplier}.
\newblock {\em ICFA Instrum. Bull.}, 23:28--41, 2001.

\bibitem{Chmill:2016ghf}
V.~Chmill, E.~Garutti, R.~Klanner, M.~Nitschke, and J.~Schwandt.
\newblock {On the characterisation of SiPMs from pulse-height spectra}.
\newblock {\em Nucl. Instrum. Meth.}, A854:70--81, 2017.

\bibitem{Otte:2006gm}
N.~Otte.
\newblock {The Silicon Photomultiplier: A New Device for High Energy Physics,
  Astroparticle Physics, Industrial and Medical Applications}.
\newblock {\em eConf}, C0604032:0018, 2006.

\bibitem{Galbiati:2005ft}
C.~Galbiati and J.F. Beacom.
\newblock {Measuring the cosmic ray muon-induced fast neutron spectrum by (n,p)
  isotope production reactions in underground detectors}.
\newblock {\em Phys. Rev.}, C72:025807, 2005.
\newblock [Erratum: Phys. Rev.C73,049906(2006)].

\bibitem{verreyken}
B.~Verreyken.
\newblock {Evaluating the SM1 calibration by use of radioactive sources and
  estimating its energy resolution for electromagnetic signals.}
\newblock Master's thesis, Universiteit Antwerp, 2016.

\bibitem{neutrinovacheret}
A.~Vacheret.
\newblock Compact segmented antineutrino detector (talk).
\newblock In {\em {XXVII International Conference on Neutrino Physics and
  Astrophysics (Neutrino 2016), London, UK, 4-9 Jul}}, 2016.

\bibitem{appryder}
N.~Ryder.
\newblock {The SoLid Experiment}.
\newblock In {\em Applied Anti-neutrino Physics, Liverpool, UK, 1 Dec}, 2016.

\bibitem{Saunders:2016gcc}
D.~Saunders.
\newblock {Muon Calibration at SoLid}.
\newblock {\em PoS}, EPS-HEP2015:086, 2015.

\bibitem{Abreu2016}
Y.~Abreu.
\newblock {SoLid: An innovative anti-neutrino detector for searching
  oscillations at the SCK\textbullet CEN BR2 reactor}.
\newblock {\em Nucl. Instrum. Meth.}, A845:467--470, 2017.

\bibitem{whitepaper}
L.~Arnold, L.~Kalousis, M.~Labare, N.~Ryder, D.~Saunders, and A.~Vacheret.
\newblock {SoLid Software and Computing Infrastructure}.
\newblock 2016.

\bibitem{Larrea:2015wra}
C.~Ghabrous~Larrea, K.~Harder, D.~Newbold, D.~Sankey, A.~Rose, A.~Thea, and
  T.~Williams.
\newblock {IPbus: a flexible Ethernet-based control system for xTCA hardware}.
\newblock {\em JINST}, 10(02):C02019, 2015.

\bibitem{Dasgupta:2015ohj}
S.~Dasgupta and D.~Cussans.
\newblock {Field Programmable Gate Arrays—Detecting Cosmic Rays}.
\newblock {\em JINST}, 10(07):C07006, 2015.

\bibitem{Arnold:2017tms}
L.~Arnold.
\newblock {Trigger for the SoLid Reactor Antineutrino Experiment}.
\newblock In {\em {Prospects in Neutrino Physics (NuPhys2016), London, UK,
  12-14 Dec}}, 2016.

\bibitem{newb}
D.~Newbold.
\newblock {SoLid Bitbucket Wiki Entry: readout system / firmware\_overview,
  \url{https://bitbucket.org/solidexperiment/readout-system/wiki/firmware_overview},
  retrieved 20 Dec 2016, non-public}, 2016.

\bibitem{trenz}
\textsc{Trenz Electronics}.
\newblock {Product Information "Micromodule Artix-7 XC7A200T-2C 4x5cm standard
  footprint (com. temp. range)",
  \url{https://shop.trenz-electronic.de/en/TE0712-02-200-2C-Micromodule-Artix-7-XC7A200T-2C-4x5cm-standard-footprint-com.-temp.-range?c=341},
  retrieved 8 Jan 2017}, 2016.

\bibitem{xilinx}
\textsc{Xilinx}.
\newblock {\em 7 Series FPGAs Data Sheet: Overview v2.2}, 2016.

\bibitem{trenz2}
\textsc{Trenz Electronics}.
\newblock {\em {Brochure All Programmable FPGA and SoC modules}}, 2016.

\bibitem{Arnold:2015ylm}
L.~Arnold.
\newblock {Irradiation Resistivity and Mitigation Measurement Design for Xilinx
  Kintex-7 FPGAs}.
\newblock Master's thesis, Fachhochschule Nordwestschweiz / University of
  Cambridge, 2015.

\bibitem{9780881335545}
J.M. Wozencraft and I.M. Jacobs.
\newblock {\em Principles of Communication Engineering}.
\newblock Waveland Pr Inc, 1965.

\bibitem{Gallager}
R.G. Gallager.
\newblock {\em {Principles of Digital Communication}}.
\newblock {Cambridge University Press}, 2008.

\bibitem{wannamaker}
R.A. Wannamaker.
\newblock {\em The Theory of Dithered Quantization}.
\newblock PhD thesis, {University of Waterloo}, 1997.

\bibitem{cordoba1989}
A.~C\'ordoba.
\newblock Dirac combs.
\newblock {\em Letters in Mathematical Physics}, 17(3):191--196, 1989.

\bibitem{lipshitz1992quantization}
S.P. Lipshitz, R.A. Wannamaker, and J.~Vanderkooy.
\newblock Quantization and dither: A theoretical survey.
\newblock {\em J. Audio Eng. Soc}, 40(5):355--375, 1992.

\bibitem{shannon1949}
C.E. Shannon.
\newblock Communication in the presence of noise.
\newblock {\em Proc. Institute of Radio Engineers}, 37(1):10--21, 1949.

\bibitem{hufschmid}
M.~Hufschmid.
\newblock {\em Information und Kommunikation. Grundlagen und Verfahren der
  Informations\"ubertragung}.
\newblock Teubner, 2006.

\bibitem{li}
B.P. Li.
\newblock {\em Modern Digital and Analog Communication Systems}.
\newblock Oxford University Press, 1998.

\bibitem{jameswitten}
G.~James, D.~Witten, T.~Hastie, and R.~Tibshirani.
\newblock {\em An Introduction to Statistical Learning}.
\newblock Springer, 2013.

\bibitem{Bishop}
C.M. Bishop.
\newblock {\em Pattern Recognition and Machine Learning 1st Edition}.
\newblock Springer, 2013.

\bibitem{mladenic}
D.~Mladeni\'{c}.
\newblock {\em Feature Selection for Dimensionality Reduction}.
\newblock Springer, 2006.

\bibitem{liumotoda}
H.~Liu and H.~Motoda.
\newblock {\em Feature Extraction, Construction and Selection: A Data Mining
  Perspective (The Springer International Series in Engineering and Computer
  Science)}.
\newblock Springer, 1998.

\bibitem{Sanchez-Marono:2007:FMF:1777942.1777962}
N.~S\'{a}nchez-Maro\~{n}o, A.~Alonso-Betanzos, and M.~Tombilla-Sanrom\'{a}n.
\newblock Filter methods for feature selection: A comparative study.
\newblock In {\em Proceedings of the 8th International Conference on
  Intelligent Data Engineering and Automated Learning, Birmingham, UK}, pages
  178--187, 2007.

\bibitem{fawcett2006introduction}
T.~Fawcett.
\newblock {An introduction to ROC analysis}.
\newblock {\em Pattern recognition letters}, 27(8):861--874, 2006.

\bibitem{citeulike:12882259}
D.M.W. Powers.
\newblock {Evaluation: From precision, recall and f-measure to ROC,
  informedness, markedness \& correlation}.
\newblock {\em Journal of Machine Learning Technologies}, 2(1):37--63, 2011.

\bibitem{uzun2005fpga}
I.S. Uzun, A.~Amira, and A.~Bouridane.
\newblock {FPGA implementations of fast Fourier transforms for real-time signal
  and image processing}.
\newblock {\em IEE Proceedings-Vision, Image and Signal Processing},
  152(3):283--296, 2005.

\bibitem{fft}
\textsc{Xilinx}.
\newblock {\em {LogiCORE IP Fast Fourier Transform v7.1}}, 2011.

\bibitem{rojas}
R.~Rojas.
\newblock {\em Neural Networks: A Systematic Introduction}.
\newblock Springer, 1998.

\bibitem{Rosenblatt58theperceptron}
F.~Rosenblatt.
\newblock The perceptron: A probabilistic model for information storage and
  organization in the brain.
\newblock {\em Psychological Review}, pages 65--386, 1958.

\bibitem{Rosenblatt62}
F.~Rosenblatt.
\newblock {\em Principles of Neurodynamics: Perceptrons and the Theory of Brain
  Mechanisms}.
\newblock Spartan Books, Washington, 1962.

\bibitem{9780521780193}
N.~Cristianini and J.~Shawe-Taylor.
\newblock {\em An Introduction to Support Vector Machines and Other
  Kernel-based Learning Methods}.
\newblock Cambridge University Press, 2000.

\bibitem{cohen}
W.~Cohen.
\newblock Linear classifiers and the perceptron.
\newblock 2008.

\bibitem{bebis1994feed}
G.~Bebis and M.~Georgiopoulos.
\newblock Feed-forward neural networks.
\newblock {\em IEEE Potentials}, 13(4):27--31, 1994.

\bibitem{Stoykov:2014uca}
A.~Stoykov, J.~B. Mosset, U.~Greuter, M.~Hildebrandt, and N.~Schlumpf.
\newblock {Use of Silicon Photomultipliers in ZnS:$^6$LiF scintillation neutron
  detectors: signal extraction in presence of high dark count rates}.
\newblock {\em JINST}, 9:P06015, 2014.

\bibitem{Eiben11parametertuning}
A.E. Eiben and S.K. Smit.
\newblock Parameter tuning for configuring and analyzing evolutionary
  algorithms.
\newblock {\em Swarm and Evolutionary Computation}, 2011.

\bibitem{pearson1895note}
K.~Pearson.
\newblock {Note on regression and inheritance in the case of two parents}.
\newblock {\em Proceedings of the Royal Society of London},
  58(347-352):240--242, 1895.

\bibitem{skahill1996vhdl}
K.~Skahill.
\newblock {\em VHDL for programmable logic}.
\newblock Addison-Wesley, 1996.

\bibitem{pichler}
S.~Brantschen, M.~Pichler, and K.~Schenk.
\newblock {\em {Digitale ASIC Schaltungstechnik. Ein Lehrgang zur Entwicklung
  von integrierten digitalen Schaltungen und FPGA}}.
\newblock Institut f\"{ur} Mikroelektronik, FHNW, 2008.

\bibitem{9780534384623}
C.H. Roth~Jr. and K.J. Lizy.
\newblock {\em Digital Systems Design Using VHDL}.
\newblock Cengage Learning, 2007.

\bibitem{nagaraj2012fpga}
Y.~Nagaraj, K.~Shrinivas, K.~Veeresh, A.~Veeresh, D.~Madhu~Patil, et~al.
\newblock {FPGA implementation of different adder architectures}.
\newblock {\em International Journal of Emerging Technology and Advanced
  Engineering}, 2(8), 2012.

\bibitem{dcdc}
\textsc{Altera}.
\newblock {\em {Datasheet Enpirion\textregistered EN63A0QI 12A PowerSoC }},
  2016.

\bibitem{andrews2002introduction}
P.B. Andrews.
\newblock {\em An introduction to mathematical logic and type theory}.
\newblock Springer, 2002.

\end{thebibliography}

\clearpage
\begin{appendices}
\newpage
\section{}
\subsection{$\widetilde{U}$}

\begin{equation}
\label{eq:fullu}
\rotatebox{90}{$
=
\left( \begin{matrix}
    c_{14}U_{e1} & c_{14}U_{e2} & c_{14}U_{e3} & s_{14} \\
    -s_{14}s_{24}U_{e1}+c_{24}U_{\mu1}& -s_{14}s_{24}U_{e2}+c_{24}U_{\mu2} &-s_{14}s_{24}U_{e3}+c_{24}U_{\mu3}  & c_{14}s_{24} \\
 -c_{24}s_{14}s_{34}U_{e1}  -s_{24}s_{34}U_{\mu1} + c_{34}U_{\tau1}  &
  -c_{24}s_{14}s_{34}U_{e2}  -s_{24}s_{34}U_{\mu2} + c_{34}U_{\tau2}&
  -c_{24}s_{14}s_{34}U_{e3}  -s_{24}s_{34}U_{\mu3} + c_{34}U_{\tau3}& c_{14}c_{24}s_{34} \\
 -c_{24}s_{34}s_{14}U_{e1}  -s_{24}c_{34}U_{\mu1} - s_{34}U_{\tau1}  &
  -c_{24}s_{34}s_{14}U_{e2}  -s_{24}c_{34}U_{\mu2} - s_{34}U_{\tau2}&
  -c_{24}s_{34}s_{14}U_{e3}  -s_{24}c_{34}U_{\mu3} - s_{34}U_{\tau3}& c_{14}c_{24}s_{34} 
        \end{matrix} \right)
$}
\end{equation}
\clearpage
\subsection{Conventions in Section~\ref{sec:trigger}}
\label{app:convention}

$\bm{B}$ -- Boolean space	\\
$\bm{C}$ -- Complex space	\\
$\bm{L}_2\bm{H}$ -- Lebesgue Hilbert or Pre-Hilbert space with $\bm{L}_2$ metrics\\
$\bm{R}$ -- Real space		\\
Superscripts indicate number of elements for vectors, matrices and tensors.\\
\\
$\mathbf{E}$ -- class answer vector/matrix \\
$\mathbf{F}$ -- feature vector/matrix \\
$\mathbf{w}$ -- weight vector/matrix \\
$x,X,\mathbf{X}$ -- discrete signal as acquired by the FPGA. Superscripts indicate the space in which the signal is represented\\
$\hat{\mathbf{Y}}$ -- predicted class\\
\\
$\operatorname {cov}(.,.)$ -- covariance	\\
$f(.)$ -- trigger function \\
$g(.)$ -- feature extraction			\\
$h(.)$ -- machine learning algorithm			\\
$\operatorname{III}(.)$ -- sampling function \\
$\delta$ -- Dirac function/comb\\
$\left|. \right|$ -- cardinality of a set/number of elements within a set\\
$\lfloor . \rfloor$ -- floor function\\
\\
$_.$ -- for matrices and tensors, ($_.$) in conjunction with one or more other indices denotes the submatrix or subvector containing all the elements at the indicated location\\

\subsection{Abbreviations}
\subsubsection{Particles}

$d$ -- down-quark	\\	
$e^-$ -- electron	\\	
$e^+$ -- positron	\\	
$\ell$ -- lepton\\	
$n$ -- neutron		\\
$p$ -- proton		\\
$u$ -- up-quark			\\
$W$ -- W boson			\\
$\alpha$ -- alpha-particle\\
$\gamma$ -- gamma/photon	\\
$\nu$ -- neutrino	\\
$\mu$ -- muon\\
$\tau$ -- tau\\

\subsubsection{Atoms/Molecules/Materials}

\LiZnS{} -- Lithium-6 Silver-doped Zinc Sulphide	\\
$\mathrm{Am}$ -- Americium\\
$\mathrm{Ar}$ -- Argon \\
$\mathrm{Be}$ -- Beryllium\\
$\mathrm{Bi}$ -- Bismuth\\
$\mathrm{Cs}$ -- Caesium\\
$\mathrm{Cf}$ -- Californium\\
$\mathrm{Cl}$ -- Chlorine\\
$\mathrm{Co}$ -- Cobalt\\
$\mathrm{Cr}$ -- Chromium\\
$\mathrm{Ga}$ -- Gallium\\
$\mathrm{Gd}$ -- Gadolinium\\
$\mathrm{Ge}$ -- Germanium\\
$\mathrm{H}$ -- Hydrogen\\
$\mathrm{He}$ -- Helium\\
$\mathrm{Li}$ -- Lithium \\
$\mathrm{Po}$ -- Polonium\\
$\mathrm{Pu}$ -- Plutonium			\\
PMMA -- polymethyl methacrylate\\
PVT -- polyvinyl toluene\\
$\mathrm{U}$ -- Uranium				\\

\subsubsection{Physical Quantities}
$E$, $e$ -- energy					\\
$L$ -- baseline\\
$m$ -- mass\\
$p$ -- momentum					\\
$P$ -- probability, power \\
$R$ -- data/bit rate\\
$t$ -- time\\
$v$ -- speed\\
$W$ -- work, energy\\

\subsubsection{Units}
$\si{\byte}$ -- Bel, Byte\\
$\mathrm{bps}$ -- bit per second\\
$\si{\clight}$ -- speed of light in vacuum\\
$\si{\electronvolt}$ -- Electronvolt		\\
$\si{\gram}$ -- gram \\
$\si{\hertz}$ -- Hertz\\
$\hbar$ -- reduced Planck constant\\
$\si{\meter}$ -- meter				\\
$\mathrm{PA}$ -- number of pixel photon avalanches	\\
$\si{\second}$ -- second			\\
$\si{\watt}$ -- Watt				\\
$\si{\degreeCelsius}$ -- degree Celsius\\
$\%$ -- percent ($\frac{1}{100}$)\\

\subsubsection{Unit Pre-fixes}
$\si{\nano}$ -- nano ($10^{-9}$)		\\
$\si{\micro}$ -- micro ($10^{-6}$)		\\
$\si{\milli}$ -- milli ($10^{-3}$)		\\
$\si{\centi}$ -- centi ($10^{-2}$)		\\
$\si{\deci}$ -- deci ($10^{-1}$)		\\
$\si{\kilo}$ -- kilo ($10^{3}$)			\\
$\si{\mega}$ -- Mega ($10^{6}$) 		\\
$\si{\giga}$ -- Giga ($10^{9}$) 		\\
$\si{\tera}$ -- Tera ($10^{12}$) 		\\
$\si{\peta}$ -- Peta ($10^{15}$) 		\\

\subsubsection{Electronics}
ADC -- Analogue-Digital Converter\\
DAQ -- Data Acquisition \\
DC -- Direct Current \\
DCR -- Dark-Count Rate\\
DDR -- Double Data Rate \\
DSP -- Digital Signal Processing \\
CLB -- Custom Logic Block \\
FPGA -- Field-Programmable Gate Array	\\
HDL -- Hardware Description Language \\
I$^2$C -- Inter-Integrated Circuit\\
I/O -- Input/Output \\
JTAG -- Joint Test Action Group \\
LVDS -- Low-Voltage Differential Signalling \\
MGT -- Multi-Gigabit Transceiver \\
MPPC -- Multi-Pixel Photon Counter\\
PDE -- Photon Detection Efficiency\\
PMT -- Photo Multiplier Tube\\
RAM -- Random -Access Memory \\
RTL -- Register-Transfer Level \\
SDRAM -- Synchronous Dynamic Random-Access Memory\\
SEU -- Single Event Upset \\
SFP -- Small Form-factor Pluggable transceiver \\
SiPM -- Silicon Photomultiplier \\
SNR -- Signal-to-Noise Ratio \\
SPI -- Serial Peripheral Interface \\
SQNR -- Signal-to-Quantisation-Noise Ratio \\
VHDL -- VHSIC Hardware Description Language \\
VHSIC -- Very High Speed Integrated Circuit \\
WLS -- Wavelength-Shifting 

\subsubsection{Machine Learning}
ANN -- Artificial Neural Network\\
FN -- False Negative(s) \\
FP -- False Positive(s) \\
FPR -- False Positive Rate \\
N -- Negatives \\
P -- Positives \\
PPV -- Positive Predictive Value \\
ROC -- Receiver Operating Characteristic\\
TN -- True Negative(s) \\
TP -- True Positive(s) \\
TPR -- True Positive Rate \\

\subsubsection{Other}
act. -- mass of active material														\\
BR2	-- Belgian Reactor 2													\\
CDHSW -- CERN  Dortmund Heidelberg Saclay Warsaw										\\
CERN -- European Organization for Nuclear Research										\\
CHORUS -- CERN Hybrid Oscillation Research apparatus										\\ 
CKM -- Cabibbo–Kobayashi–Maskawa 												\\
C.L. -- Confidence Level													\\
DANSS -- Detector of the reactor AntiNeutrino based on Solid Scintillator 							\\
e.g. -- exempli gratia -- for example												\\
GALLEX -- Gallium Experiment													\\
GNO -- Gallium Neutrino Observatory												\\
IBD	-- Inverse Beta-Decay													\\
ICARUS -- Imaging Cosmic And Rare Underground Signals										\\
i.d. -- id est -- that is													\\
ILL -- Institut Laue–Langevin													\\
KamLAND -- Kamioka Liquid Scintillator Antineutrino Detector									\\
K2K -- KEK to Kamioka 														\\
KARMEN -- Karlsruhe Rutherford Medium Energy Neutrino experiment								\\
KEK -- The High Energy Accelerator Research Organization									\\
LSND -- Liquid Scintillator Neutrino Detector											\\
MiniBooNE-- Mini Booster Neutrino Experiment											\\
MINOS -- Main Injector Neutrino Oscillation Search										\\
NEMENIX -- Neutrino Measurement Non-Income Experiment										\\
NOMAD --  Neutrino Oscillation Magnetic Detector										\\
NuLat -- Neutrino Lattice													\\
$pdf$ -- probability distribution function											\\
POSEIDON -- Position-Sensitive Detector of Neutrino										\\
PMNS -- Pontecorvo–Maki–Nakagawa–Sakata												\\
OPERA -- Oscillation Project with Emulsion-tracking Apparatus									\\
$r$ -- Pearson's correlation coefficient											\\
RAA -- Reactor Antineutrino Anomaly												\\
Ref -- Reference														\\
RENO -- Reactor Experiment for Neutrino Oscillation										\\
SAGE -- Soviet-American Gallium Experiment											\\
SCK\textbullet CEN -- Studiecentrum voor Kernenergie \textbullet\phantom{a}Centre d'\'{E}tude de l'\'{e}nergie Nucl\'{e}aire	\\
SM -- Standard Model														\\
SM1 -- SoLid Module 1														\\
SNO -- Sudbury Neutrino Observatory												\\
SoLid --  Search for oscillations with a Lithium-6 detector									\\
Super-K --  Super-Kamiokande													\\
T2K --  Tokai to Kamioka													\\
$\mathcal{O}$ -- order of													\\
$\sigma$ -- standard deviation

\end{appendices}
\end{document}